\documentclass[review,10pt,hyperref,dvipsnames]{elsarticle}

\usepackage{listings}
\usepackage{epsfig}
\usepackage{amsmath}
\usepackage{amsfonts}
\usepackage{amsbsy}
\usepackage{changes}
\usepackage{mathdots}
\usepackage{lipsum}
\usepackage[utf8]{inputenc}
\usepackage{enumerate}
\usepackage{amsmath}
\usepackage{amssymb}
\usepackage{mathabx}
\usepackage{centernot}
\usepackage[mathscr]{euscript}
\usepackage{url}
\usepackage{hyphenat}
\usepackage{verbatim}
\usepackage{accents}
\usepackage{graphicx}
\usepackage{subcaption}
\usepackage{listings}
\usepackage{graphicx}
\usepackage{grffile}
\usepackage{algorithm,algorithmicx}
\usepackage{algpseudocode}
\algnewcommand\algorithmicinput{\textbf{Input: }}
\algnewcommand\algorithmicoutput{\textbf{Output: }}
\usepackage{algpseudocode}
	\algnewcommand\algorithmicforeach{\textbf{for each}}
	\algdef{S}[FOR]{ForEach}[1]{\algorithmicforeach\ #1\ \algorithmicdo}
	\algnewcommand\algorithmictimes{\textbf{times}}
	\algdef{E}[REPEAT]{Times}[1]{#1\ \algorithmictimes}
	\algrenewcommand\textproc{\textsf}
	\algnewcommand{\IfThen}[2]{\State \algorithmicif\ #1\ \algorithmicthen\ #2}
\usepackage{afterpage}
\usepackage{epstopdf}
\usepackage{lineno}
\usepackage[normalem]{ulem}

\usepackage{mathtools}

\usepackage{hyperref}
\hypersetup{
  colorlinks=true,
}

\newif\ifdebug
\debugtrue 

\usepackage[utf8]{inputenc}
\usepackage{amsfonts, amsmath}
\usepackage{graphics}
    \graphicspath{{fig/}, {ext/}, {dat/}}
\usepackage{tikz}
	\usetikzlibrary{external, positioning, arrows.meta, calc, decorations.markings}
	\usetikzlibrary{decorations.markings} 
	\tikzexternaldisable
\usepackage{pgfplots}
	\pgfplotsset{compat=newest}
	\usepgfplotslibrary{colorbrewer, statistics, groupplots, fillbetween}
\usepackage{pgfplotstable}
	\newlength{\figurewidth}
	\newlength{\figureheight}
\usepackage{booktabs}
\usepackage[capitalise, noabbrev, nameinlink]{cleveref}
\usepackage{autonum} 
\usepackage{tcolorbox}
    \tcbuselibrary{most}
\usepackage{siunitx}

\ifdebug
\usepackage{lipsum}
\fi

\newcommand{\bse}{{\boldsymbol{e}}}

\newcommand{\bsell}{{\boldsymbol{\ell}}}
\newcommand{\bszero}{{\boldsymbol{0}}}

\newcommand{\bskappa}{{\boldsymbol{\kappa}}}

\newcommand{\bstau}{{\boldsymbol{\tau}}}

\newcommand{\bsDelta}{{\boldsymbol{\Delta}}}

\newcommand{\bbE}{{\mathbb{E}}}

\newcommand{\bbN}{{\mathbb{N}}}

\newcommand{\bbR}{{\mathbb{R}}}

\newcommand{\bbV}{{\mathbb{V}}}

\newcommand{\N}{{\mathbb{N}}} 

\DeclareSymbolFont{bbold}{U}{bbold}{m}{n}
\DeclareSymbolFontAlphabet{\mathbbold}{bbold}

\newcommand{\calB}{{\mathcal{B}}}

\newcommand{\calF}{{\mathcal{F}}}

\newcommand{\calO}{{\mathcal{O}}}

\newcommand{\calQ}{{\mathcal{Q}}}

\newcommand{\fraku}{{\mathfrak{u}}}

\tikzset{%
	>={Stealth[length=2mm, width=1.75mm]},
	default line/.style={%
		thick,
		line cap=round,
	},
	default dashed line/.style={%
		default line,
		dashed,
		mark options={solid}
	},
	default markers/.style={%
		mark=*,
		mark size=1.5pt,
	},=
}

\pgfkeys{%
    /pgf/number format/.cd,1000 sep={}
}

\pgfplotsset{%
	default axis/.style={%
		width=\figurewidth,
		height=\figureheight,
		major tick length={2pt},
		minor tick length={2pt},
		every tick/.style={black, line cap=round},
		ticklabel style={font=\scriptsize},
		legend style={%
		    draw=none,
		    font=\scriptsize,
		    at={(1.03, 1)},
		    anchor=north west,
		    fill=none,
		    legend cell align=left
		},
		cycle list/Set1,
		axis on top,
	},
	default error bar/.style={%
		only marks,
		mark size=1.25pt,
		line cap=round,
		error bars/.cd,
		y dir=both,
		y explicit,
	}
}

\tcbset{%
	outer arc=2pt,
	arc=1pt,
	attach boxed title to top left={yshift=-3mm, xshift=3mm},
	enhanced,
	boxed title style={
		top=3pt,
		bottom=3pt,
	},
  	fonttitle=\sffamily,
	top=12pt,
}

\newtcolorbox[auto counter, number within=section]{todobox}[1][]{%
	colframe=NavyBlue,
	colback=NavyBlue!20,
	boxed title style={
		colback=NavyBlue,
	},
  	title=TO DO,
}

\newtcolorbox[auto counter, number within=section]{infobox}[1][]{%
	colframe=OliveGreen,
	colback=OliveGreen!20,
	boxed title style={
		colback=OliveGreen,
	},
  	title=INFO,
}

\newcommand{\drawsquare}[3]{\draw[semithick,black,fill=#3,rounded corners=1pt] (#1,#2)--(#1,#2+1)--(#1+1,#2+1)--(#1+1,#2)--cycle;}
\newcommand{\drawthicksquare}[3]{\draw[line width=1.5pt,black,fill=#3,rounded corners=1pt] (#1,#2)--(#1,#2+1)--(#1+1,#2+1)--(#1+1,#2)--cycle;}

\colorlet{active}{WildStrawberry}
\colorlet{maximum}{NavyBlue}
\colorlet{old}{white!90!black}

\pgfplotsset{%
	index set/.style={%
		width=\figurewidth,
		height=\figureheight,
        xmin={-0.1},
        ymin={-0.1},
		major tick length={0pt},
        axis line style={{ultra thin, draw opacity=0}},
        axis equal,
        xlabel={$\ell_1$},
        ylabel={$\ell_2$},
        clip=false
    },
    3 by 3/.style={%
        xmax={4.1},
        ymax={4.1},
        xlabel={$\ell_1$},
        xtick={0.5, 1.5, 2.5, 3.5},
        xticklabels={0, 1, 2, 3},
        ytick={0.5, 1.5, 2.5, 3.5},
        yticklabels={0, 1, 2, 3},
        ylabel={$\ell_2$},
    },
    4 by 3/.style={%
        xmax={5.1},
        ymax={4.1},
        xlabel={$\ell_1$},
        xtick={0.5, 1.5, 2.5, 3.5, 4.5},
        xticklabels={0, 1, 2, 3, 4},
        ytick={0.5, 1.5, 2.5, 3.5},
        yticklabels={0, 1, 2, 3},
        ylabel={$\ell_2$},
    }
}
\usepackage[capitalise, noabbrev, nameinlink]{cleveref}
\everymath{\displaystyle}

\newcommand\E[1]{{\mathbb{E}\left[#1\right]}}
\newcommand\V[1]{{\mathbb{V}\left[#1\right]}}
\newcommand\MSE[1]{{\text{MSE}\left(#1\right)}}

\newcommand\cost[1]{{\text{cost}\left(#1\right)}}

\usepackage{amsthm}

\usepackage{smartdiagram} 
\usesmartdiagramlibrary{additions}
\usetikzlibrary{shapes.geometric,arrows,quotes}
\usepackage{pgf}
\usepackage{tikz}

\journal{Elsevier}

\begin{document}

\hypersetup{
  linkcolor=RedViolet,
  urlcolor=OliveGreen,
  citecolor=PineGreen
}

\begin{frontmatter}

\title{Multi-fidelity microstructure-induced uncertainty quantification by advanced Monte Carlo methods}



\author[add4]{Anh Tran\corref{mycorrespondingauthor}}
\author[add9]{Pieterjan Robbe}
\author[add10]{Hojun Lim}
\cortext[mycorrespondingauthor]{Corresponding author: anhtran@sandia.gov. A.T. and P.R. contributed equally to the paper.}

\address[add4]{Scientific Machine Learning Department, Sandia National Laboratories, Albuquerque, NM 87123}

\address[add9]{Plasma and Reacting Flow Science Department, Sandia National Laboratories, Livermore, CA 94550}

\address[add10]{Computational Materials and Data Science, Sandia National Laboratories, Albuquerque, NM 87123}

\begin{abstract}

Quantifying uncertainty associated with the microstructure variation of a material can be a computationally daunting task, especially when dealing with advanced constitutive models and fine mesh resolutions in the crystal plasticity finite element method (CPFEM). Numerous studies have been conducted regarding the sensitivity of material properties and performance to the mesh resolution and choice of constitutive model. However, a unified approach that accounts for various fidelity parameters, such as mesh resolutions, integration time-steps, and constitutive models simultaneously is currently lacking. This paper proposes a novel uncertainty quantification (UQ) approach for computing the properties and performance of homogenized materials using CPFEM, that exploits a hierarchy of approximations with different levels of fidelity. In particular, we illustrate how multi-level sampling methods, such as multi-level Monte Carlo (MLMC) and multi-index Monte Carlo (MIMC), can be applied to assess the impact of variations in the microstructure of polycrystalline materials on the predictions of homogenized materials properties. We show that by adaptively exploiting the fidelity hierarchy, we can significantly reduce the number of microstructures required to reach a certain prescribed accuracy. Finally, we show how our approach can be extended to a multi-fidelity framework, where we allow the underlying constitutive model to be chosen from either a phenomenological plasticity model or a dislocation-density-based model.

\end{abstract}


\end{frontmatter}


\clearpage
\tableofcontents
\clearpage



\section{Introduction}


Uncertainty quantification (UQ) plays a major role in verifying and validating many integrated computational materials engineering (ICME) models. Within the materials sciences, where the process-structure-property-performance bridge is well-established, 
quantifying uncertainty associated with microstructures is one of the most important tasks in order to predict the variability in material properties and material performance. 
The properties and performance of homogenized materials in the structure-property relationship can be computed using the crystal plasticity finite element method (CPFEM).
CPFEM considers grain scale microstructure by explicitly modeling discrete grains and their slip systems based on dislocation slip.
In CPFEM, the microstructure of a material is defined in terms of a representative volume element (RVE), that can be thought of as a stochastic sample of the entire polycrystalline microstructure. A CPFEM study then typically involves simulating multiple realizations of such an RVE.
In this paper, we propose a unified 
framework for CPFEM that exploits a hierarchy of models with different fidelity, based on multi-level Monte Carlo (MLMC) and multi-index Monte Carlo (MIMC) methods. 
As a result, the number of RVEs required to run CPFEM reduces significantly, in effect lowering the computational cost required to determine the material properties and performance.

Microstructures are known to exhibit inherent randomness both spatially and orientationally, often requiring high-dimensional representations in terms of pixels (in 2D images) and volumetric pixels or voxels (in 3D volumes).
The  variability in microstructure mainly contributes to the aleatory uncertainty of the prediction, whereas the numerical approximations in the ICME models bridging the structure-property relationship mainly contribute to the epistemic uncertainty.
This manuscript is mainly concerned with rigorously addressing the aleatory uncertainty that is induced from the microstructure perspective, while acknowledging that the epistemic uncertainty work is also addressed elsewhere~\cite{tran2022microstructure}.

In the process-structure-property-performance linkage, one tends to think of ICME models as forward models or functions that map from one space to another, for example, from process to structure or from structure to property or from process to property.
Most computational models, including ICME models, typically posses a multi-fidelity hierarchy, defined in terms of a computational accuracy versus cost trade-off. 
One of the most obvious examples is the mesh size used to represent the geometry of the microstructure RVE. The coarse-mesh CPFEM is computationally cheaper and can be thought of as a low-fidelity approximation, whereas the fine-mesh CPFEM is computationally expensive and can be regarded as a high-fidelity approximation.
Another example of a multi-fidelity hierarchy is the constitutive model in the CPFEM method: a phenomenological constitutive model can be considered as the low-fidelity approximation and more physically-based models such as a dislocation-density-based model can be thought of as a high-fidelity approximation.
Numerous mesh sensitivity analysis studies have been conducted in the literature, but none has been able to construct an approach that concurrently unifies the refinement of both mesh size and constitutive model. Furthermore, the results of these studies often depend on the material system, as well as on the numerical solver being used.
Our work is the first to rigorously address the computation of structure-homogenized material properties with CPFEM, using both a multi-fidelity approach for the constitutive model and a multi-resolution approach for the RVE simultaneously. Our method is based on an adaptive extension of the multi-level Monte Carlo and multi-index Monte Carlo sampling methods ~\cite{giles2015multilevel,haji2016multi,robbe2016dimension}. In the single-fidelity setting, our method reduces to the classic Monte Carlo (MC) method, also known as the ``ensemble of microstructure RVEs'' approach in the field of CPFEM. Therefore, this work can be seen as a generalization towards multi-fidelity CPFEM, using advanced multi-fidelity sampling methods. Such a multi-fidelity CPFEM could exploit, for example, the fidelity of the constitutive model, the integration time-step size, the order of a numerical integrator, the mesh size ($h$-refinement), and polynomial order of the element ($p$-refinement).

Given the critical importance of UQ for a wide variety of problems in materials science, several frameworks have been developed to provide robust predictions under uncertainty, see e.g., \cite{panchal2013key,mcdowell2007simulation,kalidindi2016vision}.
Comprehensive reviews of UQ applications in ICME-based simulations can be found in Honarmandi and Arr{\'o}yave \cite{honarmandi2020uncertainty}, Gabriel et al.~\cite{gabriel2020uncertainty}, and Acar~\cite{acar2021recent}.
For example,
Zhao et al.~\cite{zhao2022quantifying} incorporated measurement and parametric uncertainty to quantify the uncertainty of critical resolved \textcolor{black}{shear} stress for hexagonal close-packed (HCP) Ti alloys from nano-indentation.
Lim et al.~\cite{lim2019investigating} investigated the mesh sensitivity and polycrystalline RVE, where initial textures, hardening models, and boundary conditions are uncertain. 
\textcolor{black}{Park et al~\cite{park2021impact} investigated the effects of anisotropy, different hardening models, and grain morphology in aluminum 7079 alloy.}
Tran and Wildey~\cite{tran2020solving} applied data-consistent inversion method to infer a distribution of microstructure features from a distribution of yield stress, where the push-forward density map via a heteroscedastic Gaussian process approximation is consistent with the imposed yield stress density.
Kotha et al.~\cite{kotha2019parametrically1,kotha2019parametrically2,kotha2020uncertainty1,kotha2020uncertainty2} developed uncertainty-quantified, parametrically homogenized constitutive models to capture uncertainty in microstructure-dependent stress-strain curve, as well as stochastic yield surface, which has been broadly applied for modeling multi-scale fatigue crack nucleation in Ti alloys~\cite{ozturk2019two,ozturk2019parametrically} and for single-crystal Ni-based superalloys with support vector regression as an underlying machine learning model~\cite{weber2020machine}.
Sedighiani et al.~\cite{sedighiani2020efficient,sedighiani2022determination} applied genetic algorithm and polynomial approximation to various constitutive models, including phenomenological and dislocation-density-based models.
Tran et al.~\cite{tran2019quantifying} applied stochastic collocation (SC) method to quantify uncertainty for dendrite morphology and growth via phase-field model.
Acar et al.~\cite{acar2017stochastic} proposed a linear programming approach to maximize a mean of materials properties under the assumption of Gaussian distribution for both inputs and outputs.
Fernadez et al.~\cite{fernandez2018estimating} utilized Bayesian inference to quantify the uncertainty in stress-strain curves, where model parameters are treated as random variables.
Tallman et al.~\cite{tallman2019gaussian,tallman2020uncertainty} applied Gaussian process regression and the Materials Knowledge System framework to predict a set of homogenized materials properties with uncertainty from a distribution function for crystallograph\textcolor{red}{ic} orientations and textures.
The inductive design exploration method (IDEM) \cite{ellis2017application,mcdowell2009integrated,choi2008inductive} has been introduced as a materials design methodology to identify feasible and robust design for microstructure features, which has been broadly applied to many practical problems.
Zhang~\cite{zhang2020modern} provided a comprehensive mathematical review of advanced MC methods.
Chatterjee et al.~\cite{chatterjee2018prediction} employed a classical MC estimator to statistically study the tensile stiffness and strength of Ti-6Al-4V.
Acar and Sundararaghavan~\cite{acar2017uncertainty1,acar2017uncertainty2} quantified the uncertainty of materials properties with respect to measured pole figures and experimental variations, respectively.


In the literature, the most common method used to study microstructure-induced material properties is to consider an ensemble of micro-structure realizations, $\{\omega^{(n)}\}_{n=1}^N$, sampled from the space of microstructures $\Omega$. Two microstructures, $\omega^{(1)}$ and $\omega^{(2)}$, are said to be statistically equivalent if they are independently and identically sampled from the same space $\Omega$ using the same probability law.
The ensemble of microstructure RVEs approach is therefore mathematically equivalent to the classical MC estimator, where the structure-property map, denoted as $Q(\omega)$ and typically evaluated by running CPFEM, is fixed, and where the aleatory uncertainty associated with microstructure variation can be represented by samples from $\Omega$.
The MC method is a popular approach, because its efficiency in terms of the required number of RVE compositions does not depend on the dimensionality of the input (i.e., the number of input parameters). However, this dimension-independence comes at a price, since typically many RVE evaluations are required to reach a certain prescribed accuracy.
While the classic MC estimator 
is theoretically an unbiased estimator of $Q$, this is no longer the case if the approximation necessarily involved in numerically evaluating the structure-property map is considered. By leveraging a multi-fidelity hierarchy of these numerical approximations $\{Q_\ell\}_{\ell=0}^L$ to $Q$, the computational cost of the MC method can be reduced significantly. In particular, the high-fidelity approximation $Q_L$ for $Q$ can be replaced by a telescoping sum of canceling differences between successive fidelity levels $Q_\ell$ and $Q_{\ell-1}$, exploiting the linearity of the expectation operator. Replacing the single, expensive MC estimator for the high-fidelity approximation by multiple inexpensive MC estimators for these differences, an overall reduction of computational cost is achieved. This is the idea of the MLMC and MIMC sampling methods. Using the results of~\cite{robbe2016dimension}, we illustrate how such a multi-fidelity hierarchy can be constructed adaptively in the context of CPFEM.

The remaining of the paper is organized as follows.
Section~\ref{sec:mc_techniques} reviews the classical MC, MLMC, and MIMC methods, and outlines the adaptive MIMC method used in this study.
Section~\ref{sec:BackgroundCPFEM} provides a preliminary background for constitutive models in CPFEM.
Section~\ref{sec:Methodology} describes the integrated workflow coupling \texttt{DREAM.3D}~\cite{groeber2014dream} and \texttt{DAMASK}~\cite{roters2019damask}.
Section~\ref{sec:MLMC_CaseStudy} presents the first case study for $\alpha$-Ti with MLMC, where multiple mesh resolutions are considered.
Section~\ref{sec:MIMC_CaseStudy} presents the second case study for Al with MIMC, where multiple constitutive models (phenomenological and dislocation-density-based) and multiple mesh resolutions are considered simultaneously.
Section~\ref{sec:Discussion} discusses and Section~\ref{sec:Conclusion} concludes the paper, respectively.

\section{Monte Carlo sampling methods}
\label{sec:mc_techniques}

Multi-level and multi-index sampling methods leverage the correlation in the output of multiple models in a given model hierarchy, in order to reduce the stochastic error in the prediction of statistical quantities, such as the mean or variance of the model output. This reduction in error often leads to a significant reduction of the computational cost, as the number of model evaluations required to achieve a similar error can be reduced by several orders of magnitude. In this section, we review multi-level and multi-index sampling methods, and illustrate how these methods can be adapted to the CPFEM setting. We start by reviewing the classic approach of using ensemble averages of stochastic volume elements (SVEs) to predict mean values of the desired material property. Next, we discuss how this approach can be extended to a multi-level sampling approach, using the mesh resolution as refinement parameter. Finally, we show how the multi-level sampling approach can be extended to a multi-index sampling approach, using both the mesh resolution and the underlying constitutive model as refinement parameters. We also discuss how appropriate combinations of mesh resolution and model fidelity can be selected from a given collection of models using a greedy adaptive strategy.

\subsection{Notation}



For a given mesh resolution and a given constitutive model, we denote the uncertain microstructure of the material under consideration by $\omega \in \Omega$.
The space $\Omega$ represents 
the collection of all possible discretized microstructures $\omega$, where $\omega$ is independently and identically (i.i.d.) sampled according to a uniform law from the space of all available microstructures $\Omega$. 
The i.i.d assumption constitutes the basis for the statistical equivalence of different $\omega$s drawn from the same distribution 
of all microstructures.
In practice, $\omega$ is constructed by solving a microstructure reconstruction problem, which often leads to another optimization problem in a pure computational fashion.

Let the map from microstructure space to the homogenized material property be denoted by $Q(\omega): \Omega \to D \subseteq \bbR$.
Because the underlying microstructure $\omega$ is uncertain, 
so is any quantity derived from that same microstructure. Hence, we explicitly denote the dependency of the quantity of interest on the outcome $\omega$, i.e., $Q(\omega)$ is a random variable. For the remainder of this paper, we will be interested in computing the first-order moment or expected value of the quantity of interest $Q(\omega)$, defined as
\begin{equation}
\label{eq:integral}
    \E{Q(\omega)} \coloneqq \int_{\Omega} Q(\omega) \mathrm{d}\omega 
\end{equation}

Finally, because every material property $Q$ is based on an underlying microstructure $\omega$, which is itself associated with a certain given mesh resolution, we will use the notation $Q_L(\omega)$ to denote that the material property is obtained from an approximation of the microstructure with mesh resolution level $L$.

\subsection{The Monte Carlo method}
\label{sec:mc}

Given an ensemble of i.i.d. microstructure RVEs $\{\omega^{(n)}\}_{n=1}^N$ with corresponding predictions for the material property of interest $\{Q_L(\omega^{(n)})\}_{n=1}^N$, we can approximate~\eqref{eq:integral} by the average
\begin{equation}
\label{eq:mc_method}
\calQ_{\text{MC}} \coloneqq \frac{1}{N} \sum_{n=1}^{N} Q_L(\omega^{(n)}).
\end{equation}
The ensemble average approach in~\eqref{eq:mc_method}, also known as the Monte Carlo (MC) method, is widely used in the CPFEM literature, see, e.g., \cite{paulson2017reduced,teferra2018random,paulson2018data}. In practice, the microstructure  $\omega^{(n)}$ is often obtained from solving a microstructure reconstruction problem, which in turn is often formulated as an optimization problem. We refer to Groeber et al.~\cite{groeber2008framework1,groeber2008framework2}, Bostanabad et al.~\cite{bostanabad2018computational} and Torquato~\cite{torquato2002statistical} for comprehensive reviews of computing microstructure RVEs. CPFEM is then deployed repetitively to evaluate $Q_L(\omega^{(n)})$ for each microstructure $\omega^{(n)}$, $n = 1, 2, \ldots, N$.

It is natural to propose the average of an ensemble of material properties extracted from the microstructures $\{\omega^{(n)}\}_{n=1}^N$ to approximate the expected value in~\eqref{eq:integral}. Since the sequence \textcolor{black}{of} microstructure RVEs are i.i.d.\, we have that the expected value $\bbE[Q_L(\omega^{(1)})] = \dots = \bbE[Q_L(\omega^{(N)})] = \bbE[Q_L]$, and the strong law of large numbers guarantees that $\calQ_{\text{MC}} \to \E{Q_L}$ almost surely as the number of realizations $N$ goes  to infinity, see \cite{robert2013monte}.

There are two sources of error in the MC estimator in~\eqref{eq:mc_method}: a stochastic error, present because we approximate the expected value by an average, and a \emph{bias}, present because samples of $Q(\omega)$ are approximated by samples of $Q_L(\omega)$. These two contributions become apparent in the expression for the mean square error (MSE) of the MC estimator. We have
\begin{align}
    \MSE{\calQ_{\text{MC}}} &\coloneqq \E{ (\calQ_{\text{MC}} - \E{Q})^2 } \\
    &= \E{((\calQ_{\text{MC}} - \E{\calQ_{\text{MC}}}) + (\E{\calQ_{\text{MC}}} - \E{Q}))^2} \\
    &= \E{(\calQ_{\text{MC}} - \E{\calQ_{\text{MC}}})^2} + (\E{Q_L - Q})^2 \\
    &= \V{\calQ_{\text{MC}}} + (\E{Q_L - Q})^2 ,\label{eq:mc_mse}
\end{align}
where the cross-product term vanishes because the MC estimator is an unbiased estimator for $Q_L$, i.e., $\E{\calQ_{\text{MC}}} = \E{Q_L}$. The first term in~\eqref{eq:mc_mse} is the variance of the estimator and represents the stochastic error. Because we assume the ensemble is uncorrelated, the variance can be written as
\begin{equation}
\V{\calQ_{\text{MC}}} = \frac{1}{N^2} \sum_{n=1}^{N} \V{Q_L} = \frac{\V{Q_L}}{N}.
\end{equation}
The variance decays as $\calO(1/N)$ and can be reduced by increasing the number of microstructure RVEs $N$. The second term in~\eqref{eq:mc_mse} is the square of the bias. It can be reduced by increasing the level of resolution $L$, i.e., by decreasing the mesh size.

If we require an MSE smaller than or equal to $\varepsilon^2$, a sufficient condition is
\begin{equation}
    \frac{\V{Q_L}}{N} \leq \frac{\varepsilon^2}{2} \quad \text{and} \quad | \E{Q_L - Q} | \leq \frac{\varepsilon}{\sqrt{2}}.
\end{equation}
Hence, the number of microstructure instances $N$ should increase as $\calO(\varepsilon^{-2})$. Assuming that the cost of a single model evaluation is $C_L$, we can express the total computational cost of the MC estimator in~\eqref{eq:mc_method} as
\begin{equation}
    \mathrm{cost}(\calQ_{\text{MC}}) = N C_L.
\end{equation}
Thus, the computational cost of the MC estimator increases as $\calO(\varepsilon^{-2})$.

\subsection{The Multi-level Monte Carlo (MLMC) method}
\label{sec:mlmc}

\begin{figure}[!htbp]
\centering
\begin{subfigure}[b]{0.175\textwidth}
\includegraphics[width=\textwidth]{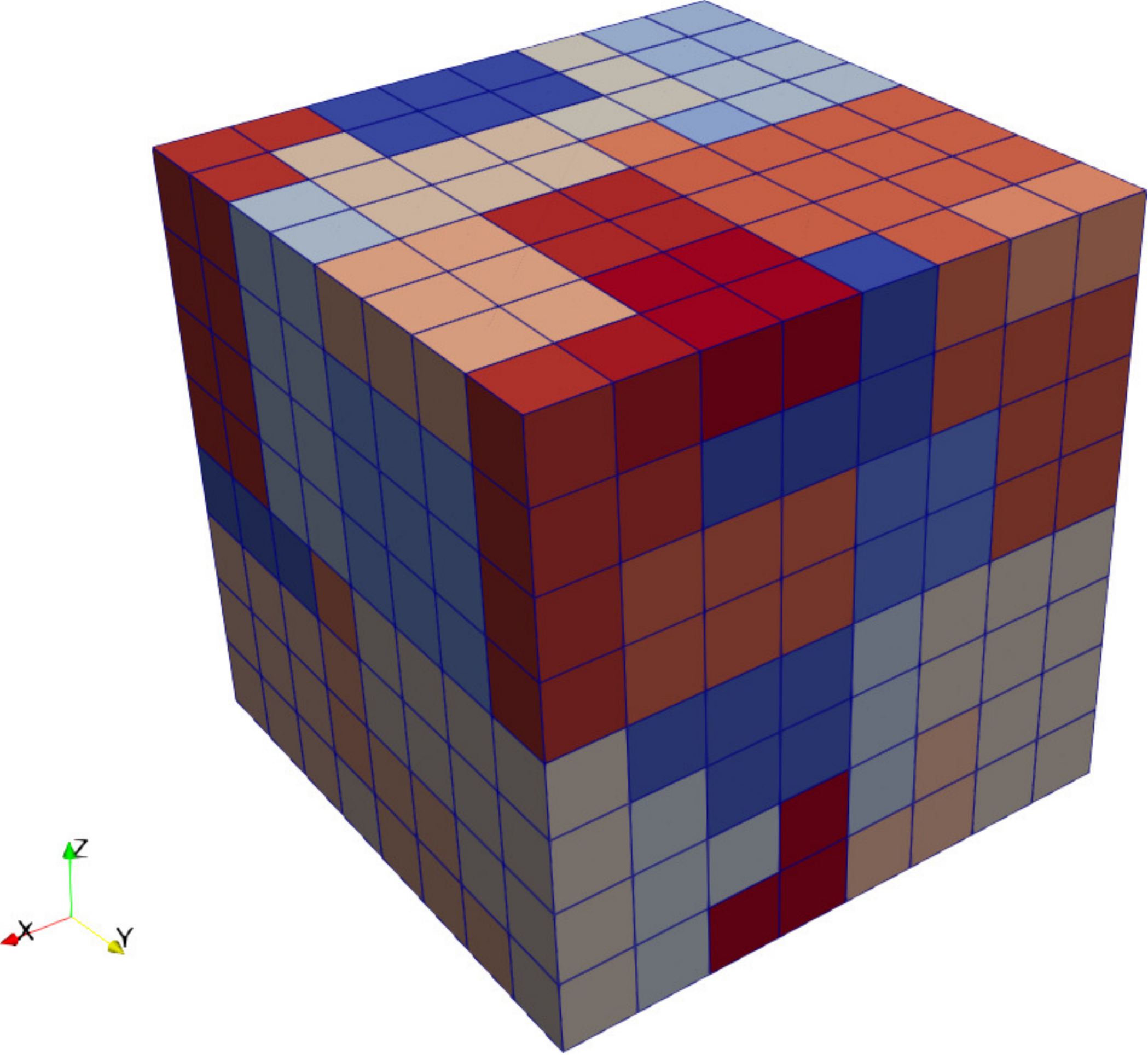}
\caption{$8 \times 8 \times 8$}
\label{fig:sve8}
\end{subfigure}
\begin{subfigure}[b]{0.175\textwidth}
\includegraphics[width=\textwidth]{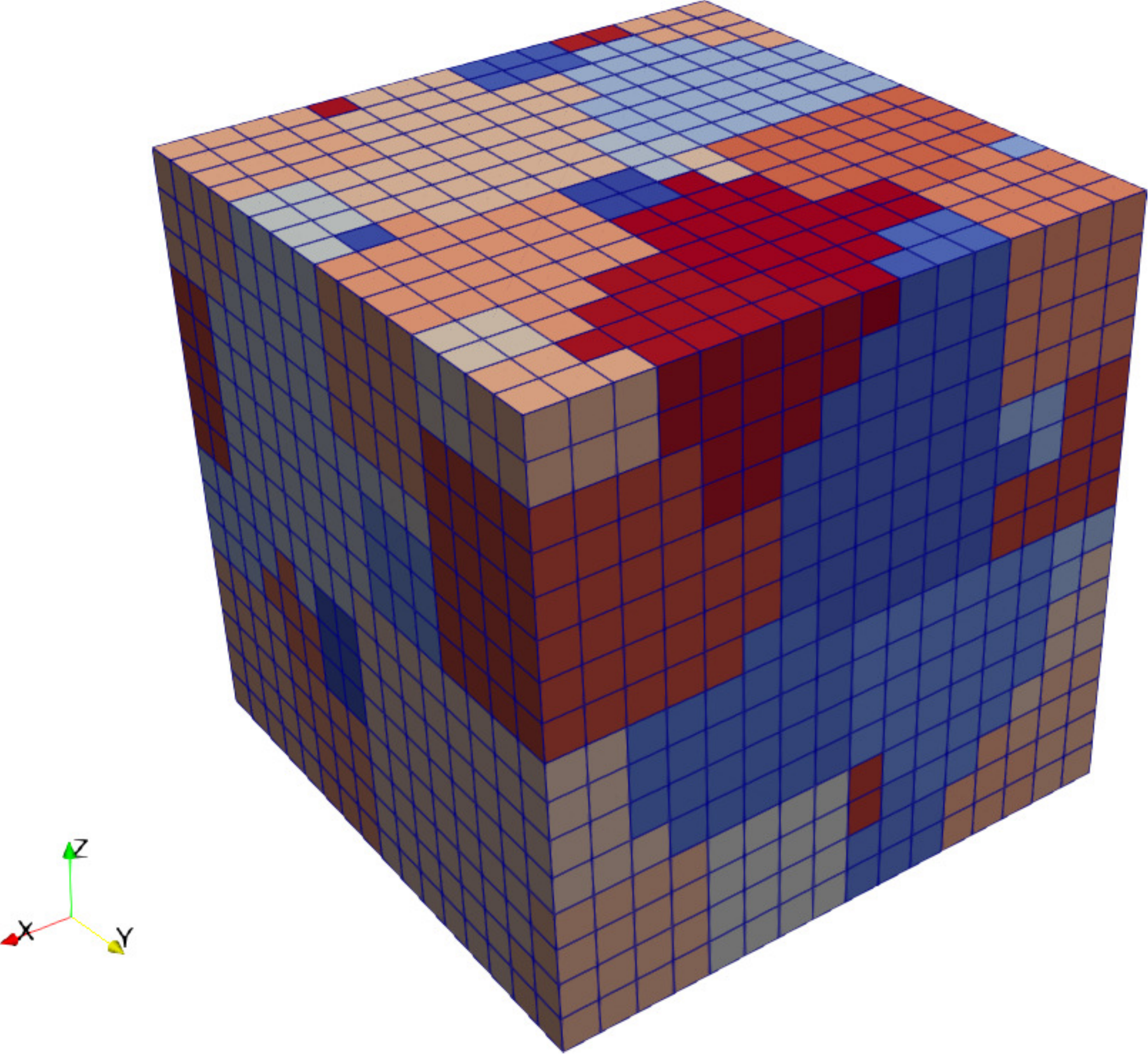}
\caption{$16 \times 16 \times 16$}
\end{subfigure}
\begin{subfigure}[b]{0.175\textwidth}
\includegraphics[width=\textwidth]{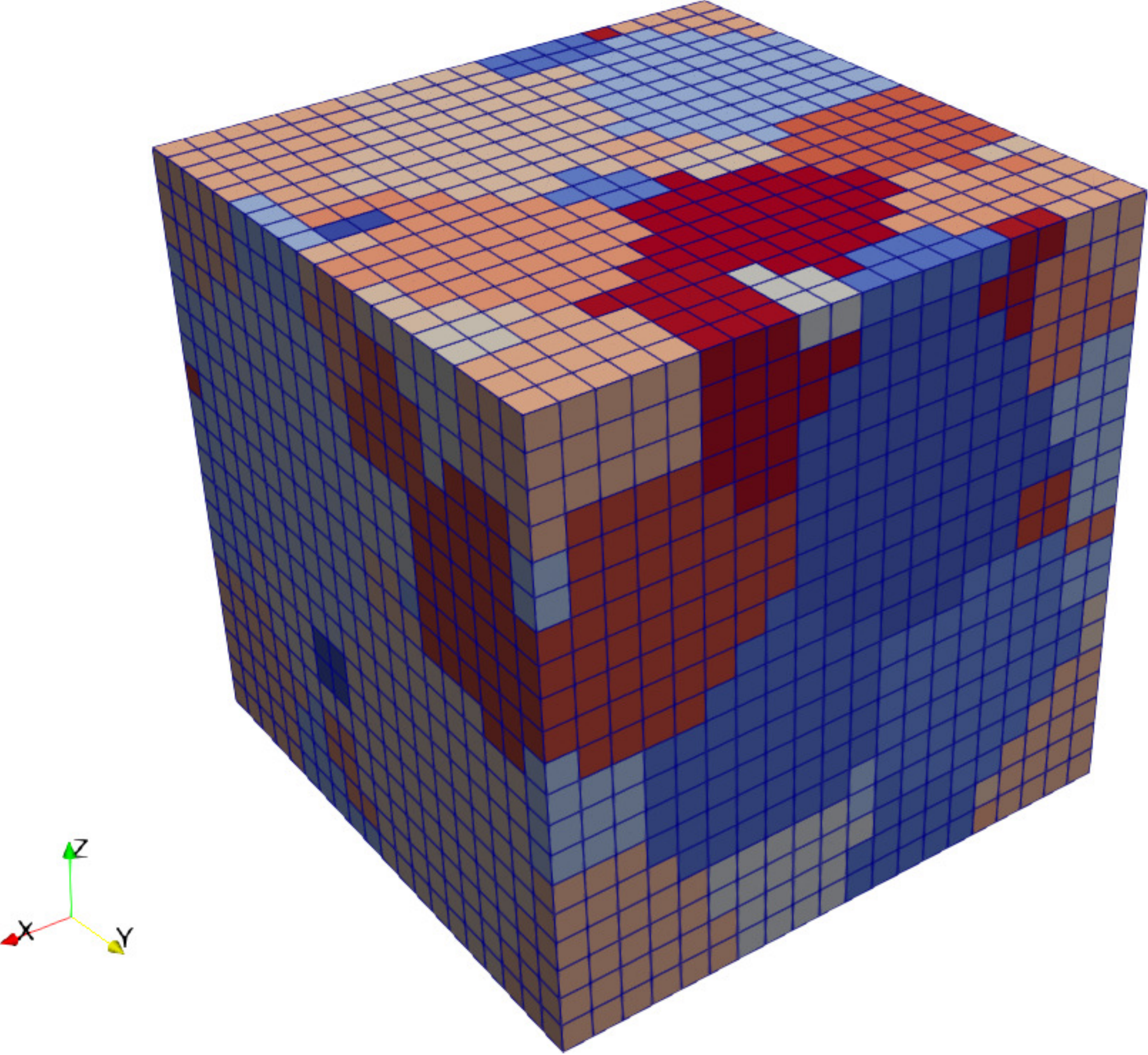}
\caption{$20 \times 20 \times 20$}
\end{subfigure}
\begin{subfigure}[b]{0.175\textwidth}
\includegraphics[width=\textwidth]{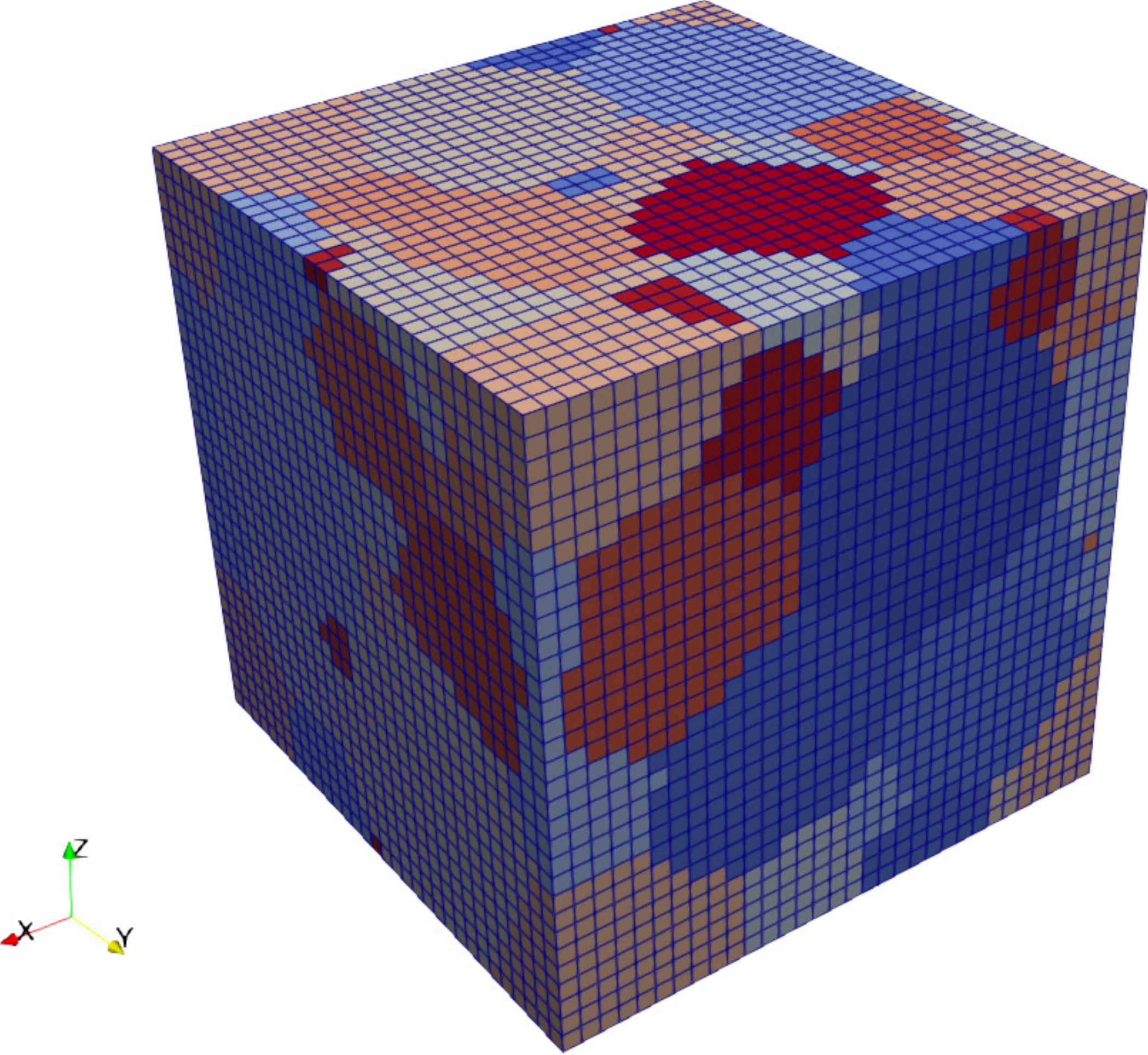}
\caption{$32 \times 32 \times 32$}
\end{subfigure}
\begin{subfigure}[b]{0.175\textwidth}
\includegraphics[width=\textwidth]{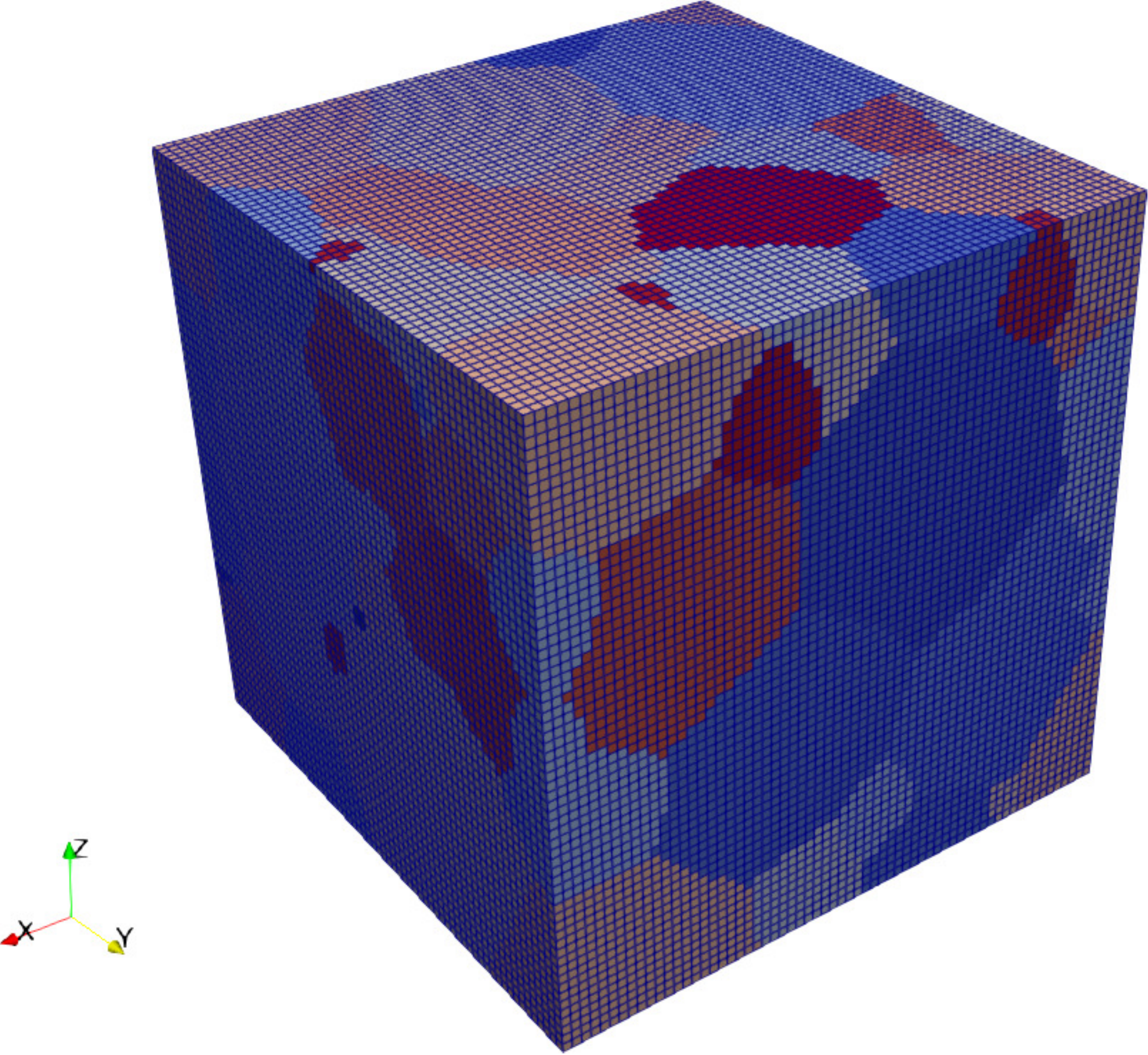}
\caption{$64 \times 64 \times 64$}
\label{fig:sve64}
\end{subfigure}
\begin{subfigure}[htbp]{1.0\textwidth}
\centering
\newcommand{\cuboid}[3]{%
    \begin{scope}[shift={#1}, scale=0.75]
        \draw[rounded corners = 0.1pt] (0,2) -- (2,2) -- (2,0) -- (0,0) -- (0,2) -- ++(135:0.71) -- ++(2,0) -- (2,2) (0,0) -- ++(135:0.71) -- ++ (0,2); 
        \node[anchor=center] at (1,1) {#2};
        \node[anchor=north] at (1,0) {#3};

    \end{scope}
}

\begin{tikzpicture}[
    every path/.style={semithick, line cap=round},
    every node/.style={inner sep=0pt, outer sep=3pt, execute at end node=\strut}
]
\foreach \mesh [count=\i] in {8, 16, 20, 32, 64}{%
    \pgfmathtruncatemacro{\k}{\i - 1}
    \cuboid{(2*\k, 0)}{$\mesh^3$}{$\ell = \k$}
}
\draw[->] (0,-1) -- node[pos=0.5, anchor=north, outer sep=6pt] {$\ell$} ++(9,0) node[anchor=north, outer sep=6pt] {mesh resolution};
\end{tikzpicture}
\caption{Schematic overview of the multi-level hierarchy obtained by varying the mesh resolution in the model.}
\label{fig:mlmc_hierarchy}
\end{subfigure}
\end{figure}

The central idea in MLMC sampling is that we do not sample from a single approximation $Q_L$ for the quantity of interest, but instead compute samples on a hierarchy of approximations $\{Q_\ell\}_{\ell=0}^L$ for the quantity of interest $Q$. In the context of CPFEM, this hierarchy corresponds to an approximation for the material parameter on a sequence of meshes with increasing resolution levels, where level $\ell=0$ corresponds to the cheapest approximation with the coarsest mesh size, and level $\ell = L$ corresponds to the most expensive approximation with the finest mesh size. An illustration of such a multi-level hierarchy is shown in \cref{fig:sve8} to \cref{fig:sve64}, and schematically in \cref{fig:mlmc_hierarchy}.

Because the expected value is a linear operator, we have that
\begin{equation}
\label{eq:tel_sum}
    \E{Q_L} = \sum_{\ell=1}^L \E{Q_\ell - Q_{\ell-1}} + \E{Q_0} = \sum_{\ell=0}^L \E{\Delta Q_\ell}
\end{equation}
where
\begin{equation}
\label{eq:mlmc_diff}
    \Delta Q_\ell \coloneqq \begin{cases}
    Q_{\ell} - Q_{\ell-1} & \text{for } \ell > 0 \\
    Q_\ell & \text{for } \ell = 0 \\
    \end{cases}.
\end{equation}
Using an independent MC estimator for each of the $L+1$ terms in the right-hand side of~\eqref{eq:tel_sum}, we obtain the MLMC estimator
\begin{equation}
\label{eq:mlmc}
    \calQ_{\text{MLMC}} \coloneqq \sum_{\ell=0}^L \frac{1}{N_\ell} \sum_{n=1}^{N_\ell} \Delta Q_\ell(\omega^{(n)}).
\end{equation}
In effect, this means that we use an ensemble of i.i.d.\ microstructure RVEs $\{\omega^{(n)}\}_{n=1}^{N_\ell}$ on each level $\ell=0, 1, \ldots, L$ to estimate the expected values on the right-hand side of~\eqref{eq:tel_sum}, where we assume that the microstructure instances on each level are mutually independent.

The MLMC estimator in~\eqref{eq:mlmc} is still an unbiased estimator for $\E{Q_L}$, i.e.,
\begin{equation}
    \E{\calQ_{\text{MLMC}}} = \sum_{\ell=0}^L \frac{1}{N_\ell} \sum_{n=1}^{N_\ell} \E{\Delta Q_\ell} = \sum_{\ell=0}^L \E{\Delta Q_\ell} = \E{\Delta Q_L}
\end{equation}
and its variance is given by
\begin{equation}
    \V{\calQ_{\text{MLMC}}} = \sum_{\ell=0}^L \frac{1}{N_\ell^2} \sum_{n=1}^{N_\ell} \V{\Delta Q_\ell} = \sum_{\ell=0}^L \frac{\V{\Delta Q_\ell}}{N_\ell}.
\end{equation}

Expanding the MSE as in~\eqref{eq:mc_mse} now yields
\begin{equation}
\label{eq:mlmc_mse}
    \MSE{\calQ_{\text{MLMC}}} = \V{\calQ_{\text{MLMC}}} + (\E{Q_L - Q})^2 = \sum_{\ell=0}^L \frac{\V{\Delta Q_\ell}}{N_\ell} + (\E{Q_L - Q})^2.
\end{equation}
Again, the MSE consists of two terms: the variance of the estimator and the square of the bias. Note that the bias of the MLMC estimator is the same as the bias of the MC estimator.

A crucial observation is that, instead of estimating the expected value $\E{Q_\ell}$ directly on level $\ell$, it is much cheaper to estimate the expected value of the difference $\E{\Delta Q_\ell}$, if the random variables $Q_\ell$ and $Q_{\ell-1}$ are strongly positively correlated, i.e.,
\begin{align}
    \V{\Delta Q_\ell} &= \bbV[Q_\ell - Q_{\ell-1}] \\
    &= \V{Q_\ell} + \V{Q_{\ell-1}} - 2 \text{cov}(Q_\ell, Q_{\ell-1}) \\
    &\ll \V{Q_\ell} + \V{Q_{\ell-1}}
\end{align}
where $\text{cov}(Q_\ell, Q_{\ell-1}) = \rho_{\ell, \ell-1} \sqrt{\V{Q_\ell} \V{Q_{\ell-1}}}$ is the covariance between $Q_\ell$ and $Q_{\ell-1}$ and $\rho_{\ell, \ell-1}$ is the Pearson correlation coefficient. In order to ensure this strong correlation, it is important to note that the difference $\Delta Q_\ell(\omega_{\ell}^{(n)})$ in~\eqref{eq:mlmc} is evaluated for the \textcolor{black}{same input microstructure} $\omega_{\ell}^{(n)}$. In the context of CPFEM, this means that the difference is computed from the material parameter prediction for the \emph{same} underlying RVE, but on two different mesh sizes in the hierarchy.

As the level parameter $\ell \rightarrow \infty$, we expect the approximations $Q_\ell$ to converge towards the true quantity of interest $Q$ in mean square sense, i.e., $\V{\Delta Q_\ell} \rightarrow 0$ as $\ell \rightarrow \infty$. In effect, this means that fewer model evaluations are required in the successive MC estimators for the difference $\Delta Q_\ell$ with increasing $\ell$. Under this assumption, we find that most samples will be taken on level $\ell=0$, where model evaluations are cheap, and fewer samples are required on the higher levels, where model evaluations are increasingly more expensive. Often, only a handful of samples with the highest resolution level are required. Compare this to the MC method outlined in \cref{sec:mc}, where all samples are taken on the same high-resolution level.

If we require an MSE smaller than or equal to $\varepsilon^2$, a sufficient condition is
\begin{equation}
\label{eq:mlmc_constraint}
    \sum_{\ell=0}^L \frac{\V{\Delta Q_\ell}}{N_\ell} \leq \frac{\varepsilon^2}{2} \quad \text{and} \quad | \E{Q_L - Q} | \leq \frac{\varepsilon}{\sqrt{2}}.
\end{equation}
An expression for the required number of SVEs $N_\ell$ on each level $\ell = 0, 1, \ldots, L$ can be obtained by minimizing the total cost of the MLMC estimator, taking into account the above constraint on the variance of the estimator. The total cost of the MLMC estimator can be expressed as
\begin{equation}
    \mathrm{cost}(\calQ_{\text{MLMC}}) = \sum_{\ell=0}^L N_\ell \Delta C_\ell,
\end{equation}
where $\Delta C_\ell$ denotes the cost to compute a sample of the multi-level difference $\Delta Q_\ell$. This yields
\begin{equation}
\label{eq:optimalSamples_mlmc}
N_{\ell} = \frac{2}{\varepsilon^2}\sqrt{\frac{\V{\Delta Q_\ell}}{\Delta C_{\ell}}} \left( \sum_{\ell=0}^L \sqrt{\V{\Delta Q_{\ell}} \Delta C_{\ell}} \right),
\end{equation}
see~\cite{giles2015multilevel} for details on the derivation. In practice, the number of samples $N_\ell$ in~\eqref{eq:optimalSamples_mlmc} should be rounded up to the nearest integer. This increases the cost of the estimator by at most one sample on each level.

In~\cite{cliffe2011multilevel}, a theoretical bound for the asymptotic cost complexity of the MLMC estimator is provided. Assuming
\begin{align}
| \E{\Delta \calQ_{\ell}} | &\leq c_1 \; 2^{-\alpha \ell}, \label{eq:C1}\tag{C1} \\
\V{\Delta \calQ_{\ell}} &\leq c_2 \; 2^{-\beta \ell} \text{ and} \label{eq:C2}\tag{C2} \\
\Delta C_\ell &\leq c_3 \; 2^{\gamma \ell} \label{eq:C3}\tag{C3}
\end{align}
with $2\alpha \geq \min(\beta,\gamma)$, we have that
\begin{equation}
\label{eq:mlmc_theorem}
\cost{\calQ_{\text{MLMC}}} \leq
\begin{cases}
c_4 \;\varepsilon^{-2} & \text{if } \beta > \gamma, \\
c_4 \;\varepsilon^{-2} (\log \varepsilon)^2 & \text{if } \beta = \gamma, \\
c_4 \;\varepsilon^{-2 - (\gamma - \beta) / \alpha} & \text{if } \beta < \gamma.
\end{cases}
\end{equation}

\subsection{The Multi-Index Monte Carlo (MIMC) method}
\label{sec:mimc}

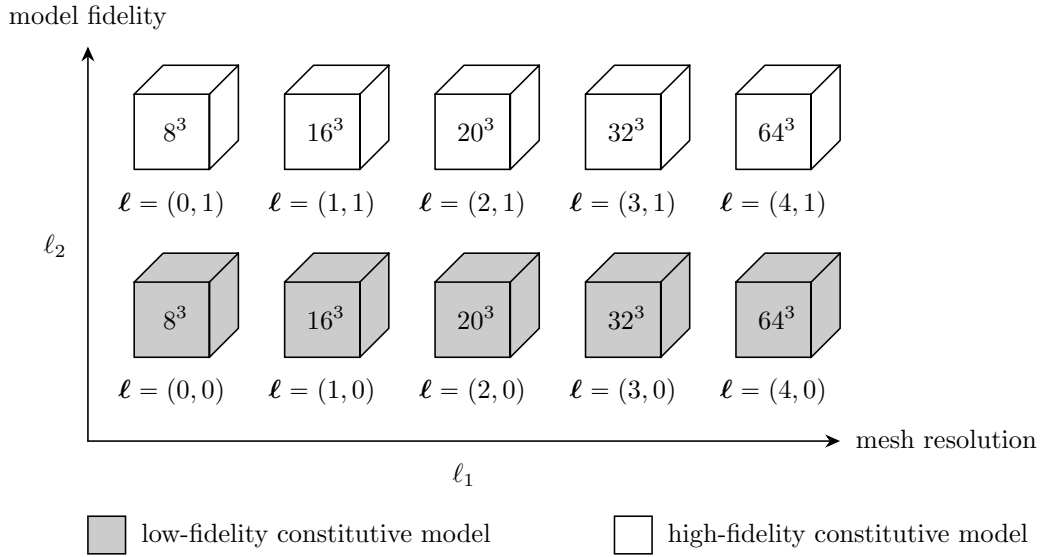
\begin{figure}[!htbp]
\centering
\newcommand{\cuboid}[4]{%
    \begin{scope}[shift={#1}, scale=0.5]
        \fill[#4] (0,2,2) -- (2,2,2) -- (2,0,2) -- (0,0,2) -- cycle;
        \fill[#4] (0,2,2) -- (0,2,0) -- (2,2,0) -- (2,2,2) -- cycle;
        \fill[#4] (2,2,0) -- (2,2,2) -- (2,0,2) -- (2,0,0) -- cycle;
        \draw[rounded corners = 0.1pt] (0,2,2) -- (2,2,2) -- (2,0,2) -- (0,0,2) -- (0,2,2) -- (0,2,0) -- (2,2,0) -- (2,0,0) -- (2,0,2) -- (2,2,2) -- (2,2,0);
        \node[anchor=center] at (1.1,1,2) {#2};
        \node[anchor=north] at (1,0,2) {#3};
    \end{scope}
}


\begin{tikzpicture}[
    every path/.style={semithick, line cap=round},
    every node/.style={inner sep=0pt, outer sep=3pt, execute at end node=\strut}
]
\foreach \fillcolor [count=\l] in {black!20, white}{%
    \pgfmathtruncatemacro{\j}{\l - 1}
    \foreach \mesh [count=\k] in {8, 16, 20, 32, 64}{%
        \pgfmathtruncatemacro{\i}{\k - 1}
        \cuboid{(2*\i, 2.5*\j)}{$\mesh^3$}{$\bsell = (\i, \j)$}{\fillcolor}
    }
}
\draw[->] (-1,-1.5) -- node[pos=0.5, anchor=north] {$\ell_1$} ++(10,0) node[anchor=west, outer sep=6pt, yshift=1pt] {mesh resolution};
\draw[->] (-1,-1.5) -- node[pos=0.5, anchor=east, outer sep=8pt] {$\ell_2$} ++(0,5.25) node[anchor=south, outer sep=3pt] {model fidelity};
\draw[fill=black!20] (-1, -3) rectangle ++(0.5, 0.5);
\node[anchor=west, yshift=1pt, outer sep=6pt] at (-0.5, -2.75) {low-fidelity constitutive model};
\draw[fill=white] (6.0, -3) rectangle ++(0.5, 0.5);
\node[anchor=west, yshift=1pt, outer sep=6pt] at (6.5, -2.75) {high-fidelity constitutive model};



\end{tikzpicture}
\caption{Schematic overview of the multi-fidelity hierarchy obtained by varying both the mesh resolution $(\ell_1)$ and the model fidelity $(\ell_2)$, where smaller $\ell$ corresponds to lower fidelity level and larger $\ell$ corresponds to higher fidelity level.}
\label{fig:mimc_hierarchy}
\end{figure}

The MIMC method is a multi-dimensional extension of the MLMC method outlined in~\cref{sec:mlmc}. Instead of using a single integer $\ell = 0, 1, \ldots, L$ to denote the resolution level in the hierarchy of models, the MIMC method uses a $d$-dimensional tuple or multi-index $\mathbf{\bsell} \in \mathbb{N}_0^d$, with $\mathbb{N}_0 = \{0,1,2,\dots \}$. In the context of CPFEM, an additional dimension of refinement could be the fidelity of the constitutive model. \cref{fig:mimc_hierarchy} illustrates this point, where the phenomenological plasticity model from \cite{roters2010overview} is treated as low-fidelity constitutive model and the non-local dislocation-based density model from \cite{kords2013role} and \cite{reuber2014dislocation} is treated as high-fidelity constitutive model.


The multi-index construction starts from a tensor product of single-direction differences, i.e.,
\begin{equation}
\label{eq:mi_diff}
\bsDelta Q_{\bsell} \coloneqq \left( \bigotimes_{j=1}^{d} \Delta_j \right) Q_\bsell \quad \text{ with } \quad
\Delta_j Q_\bsell =
\begin{cases}
Q_\bsell - Q_{\bsell - \bse_j} & \text{if } \ell_j > 0, \\
Q_\bsell & \text{if } \ell_j = 0,
\end{cases}
\end{equation}
where $\bse_j = (\delta_{ij})_{i=1}^{d}$ and $\delta_{ij}$ is the Kronecker delta. 
For example, with $d=2$ and $\bsell = (2,1)$, we have
\begin{equation}
\begin{array}{lll}
\bsDelta Q_{(2,1)} &=& \Delta_2 (\Delta_1 Q_{(2,1)}) \\
&=& \Delta_2 (Q_{(2,1)} - Q_{(1,1)}) \\
&=& Q_{(2,1)} - Q_{(1,1)} - Q_{(2,0)} + Q_{(1,0)},
\end{array}
\end{equation}
see \cref{fig:mimc_hierarchy}. In general, in the evaluation of the multi-index difference $\Delta Q_\bsell$ in $d$ dimensions, a total of $2^d$ different model approximations are involved. Near the boundary, a total of $2^{d'}$ different model approximations are involved, where $d'$ is the number of dimensions where $\ell_j > 0$, $j = 1, 2, \ldots, d$.

We note that a multi-index difference can also be written as
\begin{equation}
\label{eq:multi-index_combination}
\bsDelta Q_{\bsell} = \sum_{\substack{\fraku \subseteq \{1,\dots,d\} \\ \bsell - \bse_\fraku \in \mathbb{N}^d}} (-1)^{|\fraku|} Q_{\bsell - \fraku},
\end{equation}
where $\bse_\fraku$ is a vector with its $j$th component equal to 1 for $j \in \fraku$ and 0 everywhere else. Equation~\eqref{eq:multi-index_combination} is closer to the \emph{sparse combination technique} from~\cite{gerstner1998numerical,barthelmann2000high,bungartz2004sparse}, which inspired the construction of the MIMC method in~\cite{haji2016multi}.


The MIMC method proposed in~\cite{haji2016multi} uses an independent MC estimator to estimate each term in a finite summable subset of multi-index differences $\mathscr{I}_d$, i.e.,
\begin{equation}
\label{eq:mimc}
\calQ_{\text{MIMC}} = \sum_{\bsell \in \mathscr{I}_d} \frac{1}{N_{\bsell}} \sum_{n=0}^{N_{\bsell}} \bsDelta \calQ_{\bsell}(\omega^{(n)}).
\end{equation}
Similar to the multi-level method presented in~\cref{sec:mlmc}, the multi-index difference $\calQ_{\bsell}(\omega^{(n)})$ is based on the same outcome $\omega^{(n)}$ to ensure sufficient positive correlation between the different approximations and, hence, guarantee sufficient decay of the variance of the multi-index difference as $\bsell \rightarrow \infty$ component-wise. With sufficient variance decay as $\bsell$ increases, most of the samples will be taken on indices with low \textcolor{black}{fidelity}, while fewer samples will be required on indices with increasingly higher \textcolor{black}{fidelity}.

Note that not all multi-index sets $\mathscr{I}_d$ are suitable index sets. Specifically, we put a constraint on the index set by assuming it is \emph{downward closed} in order to be admissible. Further details of admissibility are deferred to \cref{sec:amimc}. In the case of an infinite-dimensional admissible index set $\mathscr{I}_d$, the multi-index estimator is an unbiased estimator for the expected value of the quantity of interest, as the multi-index differences satisfy the relation
\begin{equation}
\sum_{\bsell \in \mathbb{N}^{d}_0} \E{ \bsDelta Q_{\bsell}} = \E{Q}.
\end{equation}
The variance of the MIMC estimator is given by
\begin{equation}
    \V{\calQ_{\text{MIMC}}} = \sum_{\bsell \in \mathscr{I}_d} \frac{1}{N_\bsell^2} \sum_{n=1}^{N_\bsell} \V{\bsDelta Q_\bsell} = \sum_{\bsell \in \mathscr{I}_d} \frac{\V{\bsDelta Q_\bsell}}{N_\bsell}.
\end{equation}

Note that the MLMC estimator in~\eqref{eq:mlmc} is just a special case of~\eqref{eq:mimc} with $d=1$. In this case, the tuple $\bsell$ reduces to a scalar level $\ell$ and there is no tensor product of differences involved in the construction.

Expanding the MSE as in~\eqref{eq:mlmc_mse} now yields
\begin{equation}\label{eq:mimc_mse}
    \MSE{\calQ_{\text{MIMC}}} = \V{\calQ_{\text{MIMC}}} + (\E{\calQ_{\text{MIMC}} - Q})^2 = \sum_{\bsell \in \mathscr{I}_d} \frac{\V{\bsDelta Q_\bsell}}{N_\bsell} + (\bbE[\calQ_{\text{MIMC}} - Q])^2.
\end{equation}
Again, the MSE consists of two terms: the variance of the estimator and the square of the bias, and an MSE smaller than or equal to $\varepsilon^2$ can be guaranteed by choosing
\begin{equation}
\label{eq:mimc_constraint}
    \sum_{\bsell \in \mathscr{I}_d} \frac{\V{\bsDelta Q_\bsell}}{N_\bsell} \leq \frac{\varepsilon^2}{2} \quad \text{and} \quad | \bbE[\calQ_{\text{MIMC}} - Q] | \leq \frac{\varepsilon}{\sqrt{2}}.
\end{equation}
The first constraint in~\eqref{eq:mimc_constraint} yields an expression for the required number of samples on each index much similar to equation~\eqref{eq:optimalSamples_mlmc}, i.e.,
\begin{equation}
\label{eq:optimalSamples_mimc}
N_{\bsell} = \frac{2}{\varepsilon^2}\sqrt{\frac{\V{\bsDelta Q_\bsell}}{\bsDelta C_{\bsell}}} \left( \sum_{\bsell \in \mathscr{I}_d} \sqrt{\V{\bsDelta Q_{\bsell}} \bsDelta C_{\bsell}} \right),
\end{equation}
where $\bsDelta C_\ell$ denotes the cost to compute a sample of the multi-level difference $\bsDelta Q_\ell$, see~\cite{haji2016multi}. The second constraint in~\eqref{eq:mimc_constraint} will prescribe the shape of the index set $\mathscr{I}_d$. Some commonly used index sets are the total degree index set
\begin{equation}
    \mathscr{I}_d = \left\{\bsell \in \bbN_0^d : \sum_{j=1}^{d} \ell_j \leq L \right\}
\end{equation}
and the hyperbolic cross index set
\begin{equation}
    \mathscr{I}_d = \left\{\bsell \in \bbN_0^d : \prod_{j=1}^{d} (\ell_j + 1) \leq L + 1\right\},
\end{equation}
where $L$ is now a parameter that governs the size of the index set. In case of the total degree index sets, a theoretical analysis of the cost of the MIMC estimator similar to~\eqref{eq:mlmc_theorem} has been presented in \cite{haji2016multi}.

The optimal shape of the index set $\mathscr{I}_d$ is usually based on \emph{a priori} knowledge about the problem at hand. However, in most practical applications, including CPFEM, such knowledge is not readily available. In the next section, we discuss how the index set $\mathscr{I}_d$ can be constructed in an adaptive fashion, rendering the MIMC method useful in practice.

\subsection{Dimension-Adaptive Multi-Index Monte Carlo}
\label{sec:amimc}

The dimension-adaptive construction of the multi-index set $\mathscr{I}_d$ has been studied in~\cite{robbe2016dimension}. The idea of this construction is that the index set can be generated on-the-fly starting from the lowest-resolution index, using statistics of the already computed model evaluations as proxies for the true expected value and variance of the multi-index differences. A multi-index estimator that uses this adaptive construction scheme will be referred to as Dimension-Adaptive Multi-Index Monte Carlo (AMIMC). 

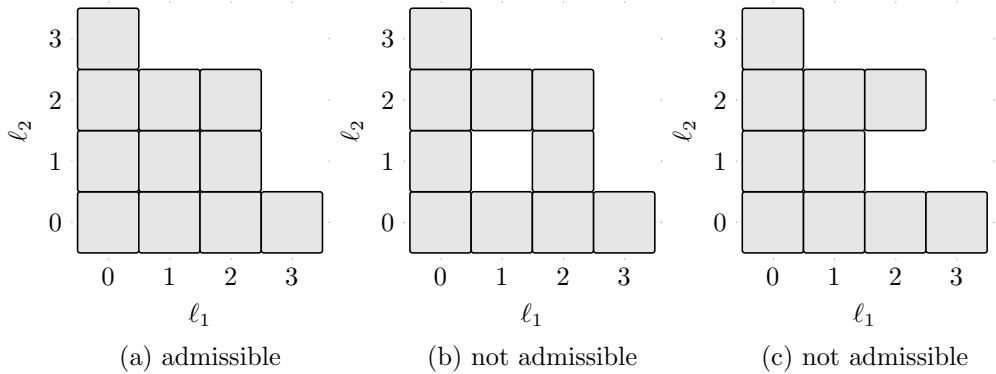
\begin{figure}[!htbp]
    \centering
\setlength{\figurewidth}{5cm}
\setlength{\figureheight}{5cm}
\tikzsetnextfilename{admissible_index_sets_example}
\begin{tikzpicture}[%
    ]
    \begin{groupplot}[%
            group style={group size={3 by 1}}
        ]
        \nextgroupplot[%
            index set,
            3 by 3,
        ]
        \drawsquare{0}{0}{old}
        \drawsquare{0}{1}{old}
        \drawsquare{0}{2}{old}
        \drawsquare{0}{3}{old}
        \drawsquare{1}{0}{old}
        \drawsquare{1}{1}{old}
        \drawsquare{1}{2}{old}
        \drawsquare{2}{0}{old}
        \drawsquare{2}{1}{old}
        \drawsquare{2}{2}{old}
        \drawsquare{3}{0}{old}
        \node[anchor=north] at (rel axis cs:0.5, -0.3) {(a) admissible};
        \nextgroupplot[%
            index set,
            3 by 3,
        ]
        \drawsquare{0}{0}{old}
        \drawsquare{0}{1}{old}
        \drawsquare{0}{2}{old}
        \drawsquare{0}{3}{old}
        \drawsquare{1}{0}{old}
        \drawsquare{1}{2}{old}
        \drawsquare{2}{0}{old}
        \drawsquare{2}{1}{old}
        \drawsquare{2}{2}{old}
        \drawsquare{3}{0}{old}
        \node[anchor=north] at (rel axis cs:0.5, -0.3) {(b) not admissible};
        \nextgroupplot[%
            index set,
            3 by 3,
        ]
        \drawsquare{0}{0}{old}
        \drawsquare{0}{1}{old}
        \drawsquare{0}{2}{old}
        \drawsquare{0}{3}{old}
        \drawsquare{1}{0}{old}
        \drawsquare{1}{2}{old}
        \drawsquare{2}{0}{old}
        \drawsquare{1}{1}{old}
        \drawsquare{2}{2}{old}
        \drawsquare{3}{0}{old}
        \node[anchor=north] at (rel axis cs:0.5, -0.3) {(c) not admissible};
    \end{groupplot}
\end{tikzpicture}
    \caption{Examples of admissible and non-admissible index sets in 2 dimensions ($d = 2$).}
    \label{fig:admissible_index_sets_examples}
\end{figure}

Before discussing the general adaptive procedure, we examine in more detail the requirements that must be satisfied for an index set to be admissible. An admissible index set is a non-empty set of multi-indices such that for all multi-indices $\bstau$ and $\bsell$, where $\bsell \in \mathscr{I}_d$, that satisfy $\bstau \leq \bsell$ component-wise, it follows that $\bstau \in \mathscr{I}_d$. Equivalently, for all $\bsell \in \mathscr{I}_d$, we have
\begin{equation}
\bsell - \bse_j \in \mathscr{I}_d \text{ for all } j=1,\ldots,d \text{ where } \ell_j > 0,
\end{equation}
with $\bse_j$ as defined in equation~\eqref{eq:mi_diff}. In other words, in an admissible index set, all indices with smaller entries in at least one direction are also included in the set. This condition ensures the validity of the telescoping sum expansion in terms of canceling differences, when defining a multi-index estimator according to equation~\eqref{eq:mimc}. Some examples of admissible and non-admissible index sets for $d=2$ are shown in \cref{fig:admissible_index_sets_examples}.

In what follows, we also require the notion of the forward neighborhood $\calF_\bsell$ of an index $\bsell$, defined as
\begin{equation}
\calF_\bsell = \{\bsell + \bse_j, 1 \leq j \leq d\}.
\end{equation}
Similarly, the backward neighborhood $\calB_\bsell$ of an index $\bsell$ is defined as
\begin{equation}
\calB_\bsell = \{\bsell - \bse_j : \ell_j > 0, 1 \leq j \leq d\}.
\end{equation}
An admissible index set contains the backward neighborhood of all indices in the set. The admissibility property is also known as \textit{downward closedness}, see~\cite{dyn2014multivariate}. An index set that is admissible is also called a downward closed index set.

The optimal shape of the index set $\mathscr{I}_d$ is the solution of a constraint optimization problem: we minimize the total cost of the MIMC estimator in~\eqref{eq:mimc} while ensuring that the bias constraint in~\eqref{eq:mimc_constraint} is satisfied. Using~\eqref{eq:optimalSamples_mimc}, the total cost of the MIMC estimator in~\eqref{eq:mimc} can be written as
\begin{equation}
\label{eq:mimc_cost}
    \mathrm{cost}(\calQ_{\text{MIMC}}) = \frac{2}{\varepsilon^2} \left( \sqrt{\bbV[\bsDelta Q_\bsell] \bsDelta C_\bsell} \right)^2.
\end{equation}
Furthermore, since the index set $\mathscr{I}_d$ is necessarily finite, the MIMC estimator is biased when estimating the expected value $\bbE[Q]$. This bias is equal to the sum of all neglected contributions, i.e., we have that
\begin{equation}
\left|\E{\calQ_{\text{MIMC}} - Q}\right| = \left|\sum_{\bsell\notin\mathscr{I}_d}\E{\bsDelta Q_\bsell}\right|
\leq \sum_{\bsell\notin\mathscr{I}_d} |\E{\bsDelta Q_\bsell}|.
\end{equation}
Hence, the optimal index set $\mathscr{I}_d$ is the solution of
\begin{equation}
\label{eq:idxset_problem}
\begin{aligned}
\min_{\mathscr{I}_d \subseteq \N_0^d} &\sum_{\bsell\in\mathscr{I}_d} \sqrt{\bbV[\bsDelta Q_\bsell] \bsDelta C_\bsell} \\
\text{subject to} &\sum_{\bsell\notin\mathscr{I}_d} |\E{\bsDelta Q_\bsell}| \leq \frac{\varepsilon}{\sqrt{2}}.
\end{aligned}
\end{equation}
This minimization problem cannot be solved analytically, unless further assumptions are made on $\bbV[\bsDelta Q_\bsell]$, $\bsDelta C_\bsell$ and $|\E{\bsDelta Q_\bsell}|$. These assumptions will directly determine the theoretically optimal shape of the index set. In practice, however, it is hard to determine \emph{a priori} which assumptions are best fit to model the problem at hand. Instead, we will reformulate optimization problem~\eqref{eq:idxset_problem} as a binary knapsack problem, similar to~\cite{gerstner1998numerical}. A binary knapsack problem is a combinatorial problem where different items with associated cost and value must be included in a collection, such that the total value is maximized, but the total cost does not exceed a certain limit. In a binary knapsack problem, there is only one item of each kind. This corresponds to the problem at hand, where the unique items (or, indices) have a certain ``value'' (bias reduction), but also a certain ``cost'' (computational cost).

The value of each index is expressed by $|\E{\bsDelta Q_\bsell}|$. The higher this value, the more the bias is reduced when this index is added to the index set. The total value $v$ of the index set $\mathscr{I}_d$ is thus
\begin{equation}
v(\mathscr{I}_d) = \sum_{\bsell\in\mathscr{I}_d} |\E{\bsDelta Q_\bsell}|.
\end{equation}

The cost of each index is expressed in terms of its contribution to the total amount of work, $\sqrt{\bbV[\bsDelta Q_\bsell] \bsDelta C_\bsell}$, see equation~\eqref{eq:mimc_cost}. The total amount of work $w$ of the index set $\mathscr{I}_d$ is thus
\begin{equation}
w(\mathscr{I}_d) = \sum_{\bsell\in\mathscr{I}_d} \sqrt{\bbV[\bsDelta Q_\bsell] \bsDelta C_\bsell}.
\end{equation}

This encourages us to construct a \textit{profit indicator} $P_\bsell>0$ for each index, defined as the ratio of its value and cost:
\begin{equation}
\label{eq:profit}
P_\bsell = \frac{|\E{\bsDelta Q_\bsell}|}{\sqrt{\bbV[\bsDelta Q_\bsell] \bsDelta C_\bsell}}.
\end{equation}
The higher this profit, the more benefit there is in including this index into the index set. An index set then consists of only those indices that have a profit indicator larger than a certain threshold $\rho$:
\begin{equation}
\mathscr{I}_d = \{\bsell\in\N_0^d : P_\bsell \geq \rho \}.
\end{equation}
This means that the optimal index set shapes are the level sets of the profit indicators.

Using the profit indicators in~\eqref{eq:profit}, the optimization problem from equation~\eqref{eq:idxset_problem} can be solved by progressive enrichment of the multi-index set $\mathscr{I}_d$. A greedy procedure would then start from the index set $\mathscr{I}_d = \{(0,\ldots,0)\}$ and successively add indices to this set, such that the bias is reduced as much as possible, whilst ensuring that the index set remains admissible during each iteration of the procedure. A possible strategy for such a greedy procedure is to partition the index set $\mathscr{I}_d$ into two disjoint subsets, $\mathscr{O}_d$ and $\mathscr{A}_d$. The admissible multi-index set $\mathscr{O}_d$ contains the \textit{old} multi-indices that have already been considered for inclusion in the index set. These indices have at least one forward neighbor in $\mathscr{I}_d=\mathscr{O}_d\cup\mathscr{A}_d$.  The set $\mathscr{A}_d$ contains the \textit{active} indices that are suitable candidates for inclusion in $\mathscr{O}_d$. These indices, by definition, have none of their forward neighbors included in the index set $\mathscr{I}_d$. The active indices form the \textit{outer boundary} of the index set, and are used to compute a bias estimate, using the heuristic
\begin{equation}
\label{eq:adaptive_bias}
\left| \sum_{\bsell\notin\mathscr{I}_d} \E{\bsDelta Q_\bsell} \right| \approx \left| \sum_{\bsell\in\mathscr{A}_d} \E{\bsDelta Q_\bsell} \right|.
\end{equation}

In every iteration of the greedy procedure, we select from $\mathscr{A}_d$ the index $\bstau$ with the largest profit indicator, where the profit is defined by equation~\eqref{eq:profit}. This index is moved from the set of active indices $\mathscr{A}_d$ to the set of old indices $\mathscr{O}_d$. The multi-index set $\mathscr{A}_d$ is then enlarged by all multi-indices $\bskappa$ in the forward neighborhood $\calF_\bstau$ of $\bstau$ for which the backward neighbors $\calB_\bskappa$ are all included in the old index set $\mathscr{O}_d$.

An example step of the greedy index set growth procedure for $d=2$ is shown in \cref{fig:adaptive_algorithm}. Suppose that, in a given iteration, the index set looks like the one shown in \cref{fig:adaptive_algorithm} (a). The index with maximum profit, $\bstau = (2,1)$, is indicated by \protect\raisebox{-2pt}{\protect\tikz[scale=0.3]{\protect\drawsquare{0}{0}{maximum}}}. First, in \cref{fig:adaptive_algorithm} (b), this index is moved from the active set $\mathscr{A}_2$ (\protect\raisebox{-2pt}{\protect\tikz[scale=0.3]{\protect\drawsquare{0}{0}{active}}}) to the old set $\mathscr{O}_2$ (\protect\raisebox{-2pt}{\protect\tikz[scale=0.3]{\protect\drawsquare{0}{0}{old}}}), and the forward neighborhood $\calF_{(2,1)} = \{ (3,1), (2,2)\}$ is considered. The forward neighborhood $\calF_{(2,1)}$ is indicated by thick black lines. Index $(3,1)$ is admissible in the old set, since both of the indices that constitute its backward neighborhood, $\calB_{(3,1)} = \{(2,1)$, $(3,0)\}$, are already included in the old set. Hence, in \cref{fig:adaptive_algorithm} (c), index $(3,1)$ is added to the active set $\mathscr{A}_2$. However, index $(2,2)$ is not admissible in $\mathscr{O}_2$, since index $(1,2)$ is part of the active set $\mathscr{A}_2$, and not of the old set $\mathscr{O}_2$. Thus, index $(2,2)$ is left untreated. See \cref{alg:amimc_construction} for a detailed description of the greedy index set growth.

\begin{figure}[!htbp]
    \centering
    \setlength{\figurewidth}{5cm}
\setlength{\figureheight}{4.5cm}
\tikzsetnextfilename{adaptive_algorithm}
\begin{tikzpicture}[%
    ]
    \begin{groupplot}[%
            group style={group size={3 by 1}}
        ]
        \nextgroupplot[%
            index set,
            4 by 3
        ]
        \drawsquare{0}{0}{old}
        \drawsquare{0}{1}{old}
        \drawsquare{0}{2}{old}
        \drawsquare{0}{3}{active}
        \drawsquare{1}{0}{old}
        \drawsquare{1}{1}{old}
        \drawsquare{1}{2}{active}
        \drawsquare{2}{0}{old}
        \drawsquare{2}{1}{maximum}
        \drawsquare{3}{0}{old}
        \drawsquare{4}{0}{active}
        \node[anchor=north] at (rel axis cs:0.5, -0.35) {(a)};
        \nextgroupplot[%
            index set,
            4 by 3
        ]
        \drawsquare{0}{0}{old}
        \drawsquare{0}{1}{old}
        \drawsquare{0}{2}{old}
        \drawsquare{0}{3}{active}
        \drawsquare{1}{0}{old}
        \drawsquare{1}{1}{old}
        \drawsquare{1}{2}{active}
        \drawsquare{2}{0}{old}
        \drawsquare{2}{1}{old}
        \drawthicksquare{2}{2}{white}
        \drawsquare{3}{0}{old}
        \drawthicksquare{3}{1}{white}
        \drawsquare{4}{0}{active}
        \node[anchor=north] at (rel axis cs:0.5, -0.35) {(b)};
        \nextgroupplot[%
            index set,
            4 by 3
        ]
        \drawsquare{0}{0}{old}
        \drawsquare{0}{1}{old}
        \drawsquare{0}{2}{old}
        \drawsquare{0}{3}{active}
        \drawsquare{1}{0}{old}
        \drawsquare{1}{1}{old}
        \drawsquare{1}{2}{active}
        \drawsquare{2}{0}{old}
        \drawsquare{2}{1}{old}
        \drawsquare{3}{0}{old}
        \drawsquare{3}{1}{active}
        \drawsquare{4}{0}{active}
        \node[anchor=north] at (rel axis cs:0.5, -0.35) {(c)};
    \end{groupplot}
\end{tikzpicture}
    \caption{Illustration of the adaptive algorithm in 2 dimensions ($d = 2$). The plots show the evolution of the multi-index set $\mathscr{I}_2$, distinguishing between the set of old indices $\mathscr{O}$ (\protect\raisebox{-2pt}{\protect\tikz[scale=0.3]{\protect\drawsquare{0}{0}{old}}}) and the set of active indices $\mathscr{A}$ (\protect\raisebox{-2pt}{\protect\tikz[scale=0.3]{\protect\drawsquare{0}{0}{active}}}). The index with maximum profit in this iteration of the algorithm is indicated by \protect\raisebox{-2pt}{\protect\tikz[scale=0.3]{\protect\drawsquare{0}{0}{maximum}}}.}
    \label{fig:adaptive_algorithm}
\end{figure}

\begin{algorithm}
\begin{algorithmic}[1]
\Statex \textbf{input:} old set $\mathscr{O}_d$ and active set $\mathscr{A}_d$
\Statex \textbf{output:} updated old set $\mathscr{O}_d$ and updated active set $\mathscr{A}_d$, set of newly added indices $\mathscr{N}$
\Statex
\Procedure{\textsf{grow\texttt{\_}index\texttt{\_}set}}{$\mathscr{O}_d$, $\mathscr{A}_d$}
\State compute profit indicators $P_\bsell$ for each index $\bsell\in\mathscr{A}_d$
\State select index $\bstau$ from $\mathscr{A}_d$ with largest profit $P_\bstau$
\State $\mathscr{A}_d\gets\mathscr{A}_d\setminus\{\bstau\}$
\State $\mathscr{O}_d\gets\mathscr{O}_d\cup\{\bstau\}$
\State $\mathscr{N}\gets\varnothing$
\ForEach {$\bskappa \in \calF_{\bstau}$}
\State $\mathsf{valid}\gets\texttt{true}$
\ForEach {$\bsell \in \calB_{\bstau_i}$}
\IfThen {$\bsell \notin \mathscr{O}_d$}
{ $\mathsf{valid}\gets\texttt{false}$ }
\EndFor
\IfThen {$\mathsf{valid} = \texttt{true}$} {$\mathscr{N}\gets\mathscr{N}\cup\{\bskappa\}$}
\EndFor
\State $\mathscr{A}_d\gets\mathscr{A}_d\cup\mathscr{N}$
\EndProcedure
\end{algorithmic}
\caption{Adaptive construction of a multi-index set}
\label{alg:amimc_construction}
\end{algorithm}

It is easy to see that the procedure indeed produces only admissible index sets. By moving the index with maximum profit from the active set to the old set, the old set remains admissible, since that index is part of the active set, and, by definition, all indices in the active set are admissible in the old set. Also, the indices in the forward neighborhood of the index with maximum profit are scanned for their admissibility in the old set, before they are added to the active set. Hence, the multi-indices that constitute the \textit{new} active set are all admissible in the new old set. This means they have all of their backward neighbors $\calB_\bsell$ included in the old index set. A set for which the backward neighbors of all indices in the set are included is, by definition, an admissible index set, as required.

\subsection{Algorithm}

A full procedure for adaptive MIMC simulation is shown in \cref{alg:amimc}. As input, the procedure requires a requested tolerance $\varepsilon$ on the \emph{root mean square error} (RMSE) of the expected value of the quantity of interest $Q$. The RMSE is defined as the square root of the MSE defined in~\eqref{eq:mimc_mse}. The outputs returned by the method are the value of the MIMC estimator, $\mathsf{E}$, and an estimate for the achieved RMSE, $\mathsf{error}$. We will now clarify some of the essential components of the algorithm.

We use the sample mean as a proxy for the true mean of the multi-index difference, i.e.,
\begin{equation}
\mathsf{E}_\bsell \coloneqq \frac{1}{N_\bsell} \sum_{n=1}^{N_\bsell} \bsDelta Q_\bsell(\omega^{(n)}) \approx \bbE[\bsDelta Q_\bsell].\label{eq:E_approx}
\end{equation}
We use the sample variance as a proxy for the true variance of the multi-index difference, i.e.,
\begin{equation}
    \mathsf{V}_\bsell \coloneqq \frac{1}{N_\bsell - 1} \sum_{n=1}^{N_\bsell} \left( \bsDelta Q_\bsell(\omega^{(n)}) - \mathsf{E}_\bsell \right)^2 \approx \bbV[\bsDelta Q_\bsell].\label{eq:V_approx}
\end{equation}
The cost $\bsDelta C_\bsell$ can be replaced by the wall-clock time $\mathsf{C}_\bsell$ needed to compute a single sample of the multi-index difference.

Starting from $\mathscr{O}_d(0)=\varnothing$ and $\mathscr{A}_d(0)=\{(0,\ldots,0)\}$, the algorithm gradually enlarges the index set $\mathscr{I}_d$ according to the procedure described in \cref{alg:amimc_construction}. For each new index $\bsell$ that is added to the index set, we compute an initial estimate for the variance contribution by taking $\widetilde{N}$ warm-up samples. When sufficient lower-resolution indices are available, we use extrapolated values for $\mathsf{V}_\bsell$ and $\mathsf{C}_\bsell$ to estimate the optimal number of samples using~\eqref{eq:optimalSamples_mimc}. We then ensure that at least 2 warm-up samples are taken on that index, to be able to compute the sample variance using~\cref{eq:V_approx}, see \cref{alg:amimc:10}. This \textit{regression} of the number of samples has been proposed in the context of MLMC, see~\cite{collier2015continuation}, but can easily be extended to the multi-index setting. Once we have estimates available for the variance and cost at each index $\bsell$, we re-evaluate equation~\eqref{eq:optimalSamples_mimc} for the quasi-optimal number of samples on each index, and perform an additional number of model evaluations accordingly.

Note that, by using the active set algorithm, no computational effort is wasted. That is, once an index is added to the active set, its samples are also used in the final evaluation of the MIMC estimator from~\eqref{eq:mimc}. Indeed, it does not make sense to take samples at these active indices, only to evaluate the profit indicator, and then to exclude these samples for the final evaluation of the estimate.

\begin{algorithm}
\begin{algorithmic}[1]
\Statex \textbf{input:} a tolerance $\varepsilon$ on the RMSE
\Statex \textbf{output:} an approximation $\mathsf{E}$ for the mean of $Q$, an error estimate $\mathsf{error}$
\Statex
\Procedure{\textsf{AMIMC}}{$\varepsilon$}
\State $\mathsf{B}\gets\infty$
\State $\mathscr{O}_d \gets \varnothing$
\State $\mathscr{A}_d \gets \{(0,\ldots,0)\}$
\Repeat
\State $\mathscr{O}_d, \mathscr{A}_d, \mathscr{N} \gets $ \textsf{grow\texttt{\_}index\texttt{\_}set}$(\mathscr{O}_d$, $\mathscr{A}_d)$
\ForEach {$\bsell\in\mathscr{N}$}
\If {$\max(\bsell) \leq 2$}
\State take warm-up samples at index $\bsell$, to have at least $\widetilde{N}$
\Else
\State estimate $N_\bsell$ by~\eqref{eq:optimalSamples_mimc}  using extrapolated values for $\mathsf{V}_\bsell$ and $\mathsf{C}_\bsell$
\State take $\max(2, \min(\widetilde{N}, N_\bsell))$ warm-up samples at index $\bsell$ \label{alg:amimc:10}
\EndIf
\State calculate $\mathsf{V}_\bsell$ using~\eqref{eq:V_approx} and compute $\mathsf{C}_\bsell$
\EndFor
\State $\mathsf{E}\gets 0$ and $\mathsf{V}\gets 0$
\ForEach {$\bsell\in\mathscr{I}_d$}
\State compute the optimal number of samples $N_\bsell$ using~\eqref{eq:optimalSamples_mimc}
\label{alg:amimc:16}
\State take additional samples at index $\bsell$, to have at least  $N_\bsell$
\State $\mathsf{E}\gets\mathsf{E}+\mathsf{E}_\bsell$ where $\mathsf{E}_\bsell$ is computed using~\eqref{eq:E_approx}
\State $\mathsf{Var}\gets\mathsf{Var}+\mathsf{V}_\bsell/N_\bsell$ where $\mathsf{V}_\bsell$ is computed using~\eqref{eq:V_approx}
\EndFor
\State compute an estimate $\mathsf{B}$ for the bias using~\eqref{eq:adaptive_bias}
\label{alg:amimc:23}
\State $\mathsf{error}\gets\sqrt{\mathsf{V} + \mathsf{B}^2}$
\Until{$\mathsf{B}\leq\varepsilon/\sqrt{2}$} \label{alg:amimc:25}
\EndProcedure
\end{algorithmic}
\caption{Adaptive Multi-Index Monte Carlo}
\label{alg:amimc}
\end{algorithm}

As with all adaptive algorithms, the algorithm could be fooled by a quantity of interest for which it seems like there is no benefit of extending the index set at some point, and for which essential contributions are hidden at an arbitrary further depth in the index set. For example, suppose that the profit indicator $P_\bsell$ for a given index $\bsell$ happens to be small, then our algorithm finds that there is no benefit in future refinement of the forward neighborhood $\calF_\bsell$. Now, there are two possibilities. Either the profit indicators of the forward neighbors of $\bsell$ are smaller than (or at most of the same magnitude as) the profit of $\bsell$, and our adaptive procedure has stopped the adaptation in that direction properly. However, it is also possible that one of the forward neighbors of $\bsell$ has a profit indicator that is considerably larger than $P_\bsell$, and thus aspires further refinement. Unfortunately, there is no way to avoid this issue, unless an \emph{a priori} analysis of the quantity of interest is performed, In effect, such an analysis would destroy the premise of the adaptive algorithm \textcolor{black}{altogether}. We refer to~\cite{kuo2016application} for an example of such an analysis for an elliptic partial differential equation (PDE) model problem. This issue could of course be avoided by actually computing the profit indicators of the indices in the forward neighborhood $\calF_\bsell$, but this just defers the problem, since we may encounter the same problem for the forward neighborhood of the forward neighbors.

An alternative profit indicator, used in the context of adaptive sparse grids, is
\begin{equation}
P_\bsell = \max \left( \zeta \frac{|\E{\bsDelta Q_\bsell}|}{|\E{\bsDelta Q_\bszero}|}, (1-\zeta) \frac{\sqrt{\bbV[\bsDelta Q_\bszero] \bsDelta C_\bszero}}{\sqrt{\bbV[\bsDelta Q_\bsell] \bsDelta C_\bsell}} \right),
\end{equation}
where $0 \leq \zeta \leq 1$ weighs the contribution of each index $\bsell$ to the bias and the computational cost. The benefit of this formulation is that it allows the user to specify the safeguard parameter $\zeta$, optionally putting more weight on the value of each multi-index and relaxing the work constraint. However, we found numerically that the profit indicator defined in~\eqref{eq:profit} yields comparable quasi-optimal index sets, without the need to calibrate an additional parameter $\zeta$.

In practice, the computation of profit indicators is based on either a set of warm-up samples, or extrapolated values from coarser levels. This means that these profit indicators, especially at the larger indices, can be extremely unreliable. To avoid that the algorithm gets stuck in a local suboptimal search direction, it may be beneficial to select suboptimal indices for further refinement. For example, one could implement an accept-reject like algorithm, that only selects the index with maximum profit with a certain acceptance rate $r$, and picks another index from the active set at random otherwise. The lower this acceptance rate, the more the adaptive algorithm will perform a global search in all coordinate directions, and may identify the \textit{hidden features} mentioned in the remark above. In the context of CPFEM, where the model fidelity as additional direction for refinement contains only two possible candidates, we deem such randomization approach unnecessary. However, when additional directions for refinement are added to the model hierarchy, e.g., by varying a time step size, an accept-reject strategy may be crucial to ensure sufficient exploration of the model search space.

\section{Constitutive models in CPFEM}
\label{sec:BackgroundCPFEM}


For small deformations, the elasto-plastic decomposition can be computed additively, whereas for large deformations, a multiplicative decomposition of deformation gradient is more appropriate, i.e.,
\begin{equation}
\mathbf{F} = \mathbf{F}_\text{e} \cdot \mathbf{F}_\text{p},
\end{equation}
following by the elasto-plastic decomposition of the velocity gradient as
\begin{equation}
\mathbf{L} = \dot{\mathbf{F}} \cdot \mathbf{F}^{-1} = \dot{\mathbf{F}}_\text{e} \cdot \mathbf{F}_\text{e}^{-1} + \mathbf{F}_\text{e} \cdot \dot{\mathbf{F}}_\text{p} \cdot \mathbf{F}_\text{p} \cdot \mathbf{F}_\text{e}^{-1}
= \mathbf{L}_\text{e} + \mathbf{F}_\text{e} \cdot \mathbf{L}_\text{p} \cdot \mathbf{F}_\text{e}^{-1},
\end{equation}
where $\mathbf{L}_\text{p}$ and $\mathbf{L}_\text{e}$ are the plastic and elastic velocity gradient, respectively.
The second Piola-Kirch\textcolor{black}{h}off stress tensor $\mathbf{S}$, which is a symmetric second-order tensor defined in the intermediate configuration, is given by
\begin{equation}
\mathbf{S}
= \frac{\mathbb{C}}{2} : (\mathbf{F}_\text{e}^T \mathbf{F}_\text{e} - \mathbf{I})
= \mathbb{C} : \bse_\text{e}
= J \mathbf{F}^{-1} \cdot \mathbf{\sigma} \cdot \mathbf{F}^{-T},
\end{equation}
where $\mathbb{C}$ is the elasticity fourth-order tensor, $\mathbf{F}_\text{e}$ is the elastic deformation gradient, $\mathbf{F}_\text{p}$ is the plastic deformation gradient \cite{roters2010overview}, $\bse_\text{e} = \frac{1}{2}\left( \mathbf{F}^T_e \mathbf{F}_\text{e} - \mathbf{I} \right)$ is the elastic Green's Lagrangian strain and $\sigma$ is the Cauchy stress tensor (cf. \cite{roters2011crystal}, Section 3.3).
The evolution of the inelastic deformation gradient $\mathbf{F}_\text{p}$
is given in terms of their respective velocity gradients $\mathbf{L}_\text{p}$ by the flow rules
\begin{align}
\dot{\mathbf{F}}_\text{p} = \mathbf{L}_\text{p} \mathbf{F}_\text{p},\\
\end{align}
The plasticity velocity gradient $\mathbf{L}_\text{p}$ in the intermediate (relaxed) configuration is determined by
\begin{equation}
\mathbf{L}_\text{p}
= \dot{\mathbf{F}}_\text{p} \cdot \mathbf{F}_\text{p}^{-1}
= \sum_{\alpha} \dot{\gamma}^{\alpha} \left( \mathbf{s}_{\text{s}}^{\alpha} \otimes \mathbf{n}_{\text{s}}^{\alpha} \right),
\end{equation}
where $\mathbf{s}_{\text{s}}^{\alpha}$ and $\mathbf{n}_{\text{s}}^{\alpha}$ are unit vectors along the slip direction and slip plane normal (cf. Section 6.2, \cite{roters2019damask}).
The driving force $\tau^{\alpha}$ for $\dot{\gamma}^{\alpha}$ is given by the Schmid law as
\begin{equation}
\tau^{\alpha} = \mathbf{M}_\text{p} \cdot \left( \mathbf{s}_{\text{s}}^{\alpha} \otimes \mathbf{n}_{\text{s}}^{\alpha} \right),
\end{equation}
\textcolor{black}{where $\mathbf{M}_\text{p}$ is the Mandel stress in the plastic configuration, calculated from the second Piola-Kirchhoff stress $\mathbf{S}$. 
}

In this section, we briefly summarize two constitutive models provided in \texttt{DAMASK}, which has been thoroughly reviewed by Roters et al.~\cite{roters2019damask} (cf. Section 6.2.2 and 6.2.3) in Section~\ref{subsec:phenomenological} and Section~\ref{subsec:dislocationdensity}, respectively, for the sake of completeness of the paper. Interested readers are referred to the work of Roters et al.~\cite{roters2010overview,roters2019damask} for a complete picture of CPFEM model in general and \texttt{DAMASK} in particular.
For spectral solver implementation, readers are referred to Eisenlohr et al.~\cite{eisenlohr2013spectral} and Shanthraj et al.~\cite{shanthraj2015numerically,shanthraj2019spectral}.

\textcolor{black}{Indeed, the multilevel method does not require a geometric structure in the number of DOF for each level. Any hierarchy that results in a decay in the variance of the multilevel difference and an increase in the computational cost as the level parameter increases, may in principle be suitable for the application of MLMC. However, the \textit{best} choice for such a hierarchy, i.e., the one that results in the lowest overall cost, is not known \textit{a priori}. 
Our motivation for choosing a geometric structure in the mesh resolution in this work is two-fold. First, in the theoretical treatment of the asymptotic cost complexity of the MLMC method, as presented in e.g., \cite{giles2015multilevel, cliffe2011multilevel}, it is customary to assume a ``power law'' for the increase in the computational cost per sample as a function of the level parameter. This corresponds to condition \eqref{eq:C3} in \cref{sec:mlmc}. The geometric structure is a natural one in the context of stochastic differential equations (SDEs), see~\cite{giles2008multilevel} and the elliptic PDE source problem, see~\cite{cliffe2011multilevel}. In the latter, the authors mention that this structure is inspired by the multigrid literature. Second, and arguably more important, it has been shown that a geometric relation is the optimal choice for the multilevel hierarchy for the elliptic source problem, see, e.g. ~\cite{haji2016optimization}.
}

\subsection{Phenomenological crystal plasticity constitutive model}
\label{subsec:phenomenological}

A phenomenological crystal plasticity constitutive model used for face-centered cubic (FCC) crystals was first proposed by Hutchinson~\cite{hutchinson1976bounds} and extended for deformation twinning by Kalidindi~\cite{kalidindi1998incorporation}.
The plastic component is parameterized in terms of resistance $\xi$ on $N_{\text{s}}$ slip and $N_{\text{tw}}$ twin systems.
The resistances on $\alpha=1,\dots,N_{\text{s}}$ slip systems evolve from $\xi_0$ to a system-dependent saturation value and depend on shear on slip and twin systems according to
\begin{equation}
\dot{\xi}^{\alpha} = h_0^{\text{s-s}} \left( 1 + c_1 \left(f^{\text{tot}}_{\text{tw}}\right)^{c_2} \right) (1 + h^{\alpha}_{\text{int}}) \left[ \sum_{\alpha' = 1}^{N_{\text{s}}} | \dot{\gamma}^{\alpha'} | \left| 1 - \frac{\xi^{\alpha'}}{\xi^{\alpha'}_{\infty}} \right|^a \text{sgn}\left( 1 - \frac{\xi^{\alpha'}}{\xi^{\alpha'}_{\infty}} \right) h^{\alpha \alpha'} \right]
+ \sum_{\beta'=1}^{N_{\text{tw}}} \dot{\gamma}^{\beta'} h^{\alpha \beta'} ,
\end{equation}
where
$f^{\text{tot}}_{\text{tw}}$ is the total twin volume fraction,
$h$ denotes the components of the slip-slip and slip-twin interaction matrices,
$h_0^{\text{s-s}}$, $h_{\text{int}}$, $c_1$, $c_2$ are model-specific fitting parameters and $\xi_{\infty}$ represents the saturated resistance.

The resistances on the $\beta = 1, \dots, N_{\text{tw}}$ twin systems evolve in a similar way,
\begin{equation}
\dot{\xi}^{\beta} = h_0^{\text{tw-s}} \left( \sum_{\alpha=1}^{N_{\text{s}}} |\gamma_{\alpha}| \right)^{c_3} \left( \sum_{\alpha'=1}^{N_{\text{s}}} |\dot{\gamma}^{\alpha'}| h^{\beta \alpha'} \right)
+ h_0^{\text{tw-tw}} \left(f^{\text{tot}}_{\text{tw}} \right)^{c_4} \left( \sum_{\beta'=1}^{N_{\text{tw}}} \textcolor{black}{\dot{\gamma}^{\beta'} h^{\beta \beta'}} \right),
\end{equation}
where $h_0^{\text{tw-s}}$, $h_0^{\text{tw-tw}}$, $c_3$, and $c_4$ are model-specific fitting parameters.
Shear on each slip system evolves at a rate of
\begin{equation}
\dot{\gamma}^{\alpha} = (1 - f^{\text{tot}}_{\text{tw}}) \dot{\gamma_0}^{\alpha} \left| \frac{\tau^{\alpha}}{\xi^{\alpha}} \right|^n \text{sgn}(\tau^{\alpha}).
\end{equation}
where slip due to mechanical twinning accounting for the unidirectional character of twin formation is computed slightly differently,
\begin{equation}
\dot{\gamma} = (1 - f^{\text{tot}}_{\text{tw}}) \dot{\gamma_0} \left| \frac{\tau}{\xi} \right|^n \mathcal{H}(\tau),
\end{equation}
where $\mathcal{H}$ is the Heaviside step function.
The total twin volume is calculated as
\begin{equation}
f^{\text{tot}}_{\text{tw}} = \max\left(1.0, \sum_{\beta=1}^{N_{\text{tw}}} \frac{\gamma^{\beta}}{\gamma^{\beta}_{\text{char}}} \right),
\end{equation}
where $\gamma_{\text{char}}$ is the characteristic shear due to mechanical twinning and depends on the twin system.

\subsection{Dislocation-density-based constitutive model}
\label{subsec:dislocationdensity}

A model for the plastic velocity gradient with contribution of mechanical twinning and phase transformation was developed in Kalidindi~\cite{kalidindi1998incorporation} and is given by
\begin{equation}
\mathbf{L}_\text{p} = (1 - f^{\text{tot}}_{\text{tw}} -f^{\text{tot}}_{\text{tr}}) \sum_{\alpha=1}^{N_{\text{s}}} \dot{\gamma}^{\alpha} \mathbf{s}_{\text{s}}^{\alpha} \otimes \mathbf{n}_{\text{s}}^{\alpha} 
+ \sum_{\beta=1}^{N_{\text{tw}}} \dot{\gamma} \mathbf{s}_{\text{tw}}^{\textcolor{black}{\beta}} \otimes \mathbf{n}_{\text{tw}}^{\beta} 
+ \sum_{\chi=1}^{N_{\text{tr}}} \dot{\gamma}^{\chi} \mathbf{s}_{\text{tr}}^{\textcolor{black}{\chi}} \otimes \mathbf{n}_{\text{tr}}^{\textcolor{black}{\chi}},
\end{equation}
where $\chi = 1, \dots,N_{\text{tr}}$ is the $\varepsilon$-martensite with volume fraction $f_{\text{tr}}$ on $N_{\text{tr}}$ transformation systems, $\mathbf{s}_{\text{s}}^{\alpha}$ and $\mathbf{n}_{\text{s}}^{\alpha}$ are unit vectors along the shear direction and shear plane normal of $N_{\text{s}}$ slip systems $\alpha$, $\mathbf{s}_{\text{tw}}^{\textcolor{black}{\beta}}$ and $\mathbf{n}_{\text{tw}}^{\textcolor{black}{\beta}}$ are those of $N_{\text{tw}}$ twinning systems $\beta$, and $\mathbf{s}_{\text{tw}}^{\textcolor{black}{\chi}}$ and $\mathbf{n}_{\text{tr}}^{\textcolor{black}{\chi}}$ are those of $N_{\text{tr}}$ transformation systems $\chi$.
The Orowan equation models the shear rate on the slip system $\alpha$ as
\begin{equation}
\dot{\gamma}^{\alpha} = \rho_e b_{\text{s}} \nu_0 \exp \left[ - \frac{Q}{k_B T} \left\{ 1 - \left( \frac{\tau_{\text{eff}}^{\alpha}}{\tau_{\text{sol}}} \right)^p \right\}^q \right],
\end{equation}
where $b_{\text{s}}$ is the length of the slip Burgers vector, $\nu_0$ is a reference velocity, $Q_{\text{s}}$ is the activation energy for slip, $k_B$ is the Boltzmann constant, $T$ is the temperature, $\tau_{\text{eff}}$ is the effective resolved shear stress, $\tau_{\text{sol}}$ is the solid solution strength, $0 < \rho \leq 1$ and $1 \leq q \leq 2 $ are fitting parameters controlling the glide resistance profile.
Blum and Eisenlohr~\cite{blum2009dislocation} models the evolution of dislocation densities, particularly the generation of unipolar dislocation density and formation of dislocation dipoles, respectively, as
\begin{equation}
\dot{\varrho} = \frac{|\dot{\gamma}|}{b_\text{s} \textcolor{black}{\Lambda_\text{s}}} - \frac{2\hat{d}}{b_\text{s}} \varrho |\dot{\gamma}|, \quad
\dot{\varrho}_\text{di} = \frac{2(\hat{d} - \widecheck{d})}{b_\text{s}} \varrho |\dot{\gamma}| - \frac{2 \widecheck{d}}{b_\text{s}} \varrho_\text{di} |\dot{\gamma}| - \varrho_\text{di} \frac{4\nu_\text{cl}}{\hat{d} - \widecheck{d}},
\end{equation}
where the dislocation climb velocity is
$\nu_\text{cl} = \frac{G D_0 V_\text{cl}}{\pi (1 - \nu) k_\text{B} T} \frac{1}{\hat{d} + \widecheck{d}}\exp \left( - \frac{\calQ_\text{cl}}{k_\text{B} T} \right)$,
Strain hardening is described in terms of a dislocation mean free path, where the mean free path is denoted by $\textcolor{black}{\Lambda}$. 
$D_0$ is the pre-factor of self-diffusion coefficient, $V_\text{cl}$ is the activation volume for climb, $\calQ_\text{cl}$ is the activation energy for climb, $\hat{d} = \frac{3G \textcolor{black}{b}_\text{s}}{16\pi |\tau|}$ is the glide \textcolor{black}{plane separation} below which two dislocations form a stable dipole, $\widecheck{d} = D_a b_\text{s}$ is the distance below which two dislocations \textcolor{black}{spontaneously} annihilate.
The mean free path for slip is modeled as
\begin{equation}
\frac{1}{\textcolor{black}{\Lambda}_\text{s}} = \frac{1}{D} + \frac{1}{\lambda_\text{s}} + \frac{1}{\lambda_\text{tw}} + \frac{1}{\lambda_\text{tr}}
\end{equation}
where
\begin{equation}
\frac{1}{\lambda_\text{s}^{\alpha}} = \frac{1}{i_\text{s}} \left( \sum_{\alpha'=1}^{N_\text{s}} p^{\alpha \alpha'} (\varrho^{\alpha'} + \varrho_\text{di}^{\alpha'}) \right)^{1/2}, \quad
\frac{1}{\lambda_{\text{tw}}^{\alpha}} = \sum_{\beta = 1}^{N_{\text{tw}}} h^{\alpha \beta} \frac{f^\beta_{\text{tw}}}{t_\text{tw} (1 - f^\text{tot}_\text{tw})}, \quad
\frac{1}{\lambda_\text{tr}^{\alpha}} = \sum_{\chi=1}^{N_\text{tr}} h^{\alpha \chi} \frac{f^{\chi}_\text{tr}}{t_\text{tr}(1 - f^\text{tot}_\text{tr})},
\end{equation}
where $D$ is the average grain size, $i_\text{s}$ is a fitting parameter, $t_\text{tw}$ is the average twin thickness, and $t_\text{tr}$ is the average $\varepsilon$-martensite thickness.
The mean free path for twinning and for transformation are computed, respectively, as
\begin{equation}
\frac{1}{\textcolor{black}{\Lambda}^\beta_\text{tw}} = \frac{1}{i_\text{tw}} \left( \frac{1}{D} + \sum_{\beta'=1}^{N_\text{tw}} h^{\beta \beta'} f_{\text{tw}}^{\beta'} \frac{1}{t_\text{tw} ( 1 - f^\text{tot}_\text{tw})} \right), \quad
\frac{1}{\textcolor{black}{\Lambda}^\chi_\text{tr}} = \frac{1}{i_\text{tr}} \left( \frac{1}{D} + \sum_{\chi'=1}^{N_\text{tr}} h^{\chi \chi'} f_{\text{tr}}^{\chi'} \frac{1}{t_\text{tr} ( 1 - f^\text{tot}_\text{tr})} \right),
\end{equation}
$i_\text{w}$ and $i_\text{tr}$ are fitting parameters.
The nucleation rates for twins and $\varepsilon$-martensite are $\dot{N} = \dot{N}_0 P_\text{ncs} P$. \textcolor{black}{$\dot{N}_0$ is the number density of potential twin or $\varepsilon$-martensite nuclei per unit time. The probability to form a twin or $\varepsilon$-martensite nucleus is modeled as
\begin{equation}
P_\text{ncs} = 1 - \exp\left[ -\frac{V_{\text{cs}}}{k_\text{B}T} (\tau_\text{r} -\tau) \right],
\end{equation}
where $V_\text{cs}$ is the cross-slip activation volume.
}

\textcolor{black}{
The stress required to form the twin nucleus from an external applied shear stress amounts to
\begin{equation}
\hat{\tau}_\text{r} = \frac{G b_s}{2 \pi (x_0 + x_\text{c})} + \frac{G b_\text{s} \cos(\pi/3)}{2 \pi x_0},
\end{equation}
where the equilibrium separation $x_0$ of Shockley partials in fcc metals is calculated as
\begin{equation}
x_0 = \frac{G}{\Gamma_{\text{sf}}} \frac{b_\text{s}^2}{8\pi} \frac{2+\nu}{1-\nu},
\end{equation}
where $\Gamma_\text{sf}$ is the stacking fault energy and $\nu$ is the Poisson ratio. 
}

\textcolor{black}{
The probability $P$ that a nucleus bows out to form a twin or $\varepsilon$-martensite is
\begin{equation}
P_\text{tw} = \exp \left[ - \left( \frac{\hat{\tau}_\text{tw}}{\tau} \right)^{p_\text{tw}} \right], \quad
P_\text{tr} = \exp \left[ - \left( \frac{\hat{\tau}_\text{tr}}{\tau} \right)^{p_\text{tr}} \right],
\end{equation}
$p_\text{tw}$ and $p_\text{tr}$ are fitting parameters.
The critical stresses for twin and $\varepsilon$-martensite growth are 
\begin{eqnarray}
\hat{\tau}_\text{tw} = \frac{\Gamma_\text{sf}}{3 b_\text{tw}} + \frac{3G b_\text{tw}}{L_\text{tw}}, \\
\hat{\tau}_\text{tr} = \frac{2 \sigma^{\gamma/\varepsilon}}{3b_\text{tr}} + \frac{3 G b_\text{tr}}{L_\text{tr}} + \frac{h \Delta G^{\gamma \to \varepsilon}}{3 b_\text{tr}},
\end{eqnarray}
where $b_\text{tw}$ and $b_\text{tr}$ are the magnitudes of the Burgers vectors for twinning and transformation, respectively, $L_\text{tw}$ and $L_\text{tr}$ are the widths of the respective nuclei, $\sigma^{\gamma/\varepsilon}$ is the interface energy between $\gamma-$ and $\varepsilon-$ phase, $\Delta G^{\gamma \to \varepsilon}$ is the change in Gibbs free energy per unit volume from fcc to the hcp phase. 
}

\textcolor{black}{
The evolution of the twin and $\varepsilon$-martensite volume fractions follows a rate
\begin{equation}
\dot{f} = (1 - f^\text{tot}_\text{tw} - f^\text{tot}_\text{tr}) V \dot{N},
\end{equation}
where their volumes $V$ are assumed of thin discs $V = \frac{\pi}{4} \Gamma^2 t$.
The shearing rates of the $\beta$ twin system and the $\chi$ transformation system are 
\begin{equation}
\dot{\gamma} = \gamma_\text{char} \dot{f}.
\end{equation}
We note that the description of the dislocation-density-based constitutive model is fully described in Section 6.2.3 of Roters et al~\cite{roters2019damask}.
}



\section{Methodology}
\label{sec:Methodology}

\begin{figure}[!htbp]
\centering
\input{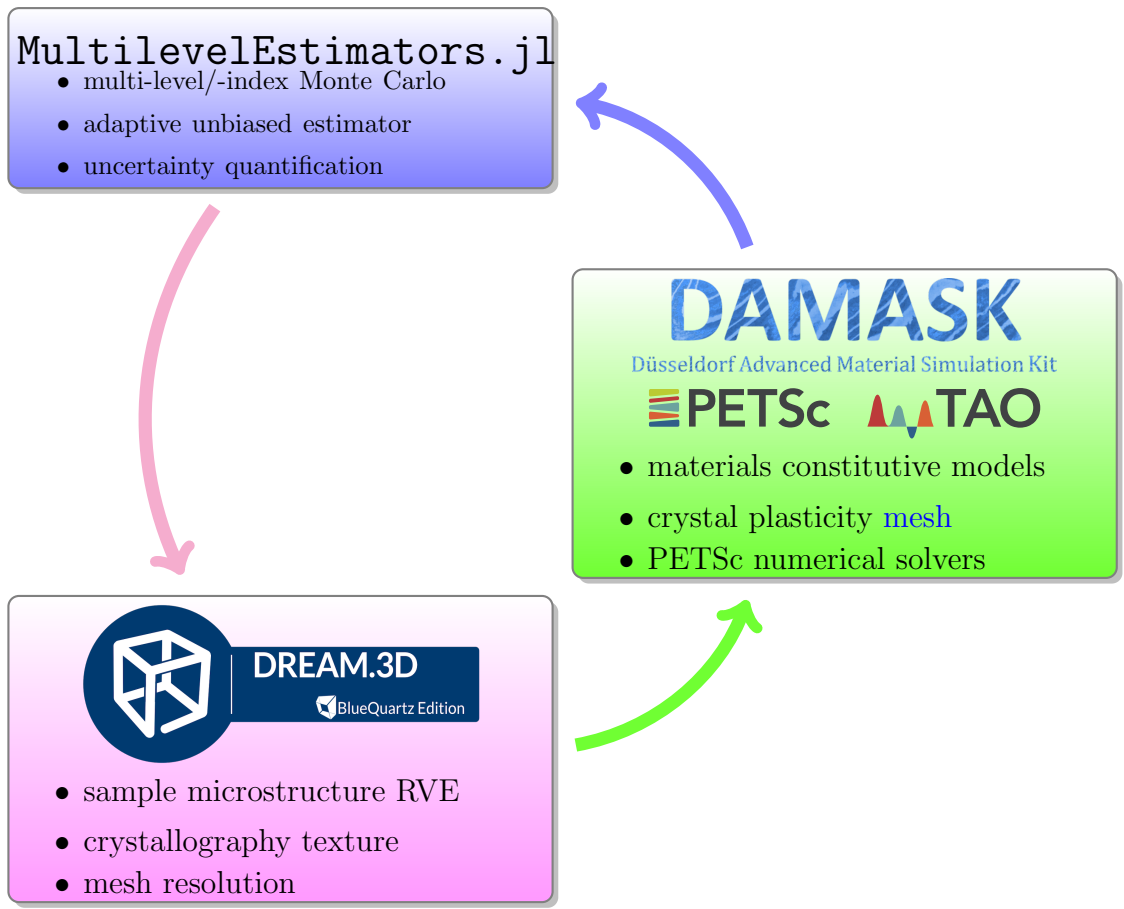}
\caption{Multi-fidelity uncertainty quantification workflow for CPFEM. At each iteration, \texttt{MultilevelEstimators.jl} requests an evaluation of the user code at different fidelity levels for a \textit{fixed} stochastic sample, i.e. a fixed microstructure RVE realization. \texttt{DREAM.3D} is then employed to generate a microstructure RVE on multiple mesh resolutions. \texttt{DAMASK} uses the generated microstructure geometries and subsequently evaluates the quantity of interest using one or more combinations of constitutive model and numerical configurations.}
\label{fig:uq_workflow}
\end{figure}

In this section, we describe the automatic workflow that couples \texttt{DREAM.3D}~\cite{groeber2014dream}, a tool for generating the required microstructures, \texttt{DAMASK}~\cite{roters2012damask}, a unified multi-physics CPFEM simulation package, and \texttt{MultilevelEstimators.jl}, a UQ software package that implements the adaptive MIMC method outlined in \cref{sec:amimc}.
Python scripts are developed to transfer information from \texttt{DREAM.3D} to \texttt{DAMASK}, and from \texttt{DAMASK} to \texttt{MultilevelEstimators.jl}.
We adopted the automatic workflow to couple \texttt{DREAM.3D} and \texttt{DAMASK}~\cite{roters2012damask} from Diehl et al~\cite{diehl2017identifying}. The overall UQ workflow is controlled by \texttt{MultilevelEstimators.jl}.
At each iteration, the package requests an evaluation of the user code with a specific index $\ell$ (in the multi-level setting) or tuple $\mathbf{\bsell}$ (in the multi-index setting).
The request is sent to \texttt{DREAM.3D}, in order to sample one unique microstructure RVE, which is then subsequently coarsened from fine mesh-resolution to coarse mesh-resolution, creating multiple geometries that approximate the same microstructure RVE, as shown in~\cref{fig:sve8} to \cref{fig:sve64}.
As \texttt{DREAM.3D} does not currently support reproducibility for microstructure reconstruction, it is important to save the generated microstructure geometries in order to evaluate the multi-level or multi-index difference.
In the multi-index setting, also an appropriate constitutive model is chosen, before  \texttt{DAMASK} is invoked to run the CPFEM simulation.
The quantity of interest is obtained from a post-process and finally returned to \texttt{MultilevelEstimators.jl}.
The algorithm iterates until a user-specified convergence criterium is met.
The UQ package allows parallelized evaluations of the user code, in order to exploit computational resources on high-performance computing systems. However, in this work, we limit the scope of the demonstration with sequential MC sampling.
Figure~\ref{fig:uq_workflow} shows a schematic illustration for the coupled workflow, which integrates \texttt{MultilevelEstimators.jl} as the UQ toolbox, \texttt{DREAM.3D} as the microstructure generator, and \texttt{DAMASK} as the forward CPFEM package.
It should be noted that, since \texttt{DAMASK} is built upon \texttt{PETSc}, see~\cite{abhyankar2018petsc,balay2019petsc}, it is possible to consider other numerical parameters as fidelity parameters, such as a time step.








\section{Case study 1: MLMC for $\alpha$-Titanium }
\label{sec:MLMC_CaseStudy}

\subsection{CPFE model of $\alpha$-Ti}
\label{subsec:ConstitutiveModelAlphaTi}

\begin{figure}[!htbp]
\centering
\begin{subfigure}[b]{0.175\textwidth}
\centering
\includegraphics[width=\textwidth]{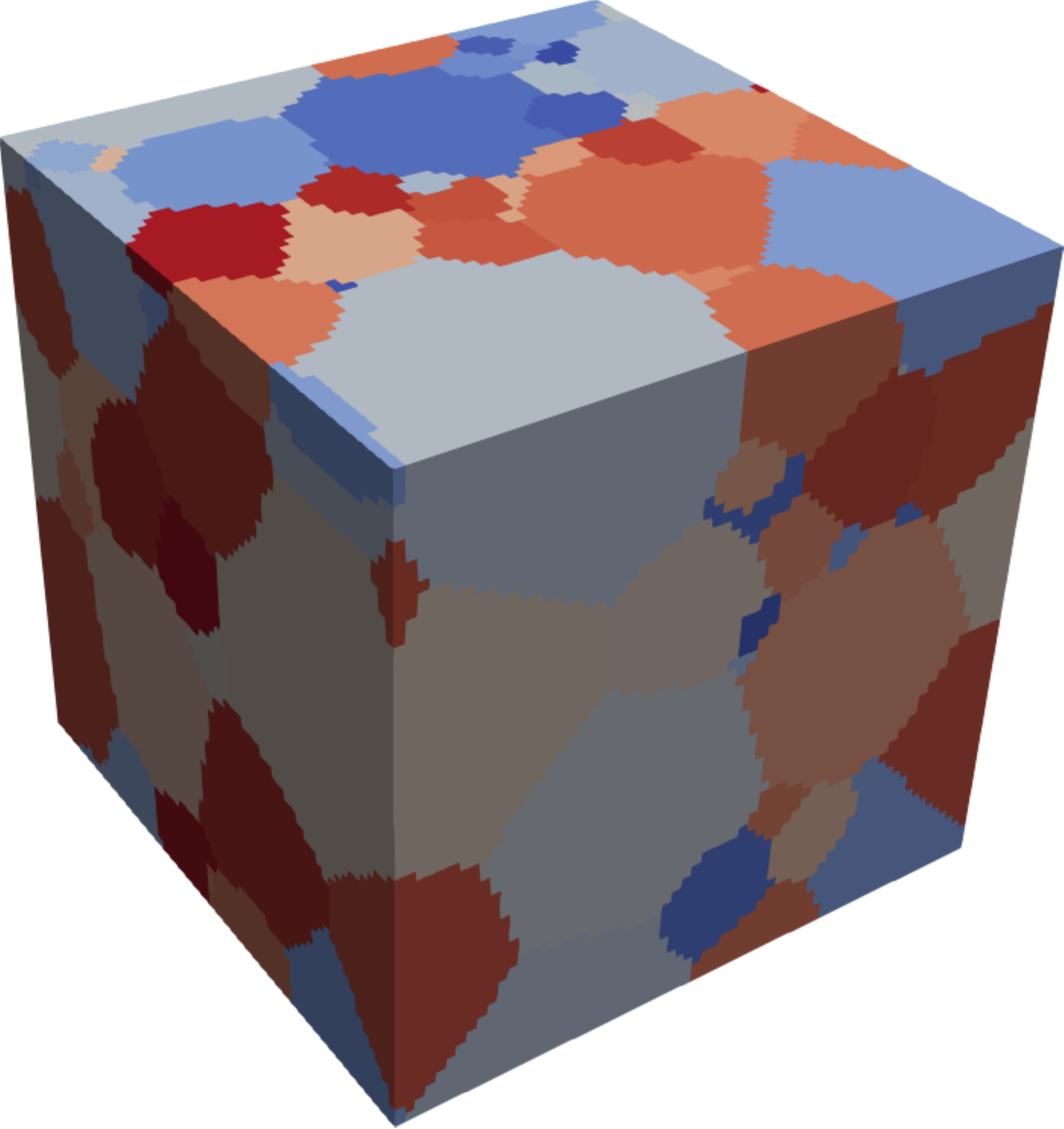}
\caption{\textcolor{black}{Ti RVE 1.}}
\end{subfigure}
\begin{subfigure}[b]{0.175\textwidth}
\centering
\includegraphics[width=\textwidth]{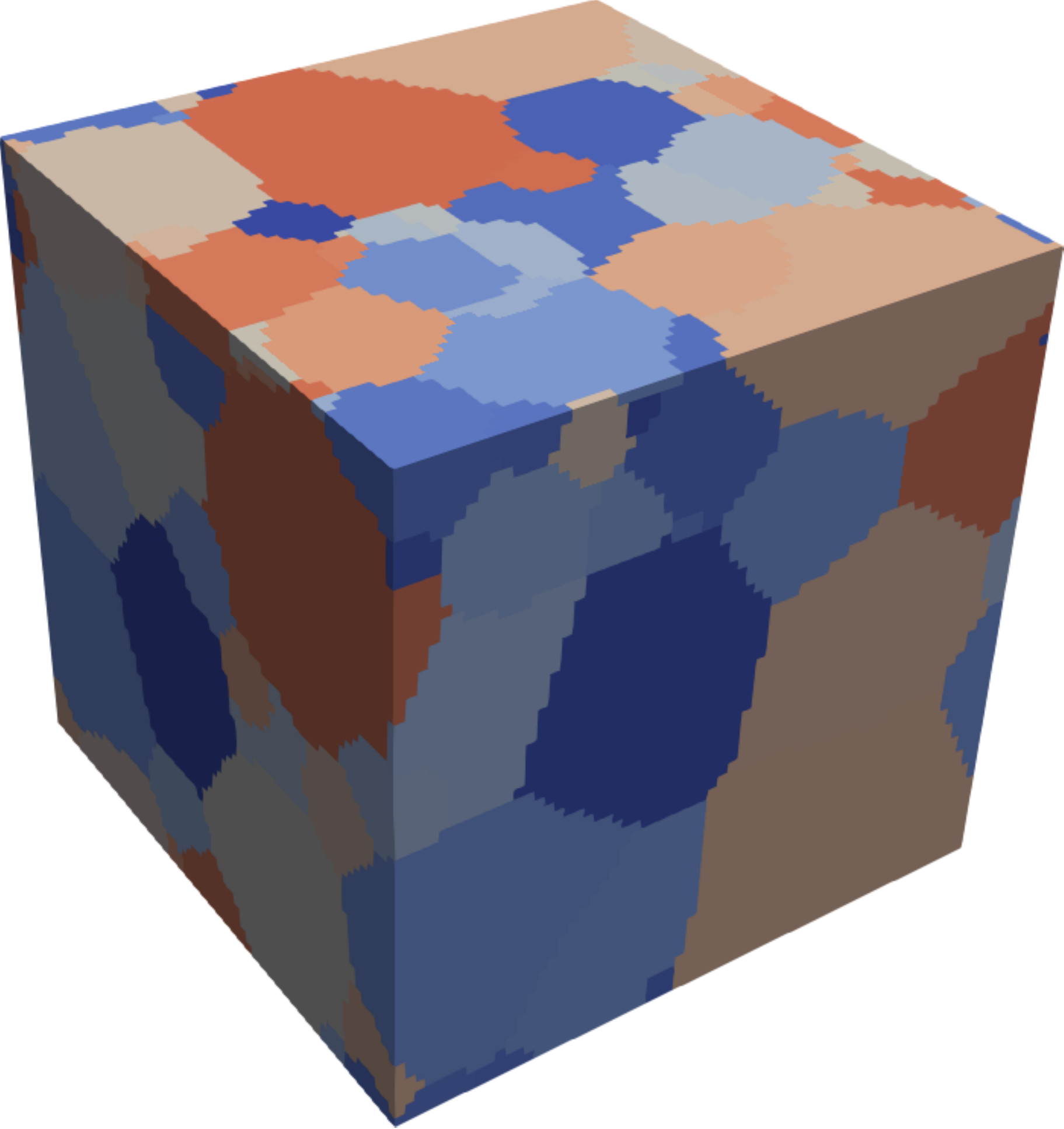}
\caption{\textcolor{black}{Ti RVE 2.}}
\end{subfigure}
\begin{subfigure}[b]{0.175\textwidth}
\centering
\includegraphics[width=\textwidth]{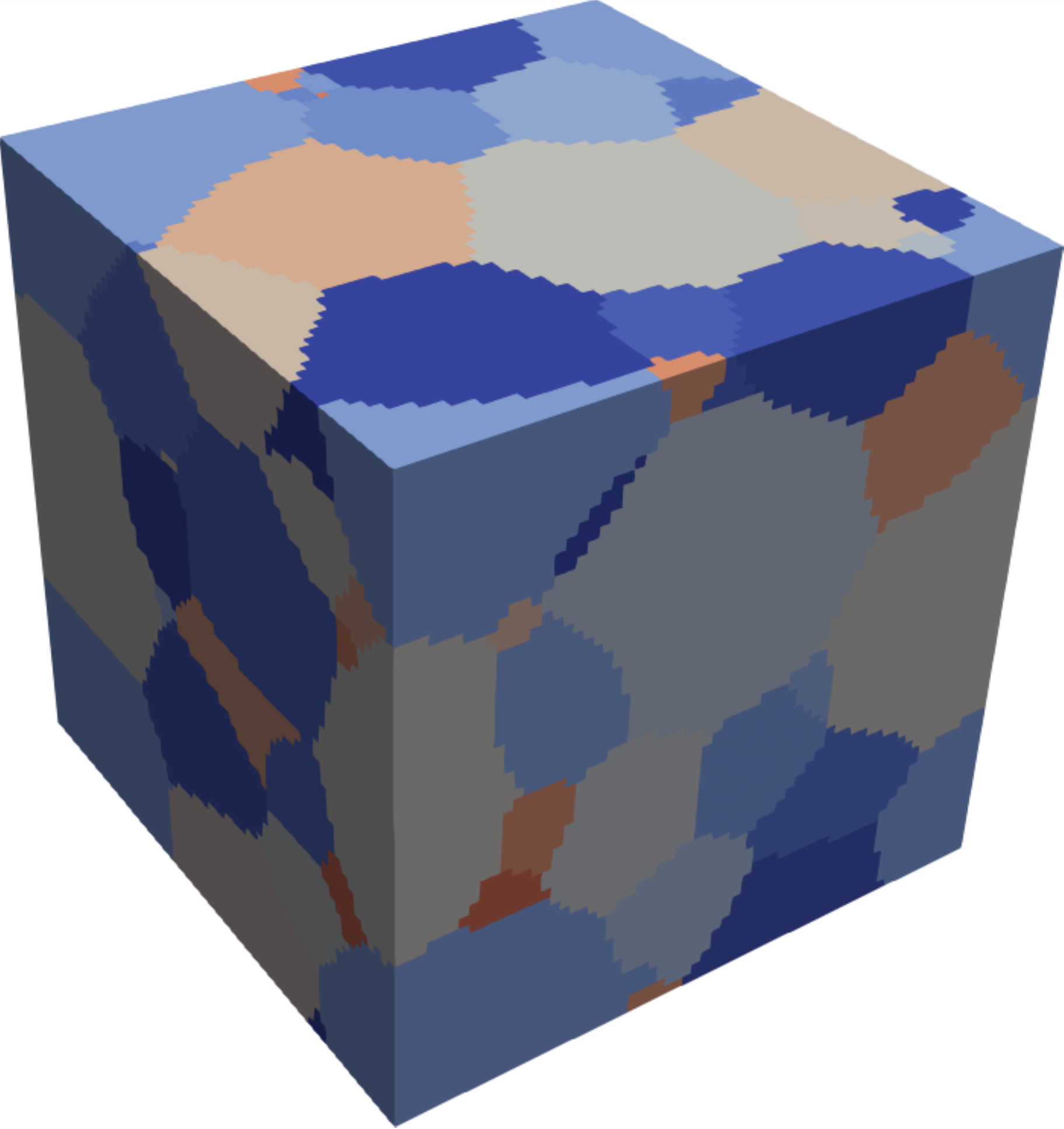}
\caption{\textcolor{black}{Ti RVE 3.}}
\end{subfigure}
\begin{subfigure}[b]{0.175\textwidth}
\centering
\includegraphics[width=\textwidth]{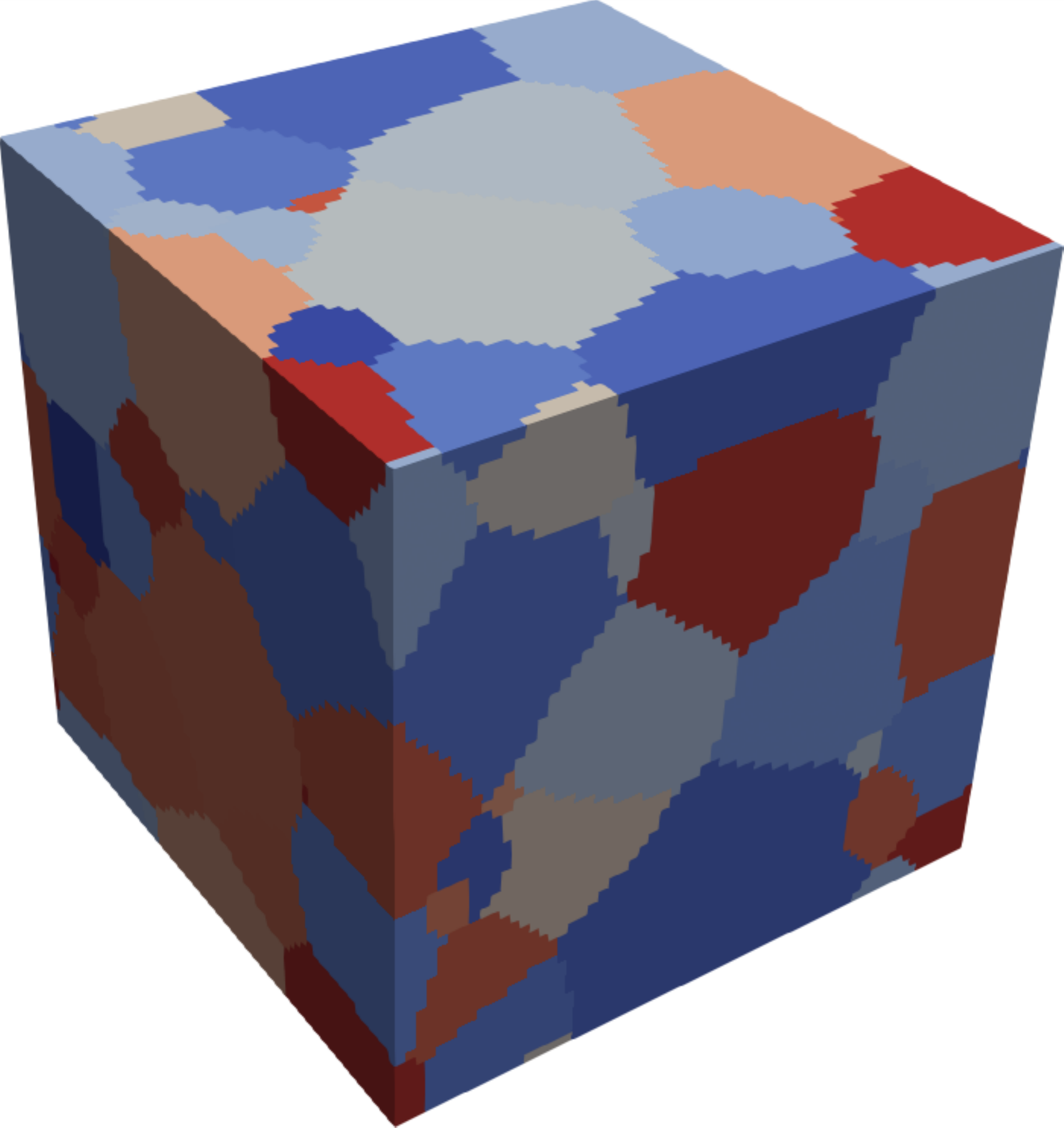}
\caption{\textcolor{black}{Ti RVE 4.}}
\end{subfigure}
\begin{subfigure}[b]{0.175\textwidth}
\centering
\includegraphics[width=\textwidth]{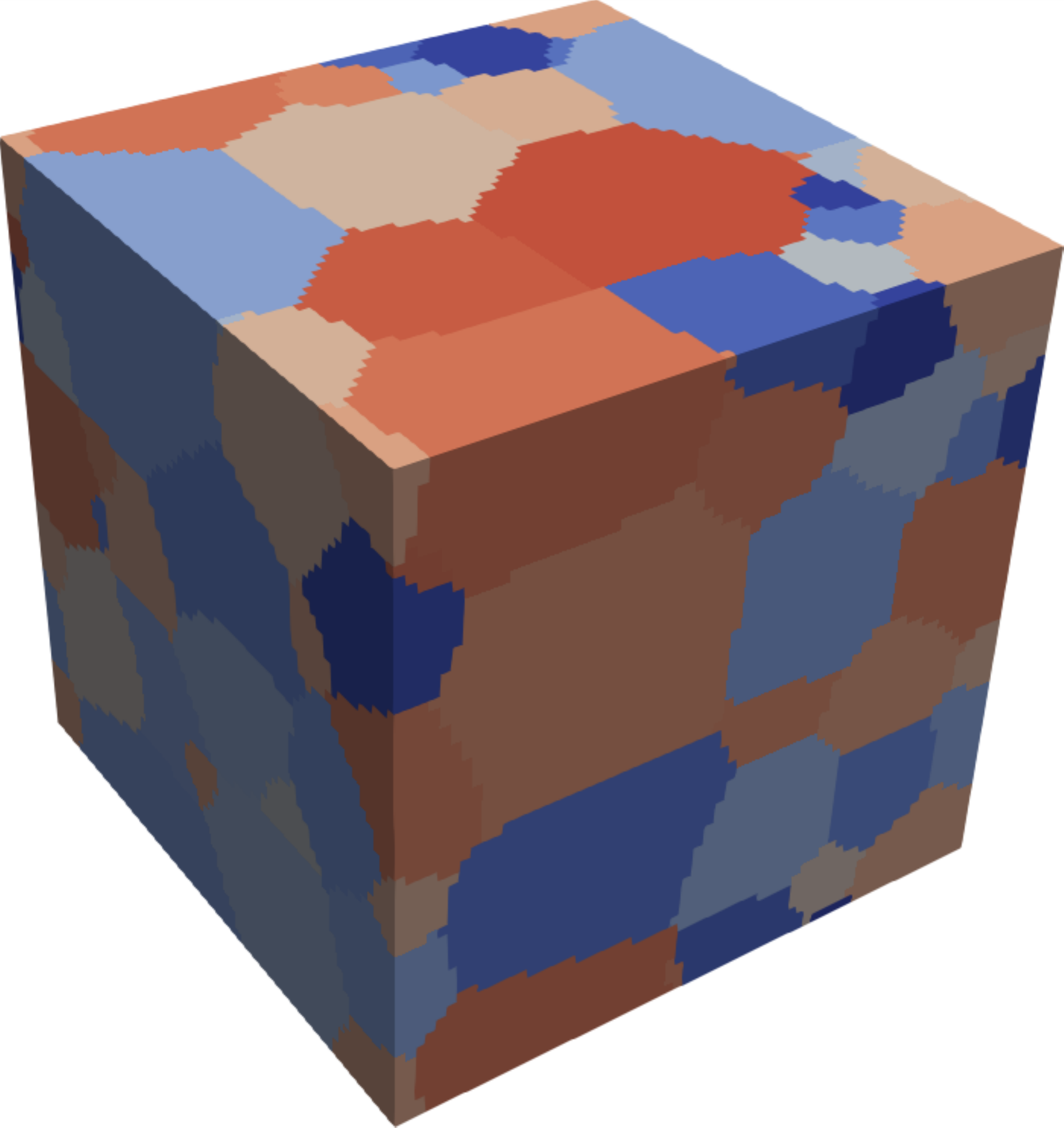}
\caption{\textcolor{black}{Ti RVE 5.}}
\end{subfigure}
\caption{\textcolor{black}{An illustrative microstructure ensemble of 5 $\alpha$-Ti RVEs.}}
\label{fig:5TiEnsemble}
\end{figure}

\begin{figure}[!htbp]
\centering
\includegraphics[width=\textwidth]{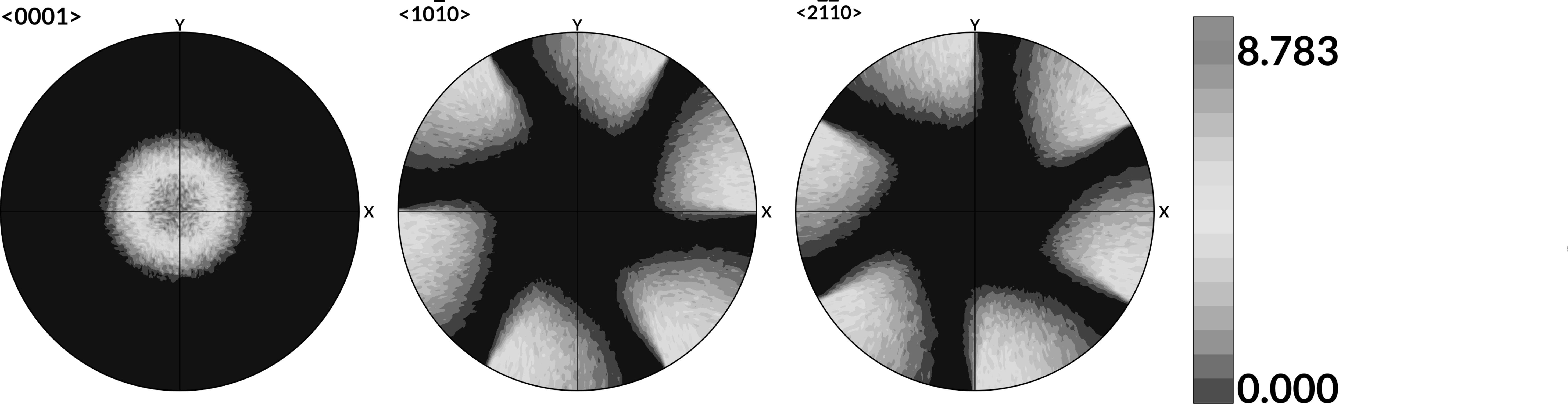}
\caption{Microstructure crystallography texture of $\alpha$-Ti with $(\phi_1, \theta, \phi_2) = (90, 0, 0)$.}
\label{fig:cropped_hcpTexture-90-0-0-Masked-eps-converted-to.pdf}
\end{figure}

In this section, we present the first case study considering MLMC and CPFEM with multiple mesh resolutions, where the material system of interest is hexagonal-closed packed (HCP) $\alpha$-Titanium. The phenomenological constitutive model parameters are listed in \cref{tab:ConstitutiveAlphaTi}.
The constitutive model captures dislocation slip contribution\textcolor{black}{s} to plasticity behavior of $\alpha$-Ti.
The grain size is described by a log-normal distribution, i.e.,
\begin{equation}
p_D(d; \mu_D, \sigma_D) = \frac{1}{d \sigma_D \sqrt{2\pi}} \exp\left({ - \frac{(\ln d - \mu_D)^2}{2\sigma_D^2} }\right),
\label{eq:GrainLogNormalDistribution}
\end{equation}
where $\mu_D$ and $\sigma_D$ are 4.0 and 1.2, respectively, \textcolor{black}{$d$ is in $\mu m$}.
The crystallographic texture for $\alpha$-Ti is shown in \cref{fig:cropped_hcpTexture-90-0-0-Masked-eps-converted-to.pdf}, with the Euler angles of $(\phi_1, \theta, \phi_2) = (90, 0, 0)$.
Microstructure RVEs of 320$\mu$m$^3$ are considered at multiple mesh resolutions. 
Uniaxial loading condition is applied \textcolor{black}{with $\dot{F}_{11} = 10^{-3}$s$^{-1}$.}
\textcolor{black}{Figure~\ref{fig:5TiEnsemble} presents an illustrative microstructure ensemble consisting of five $\alpha$-Ti microstructure RVEs, with the aforementioned grain size and crystallographic texture. In this case study, the quantity of interest is the effective yield stress, calculated by offsetting the effective strain at 0.2\%. }
\textcolor{black}{Readers interested in CPFEM modeling of $\alpha$-Ti are kindly referred to prior works in dislocation-density-based constitutive model~\cite{alankar2011dislocation}, anisotropic indentation response~\cite{zambaldi2012orientation}, influence of grain boundaries on plastic deformation~\cite{su2016quantifying}. 
Twinning is not considered in this constitutive model because it was not observed in nanoindentation experiments~\cite{zambaldi2012orientation,su2016quantifying}, even though later experimental work on electron backscattered diffraction and Laue microdiffraction~\cite{wang2013study} would confirm two tensile twinning modes T1 $\{10\overline{1}2\}\langle\overline{1}011\rangle$ and T2 $\{11\overline{2}1\}\langle\overline{1}\overline{1}26\rangle$, besides the other two compressive twinning modes C1 $\{11\overline{2}2\}\langle11\overline{2}\overline{3}\rangle$ and C2 $\{10\overline{1}1\}\langle10\overline{1}\overline{2}\rangle$. 
}

\begin{table}[!htbp]
\caption{Parameters for $\alpha$-Ti used in this case study ~\cite{zambaldi2012orientation,su2016quantifying}.}
\label{tab:ConstitutiveAlphaTi}
\begin{tabular*}{\textwidth}{c @{\extracolsep{\fill}} cccc} \toprule
variable                              & description                       & units     &  reference value          \\ \midrule
$c/a$                                 & lattice parameter ratio                 &   --    &  1.587            \\
$C_{11}$                              & elastic constant                    &   GPa     &  160.9              \\
$C_{12}$                              & elastic constant                    &   GPa     &   90.0              \\
$C_{13}$                              & elastic constant                    &   GPa     &   66.0              \\
$C_{33}$                              & elastic constant                    &   GPa     &  181.7              \\
$C_{44}$                              & elastic constant                    &   GPa     &   46.5              \\
$\dot{\gamma}_0$                      & slip reference shear rate           &  s$^{-1}$   &  0.001        \\
$\tau_{0,\text{basal} \langle a \rangle }$    & basal $\langle a \rangle$ slip resistance                       &   MPa     & 349.3           \\
$\tau_{0,\text{pris} \langle a \rangle}$      & prismatic $\langle a \rangle$ slip resistance                   &   MPa     & 568.6           \\
$\tau_{0,\text{pyr} \langle c+a \rangle}$     & pyramidal $\langle c+a \rangle$ slip resistance     &   MPa     & 1107.9    \\
$\tau_{\infty,\text{basal} \langle a \rangle }$   & basal $\langle a \rangle$ saturation stress                               &   MPa     &  568.6            \\
$\tau_{\infty,\text{pris} \langle a \rangle}$   & prismatic $\langle a \rangle$ saturation stress                           &   MPa     & 1505.2            \\
$\tau_{\infty,\text{pyr} \langle c+a \rangle}$  & pyramidal $\langle c+a \rangle $ saturation stress    &   MPa     & 3420.1  \\
$h_{0}^{\text{s}-\text{s}}$                 & slip-slip hardening parameter             &   MPa     &  15               \\
$n_\text{s}$                            & slip strain rate sensitivity parameter        &   --      &  20               \\
$a$                                   & slip hardening parameter                &   --  &  2.0                \\ \bottomrule
\end{tabular*}
\normalsize
\end{table}


\subsection{Application of MLMC for $\alpha$-Ti}
\label{sec:mlmc_results}

\begin{figure}[!htbp]
    \centering
    \setlength{\figurewidth}{7cm}
\setlength{\figureheight}{6cm}
\tikzsetnextfilename{cost_per_sample_vs_level}
\begin{tikzpicture}
\begin{axis}[ticklabel style={{font=\small}}, major tick length={2pt}, every tick/.style={{black, line cap=round}}, axis on top, legend style={{draw=none, font=\small, at={(0.03,0.97)}, anchor=north west, fill=none, legend cell align=left}}, xlabel={level $\ell$}, xtick distance={1}, ylabel={$\log_2(\textrm{cost}(\;\cdot\;))$ [seconds]}, xmin=0, xmax=4]
    \addplot[default line, default markers, color={ForestGreen}]
        table[row sep={\\}]
        {
            \\
            0.0  6.08321  \\
            1.0  9.089159131911238  \\
            2.0  9.955126781261366  \\
            3.0  10.754887502163468  \\
            4.0  12.129283016944966  \\
        }
        ;
    \addlegendentry {$Q_{\ell}$}
    \addplot [thick, default dashed line]
        table[y={create col/linear regression={y=1}}, row sep={\\}] 
        {
            \\
            0.0  6.08321  \\
            1.0  9.089159131911238  \\
            2.0  9.955126781261366  \\
            3.0  10.754887502163468  \\
            4.0  12.129283016944966  \\
        }
        ;
    \addlegendentry {$\gamma \approx 1.38$}
\end{axis}
\end{tikzpicture}
    \caption{Increase of the average computational cost per sample expressed in seconds as a function of the level parameter $\ell$ using the mesh refinement as a fidelity parameter in the MLMC experiment. The dashed line corresponds to a fit of the computational cost proportional to $2^{\gamma}$ with $\gamma\approx1.38$\textcolor{black}{, see condition \eqref{eq:C3}}.}
    \label{fig:cost_per_sample_vs_level}
\end{figure}
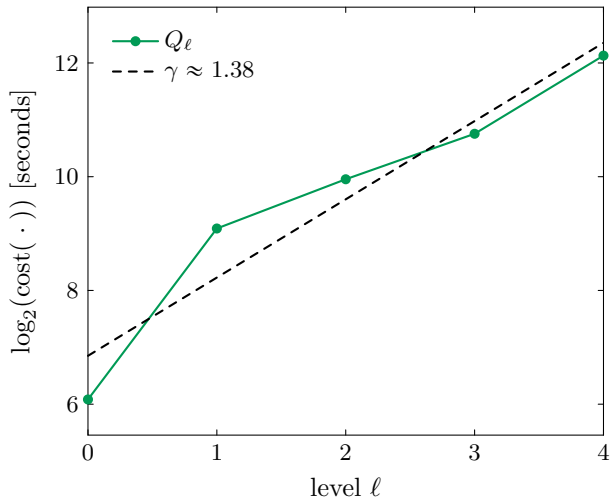

A hierarchy of low-fidelity models is constructed by varying the mesh size of the microstructure RVE, see \cref{fig:sve8} to \cref{fig:sve64} for an illustration. The coarsest microstructure RVE (level $\ell = 0$) is constructed on a $8 \times 8 \times 8$ mesh. The finest microstructure RVE (level $\ell = 4$) is constructed on a $64 \times 64 \times 64$ mesh. Intermediate levels use 16 ($\ell = 1$), 20 $(\ell = 2)$, and 32 ($\ell = 3$) voxels in each dimension, respectively. With this choice of low-fidelity models, the computational cost per sample approximately doubles with increasing level parameter $\ell$, see \cref{fig:cost_per_sample_vs_level}. \textcolor{black}{This means that the model hierarchy satisfies constraint~\eqref{eq:C3} with $\gamma\approx1.38$. This choice for a geometric structure in the number of degrees of freedom per level is a natural one, inspired by the multigrid literature, and has been proposed in other settings, such as the PDE problem in~\cite{cliffe2011multilevel}.}

\begin{figure}[!htbp]
    \centering
    \begin{tikzpicture}
\begin{axis}[ticklabel style={{font=\small}}, major tick length={2pt}, every tick/.style={{black, line cap=round}}, axis on top, legend style={{draw=none, font=\small, at={(1.03,0.97)}, anchor=north west, fill=none, legend cell align=left, /tikz/every odd column/.append style={column sep=3pt}}}, xlabel={level $\ell$}, xtick distance={1}, ylabel={number of model evaluations $N_\ell$}, ymode={log}, xmin=0, xmax=4, ymin=1e0, ymax=1e3]
    \addplot[default line, default markers, color={rgb,1:red,0.5;green,0.0;blue,0.0}]
        table[row sep={\\}]
        {
            \\
            0  677  \\
            1  75  \\
            2  23  \\
            3  4  \\
            4  2  \\
        }
        ;
    \addlegendentry {$\varepsilon^{(10)}=\SI[retain-zero-exponent=true]{5.000e+00}{}$}
    \addplot[default line, default markers, color={rgb,1:red,0.9444;green,0.0;blue,0.0}]
        table[row sep={\\}]
        {
            \\
            0  392  \\
            1  46  \\
            2  14  \\
            3  2  \\
        }
        ;
    \addlegendentry {$\varepsilon^{(9)}=\SI[retain-zero-exponent]{6.500e+00}{}$}
    \addplot[default line, default markers, color={rgb,1:red,1.0;green,0.3889;blue,0.0}]
        table[row sep={\\}]
        {
            \\
            0  219  \\
            1  30  \\
            2  10  \\
            3  2  \\
        }
        ;
    \addlegendentry {$\varepsilon^{(8)}=\SI[retain-zero-exponent]{8.450e+00}{}$}
    \addplot[default line, default markers, color={rgb,1:red,1.0;green,0.8333;blue,0.0}]
        table[row sep={\\}]
        {
            \\
            0  124  \\
            1  18  \\
            2  10  \\
            3  2  \\
        }
        ;
    \addlegendentry {$\varepsilon^{(7)}=\SI{1.099e+01}{}$}
    \addplot[default line, default markers, color={rgb,1:red,0.7222;green,1.0;blue,0.2778}]
        table[row sep={\\}]
        {
            \\
            0  84  \\
            1  14  \\
            2  10  \\
            3  2  \\
        }
        ;
    \addlegendentry {$\varepsilon^{(6)}=\SI{1.428e+01}{}$}
    \addplot[default line, default markers, color={rgb,1:red,0.2778;green,1.0;blue,0.7222}]
        table[row sep={\\}]
        {
            \\
            0  46  \\
            1  10  \\
            2  10  \\
            3  2  \\
        }
        ;
    \addlegendentry {$\varepsilon^{(5)}=\SI{1.856e+01}{}$}
    \addplot[default line, default markers, color={rgb,1:red,0.0;green,0.8333;blue,1.0}]
        table[row sep={\\}]
        {
            \\
            0  31  \\
            1  10  \\
            2  10  \\
            3  2  \\
        }
        ;
    \addlegendentry {$\varepsilon^{(4)}=\SI{2.413e+01}{}$}
    \addplot[default line, default markers, color={rgb,1:red,0.0;green,0.3889;blue,1.0}]
        table[row sep={\\}]
        {
            \\
            0  19  \\
            1  10  \\
            2  10  \\
            3  2  \\
        }
        ;
    \addlegendentry {$\varepsilon^{(3)}=\SI{3.137e+01}{}$}
    \addplot[default line, default markers, color={rgb,1:red,0.0;green,0.0;blue,0.9444}]
        table[row sep={\\}]
        {
            \\
            0  12  \\
            1  10  \\
            2  10  \\
            3  2  \\
        }
        ;
    \addlegendentry {$\varepsilon^{(2)}=\SI{4.079e+01}{}$}
    \addplot[default line, default markers, color={rgb,1:red,0.0;green,0.0;blue,0.5}]
        table[row sep={\\}]
        {
            \\
            0  12  \\
            1  10  \\
            2  10  \\
            3  2  \\
        }
        ;
    \addlegendentry {$\varepsilon^{(1)}=\SI{5.302e+01}{}$}
\end{axis}
\end{tikzpicture}
    \caption{Required number of samples $N_\ell$ at each level $\ell = 0, 1, \ldots, 4$ to reach a target accuracy of $\varepsilon^{(k)}$, $k=1, 2, \ldots, 10$, in the MLMC experiment. On levels $\ell=0, 1$ and $2$, a minimum number of warm-up samples $N_\ell = 10$ is imposed to get an initial estimate of the variance of the multi-level differences $\bbV[\Delta Q_\ell]$. For the target tolerance $\varepsilon^{\text{target}} = \varepsilon^{(10)}$, only two evaluations of the high-fidelity model are required. \textcolor{black}{Note that the number of samples $N_\ell$ for $\varepsilon^{(1)}$ and $\varepsilon^{(2)}$ coincide on the figure.}}
    \label{fig:mlmc_number_of_samples}
\end{figure}
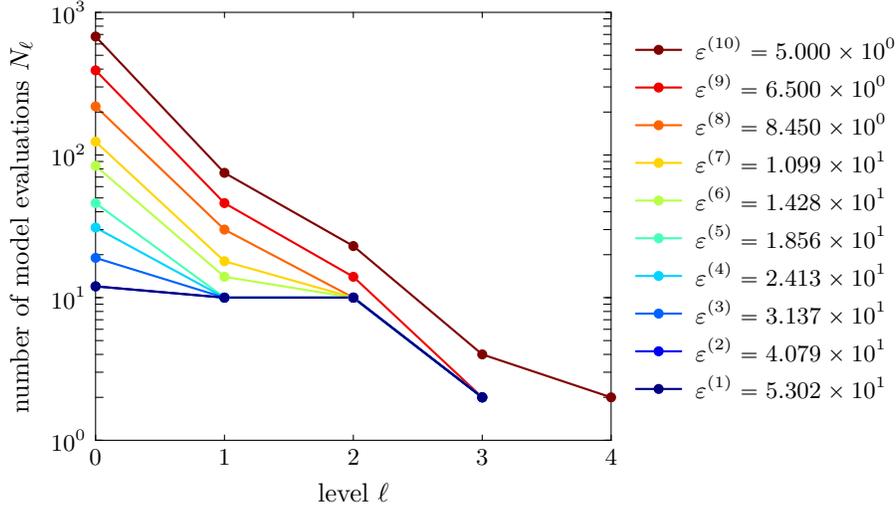

With this hierarchy of low-fidelity models, we set up an MLMC experiment for a sequence of decreasing absolute tolerances on the RMSE defined as  $\varepsilon^{(k)} = \frac{\varepsilon^{(K)}}{f^{K - k}}$, $k=1, 2, ..., K$, with $f = 1.3$, $K = 10$ and a target absolute tolerance of $\varepsilon^{(K)} = 5$. In \cref{fig:mlmc_number_of_samples}, we plot the number of model evaluations taken on each level $\ell = 0, 1, \ldots, 4$, for different target tolerances $\varepsilon^{(k)}$. Notice how most model evaluations are taken on the coarser levels, where samples are cheap, and only a handful of model evaluations are required on the finer levels. For example, on the finest level $\ell = 4$ and for the target tolerance $\varepsilon^{(K)}$, only two model evaluations are required. Notice that for levels $\ell=0, 1$ and $2$, a minimum number of samples $N_\ell = 10$ is required. These \emph{warm-up} samples are used to estimate the variance of the multi-level differences $\bbV[\Delta Q_\ell]$, as they appear in expression \eqref{eq:optimalSamples_mlmc} for the optimal number of samples $N_\ell$. On levels $\ell = 3$ and $\ell = 4$, the variance of the multi-level difference is estimated by linear extrapolation through the already available estimates for $\bbV[\Delta Q_\ell]$ on previous levels.

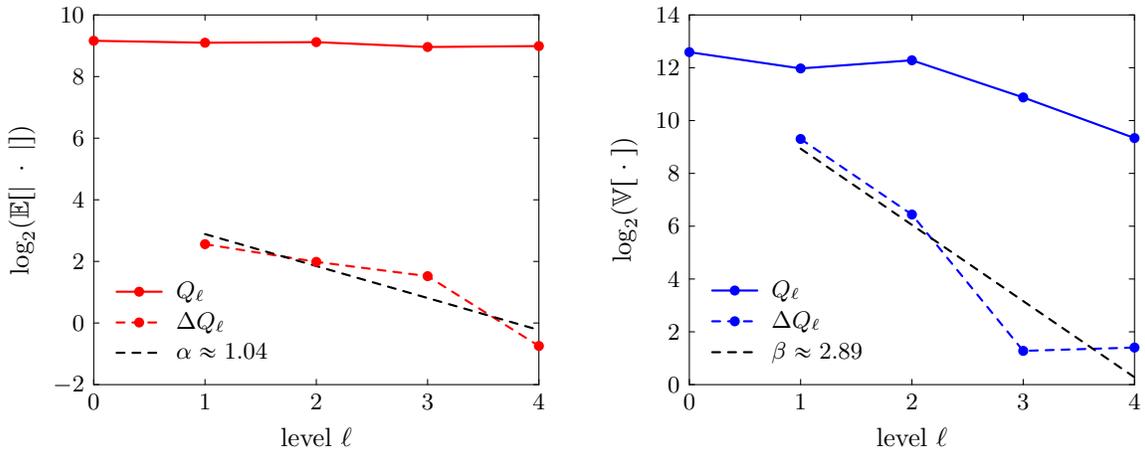
\begin{figure}[!htbp]
    \centering
    \setlength{\figurewidth}{7.5cm}
\setlength{\figureheight}{6.5cm}
\begin{tikzpicture}
\begin{groupplot}[group style={group size={2 by 1}, horizontal sep=2cm}, width=\figurewidth, height=\figureheight]
\nextgroupplot[ticklabel style={{font=\small}}, major tick length={2pt}, every tick/.style={{black, line cap=round}}, axis on top, legend style={{draw=none, font=\small, at={(0.03,0.03)}, anchor=south west, fill=none, legend cell align=left, /tikz/every odd column/.append style={column sep=3pt}}}, xlabel={level $\ell$}, xtick distance={1}, ylabel={$\log_2(\mathbb{E}[|\;\cdot\;|])$}, xmin=0, xmax=4, ytick distance=2, ymax=10, ymin=-2]
    \addplot[default line, color={red}, default markers]
        table[row sep={\\}]
        {
            \\
            0.0  9.161760162764576  \\
            1.0  9.09832648907502  \\
            2.0  9.117442590503302  \\
            3.0  8.959338154156868  \\
            4.0  8.989980542188398  \\
        }
        ;
    \addlegendentry {$Q_{\ell}$}
    \addplot[default dashed line, color={red}, default markers]
        table[row sep={\\}]
        {
            \\
            1.0  2.55926860036959  \\
            2.0  1.9852216217155736  \\
            3.0  1.5210004679052862  \\
            4.0  -0.7452850320230633  \\
        }
        ;
    \addlegendentry {$\Delta Q_{\ell}$}
    \addplot [thick, default dashed line]
        table[y={create col/linear regression={y=1}}, row sep={\\}]
        {
            \\
            1.0  2.55926860036959  \\
            2.0  1.9852216217155736  \\
            3.0  1.5210004679052862  \\
            4.0  -0.7452850320230633  \\
        }
        ;
    \addlegendentry {$\alpha \approx 1.04$}
    \nextgroupplot[ticklabel style={{font=\small}}, major tick length={2pt}, every tick/.style={{black, line cap=round}}, axis on top, legend style={{draw=none, font=\small, at={(0.03,0.03)}, anchor=south west, fill=none, legend cell align=left, /tikz/every odd column/.append style={column sep=3pt}}}, xlabel={level $\ell$}, xtick distance={1}, ylabel={$\log_2(\mathbb{V}[\;\cdot\;])$}, xmin=0, xmax=4, ytick distance=2, ymax=14, ymin=0]
    \addplot[default line, color={blue}, default markers]
        table[row sep={\\}]
        {
            \\
            0.0  12.590698525787603  \\
            1.0  11.974038054037813  \\
            2.0  12.282704630313026  \\
            3.0  10.876809129598959  \\
            4.0  9.337540767001641  \\
        }
        ;
    \addlegendentry {$Q_{\ell}$}
    \addplot[default dashed line, color={blue}, default markers]
        table[row sep={\\}]
        {
            \\
            1.0  9.303469303092847  \\
            2.0  6.441346490247995  \\
            3.0  1.2782005623012813  \\
            4.0  1.4043965671769456  \\
        }
        ;
    \addlegendentry {$\Delta Q_{\ell}$}
    \addplot [thick, default dashed line]
        table[y={create col/linear regression={y=1}}, row sep={\\}]
        {
            \\
            1.0  9.303469303092847  \\
            2.0  6.441346490247995  \\
            3.0  1.2782005623012813  \\
            4.0  1.4043965671769456  \\
        }
        ;
    \addlegendentry {$\beta \approx 2.89$}
\end{groupplot}
\end{tikzpicture}
    \caption{Behaviour of the expected value (\emph{left}) and variance (\emph{right}) of the quantity of interest $Q_\ell$ and the multi-level difference $\Delta Q_\ell$ as a function of the level $\ell$ using the mesh refinement as a fidelity parameter. We numerically fitted the values $\bbE[\Delta Q_\ell] \;\propto\; 2^{-\alpha \ell}$ with $\alpha \approx 1.04$ and $\bbV[\Delta Q_\ell] \;\propto\; 2^{-\beta \ell}$ with $\beta \approx 2.89$.}
    \label{fig:mlmc_expected_value_and_variance}
\end{figure}

In \cref{fig:mlmc_expected_value_and_variance}, we illustrate the decay of the expected value of the multi-level differences $\bbE[\Delta Q_\ell]$ and the decay of the variance of the multi-level differences $\bbV[\Delta Q_\ell]$. The latter quantity expresses the efficiency of the low-fidelity models as a control variate for the quantity of interest. 
The faster the decay of the variances $\bbV[\Delta Q_\ell]$, the more efficient the MLMC estimator will be. In this experiment, we numerically fitted the values $\bbE[\Delta Q_\ell] \;\propto\; 2^{-\alpha \ell}$ with $\alpha \approx 1.04$ and $\bbV[\Delta Q_\ell] \;\propto\; 2^{-\beta \ell}$ with $\beta \approx 2.89$. Notice how the value for $\bbV[\Delta Q_\ell]$ and $\bbE[\Delta Q_\ell]$ at level $\ell=4$ is extract from only two model evaluations, so the predicted value for the expected value and variance of the multi-level difference may be inaccurate. As a consequence, the rate $\beta$ might be an underestimation of the actual value. In either case, with $2\alpha \geq \min(\beta, \gamma)$ and $\beta > \gamma$, we expect the cost of the MLMC estimator to scale as $O(\varepsilon^{-2})$, where $\varepsilon$ is the imposed tolerance on the RMSE, i.e., the most optimistic scenario from \eqref{eq:mlmc_theorem}. This is indeed confirmed in \cref{fig:mlmc_cost_comparison}, where we show the cost of the MLMC estimator, expressed in wall clock time (seconds), as a function of the imposed tolerance $\varepsilon$. For comparison, and also on \cref{fig:mlmc_cost_comparison}, we indicate the cost of an equivalent single-fidelity MC simulation. We did not actually perform these simulations, because of their excessive computational requirements, but estimated the cost of the corresponding MC simulations using the average cost of a high-fidelity simulation and the estimated variance of the quantity of interest. Notice how the MLMC simulation for the target tolerance $\varepsilon^{\text{target}} = 5$ is about 12 times faster than an equivalent MC simulation, as the computational cost is reduced from 31 days to 2 and a half days. The numerically observed cost-complexity rate of the MLMC method is $\calO(\varepsilon^{-2})$, asymptotically for $\varepsilon \rightarrow 0$, as predicted.

\begin{figure}[!htbp]
\centering
\includegraphics[width=\textwidth]{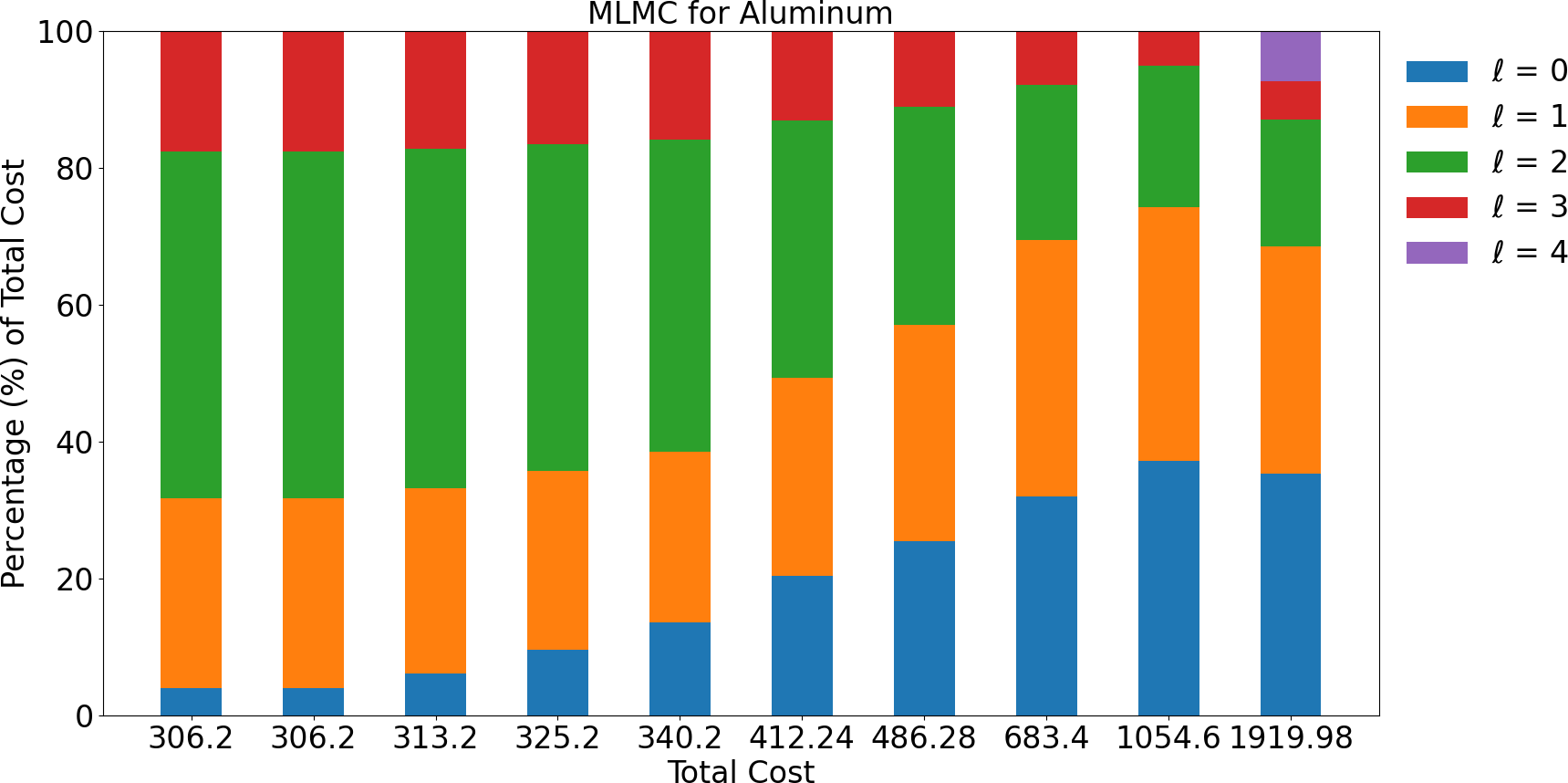}
\caption{\textcolor{black}{The distribution of the total cost across different levels. The level of fidelity corresponds to the discretization of the mesh: $8^3$ ($\ell=0$), $16^3$ ($\ell=1$), $20^3$ ($\ell=2$), $32^3$ ($\ell=3$), $64^3$ ($\ell=4$).}}
\label{fig:cropped_mlmcCostDecomp}
\end{figure}

\textcolor{black}{Figure~\ref{fig:cropped_mlmcCostDecomp} show the distribution of the total cost across different levels, as a function of the tolerance $\varepsilon$ on the RMSE of the MLMC estimator. As the tolerance $\varepsilon$ decreases, the cost of the MLMC estimator increases, a larger fraction of the cost is spent on the coarser levels, and only a minor fraction of the cost is spent on the finer levels. The total cost here is measured in computational time (in seconds) spent on these levels.}

\section{Case study 2: MIMC for Aluminum}
\label{sec:MIMC_CaseStudy}

\subsection{CPFE models of Aluminum}
\label{subsec:ConstitutiveModelAluminum}

\begin{table}[!htbp]
\caption{Parameters for phenomenological constitutive model for Aluminum (cf. Table 2 in Roters et al~\cite{roters2019damask}).}
\label{tab:AlumParamsPhenomenological}
\begin{tabular*}{\textwidth}{c @{\extracolsep{\fill}} cccc} \hline
variable                              & description                         & units     &  reference value          \\ \hline
$C_{11}$                              & elastic constant                    &   GPa     &  106.75             \\
$C_{12}$                              & elastic constant                    &   GPa     &   60.41             \\
$C_{44}$                              & elastic constant                    &   GPa     &   28.34             \\
$\dot{\gamma}_0$                      & reference shear rate                &  s$^{-1}$ &   0.001             \\
$\tau_0$                              & slip resistance                     &   MPa     &  31.0               \\
$\tau_{\infty}$                       & saturation stress                   &   MPa     &  63.0               \\
$h_{0}$                               & slip hardening parameter            &   MPa     &  75                 \\
$n$                                   & strain rate sensitivity parameter   &   --      &  20                 \\
$a$                                   & slip hardening parameter            &   --      &  2.25               \\
$h^{\alpha \beta'}$                   & slip-slip interaction matrix component     &   --      &  1.0 or 1.4         \\ \hline
\end{tabular*}
\normalsize
\end{table}

\begin{table}[!htbp]
\caption{Parameters for dislocation-density-based constitutive model for Aluminum (cf. Table 7.1 in Kords~\cite{kords2013role}).}
\label{tab:AlumParamsDislocationDensity}
\begin{tabular*}{\textwidth}{c @{\extracolsep{\fill}} cccc} \hline
variable                              & description                         & units             &  reference value          \\ \hline
$C_{11}$                              & elastic constant                    &   GPa             &  106.75             \\
$C_{12}$                              & elastic constant                    &   GPa             &   60.41             \\
$C_{44}$                              & elastic constant                    &   GPa             &   28.34             \\
$\mu$                                 & isotropic shear modulus             &   GPa             &   26.27             \\
$\nu$                                 & Poisson ratio                       &                   &   0.345             \\
$b$                                   & length of Burgers vector            &   nm              &   0.286             \\
$\Omega$                              & atomic volume                       &   nm$^3$          &   0.017             \\
$\widecheck{d}_{\text{e}}$            & minimum edge \textcolor{black}{di}pole separation        &   nm              &   1.6               \\
$\widecheck{d}_{\text{s}}$            & minimum screw \textcolor{black}{di}pole separation       &   nm              &   10                \\
$\lambda_0$                           & dislocation multiplication constant &   --              &   60                \\
$k_1$                                 & edge contribution to multiplication &   --              &   0.1               \\
$\rho_0$                              & initial overall dislocation density &   m$^{-2}$        &   $6 \cdot 10^{10}$ \\
$D_{\text{SD}}$                       & self-diffusivity (at $T=300K$)      &   m$^{2}$s$^{-1}$ &   $7 \cdot 10^{-29}$ \\
$Q_\text{S}$                          & solid-solution activation energy    &   eV              &   1.25               \\
$c_\text{at}$                         & solid-solution concentration        &   --              &   $1.5 \cdot 10^{-6}$ \\
$d_\text{obst}$                       & solid-solution size                 &   nm              &   $0.572$           \\
$\tau_\text{Peierls}$                 & Peierls stress                      &   MPa             &   0.1                \\
$w_\text{k}$                          & double kink width                   &   nm              &   2.86               \\
$ p $                                 & energy barrier profile constants    &   --              &   1.0               \\
$ q $                                 & energy barrier profile constants    &   --              &   1.0               \\
$\nu_{\alpha}$                        & attack frequency                    &   GHz             &   50                \\
$\eta        $                        & dislocation viscosity               &   Pa s            &   0.01              \\
$k_3$                                 & edge jog formation factor           &   --              &   1.0               \\ \hline
\end{tabular*}
\normalsize
\end{table}

In this case study, we consider a MIMC case study with multiple mesh resolutions and multiple constitutive models, simultaneously.
The first index $\ell_1$ of $\bsell = (\ell_1, \ell_2)$ corresponds to the mesh resolution index, whereas the second index $\ell_2$ corresponds to the constitutive model index, respectively.
The phenomenological constitutive model (i.e. $\ell_2 = 0$) is considered as the low-fidelity model, whereas the dislocation-density-based (i.e. $\ell_2 = 1$) is considered as the high-fidelity constitutive model.
Again, multiple mesh resolutions for microstructure RVEs are considered in this case study, varying at $8^3$, $16^3$, $20^3$, $32^3$, and $64^3$, which corresponds to $\ell_1 = 0,1,2,3,4$, respectively, and similar to the first case study (\cref{sec:MLMC_CaseStudy}).
A schematic illustration of the multi-fidelity hierarchy is shown in~\cref{fig:mimc_hierarchy}.
Following previous studies~\cite{roters2019damask,wicke2019mixed,han2020using,zhao2008investigation}, we utilize the values of the model parameters listed in Table~\ref{tab:AlumParamsPhenomenological} and Table~\ref{tab:AlumParamsDislocationDensity} for phenomenological and dislocation-density-based constitutive models, respectively.

\begin{figure}[!htbp]
\centering
\includegraphics[width=\textwidth]{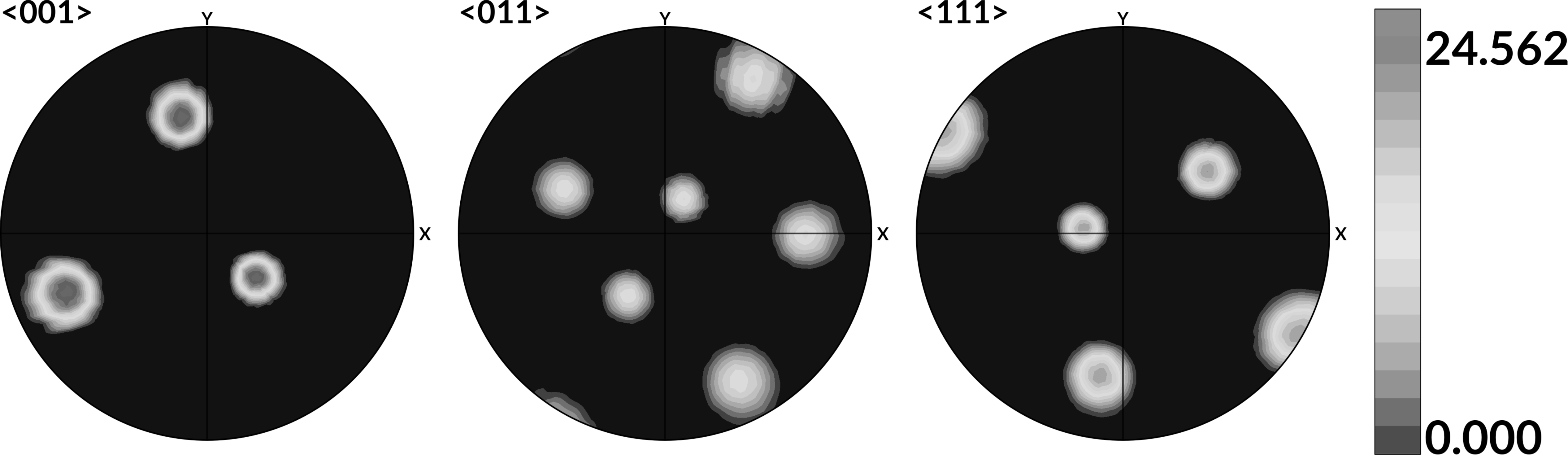}
\caption{Microstructure crystallography texture of Aluminum with $(\phi_1, \theta, \phi_2) = (45, 35, 65)$.}. 
\label{fig:cropped_fccTexture-45-35-65-Masked-eps-converted-to.pdf}
\end{figure}

\begin{figure}[!htbp]
\centering
\begin{subfigure}[b]{0.175\textwidth}
\centering
\includegraphics[width=\textwidth]{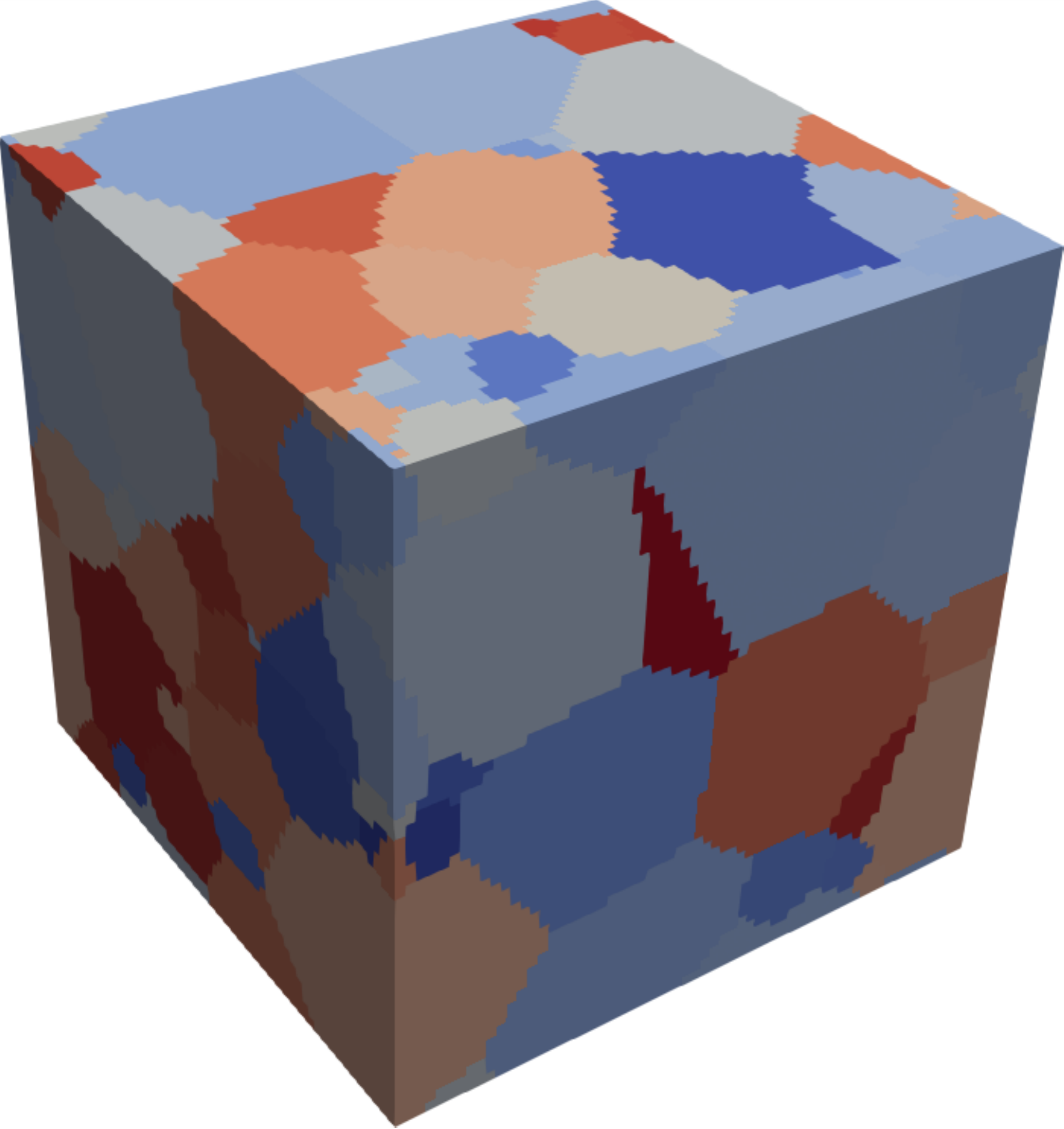}
\caption{\textcolor{black}{Al RVE 1.}}
\end{subfigure}
\begin{subfigure}[b]{0.175\textwidth}
\centering
\includegraphics[width=\textwidth]{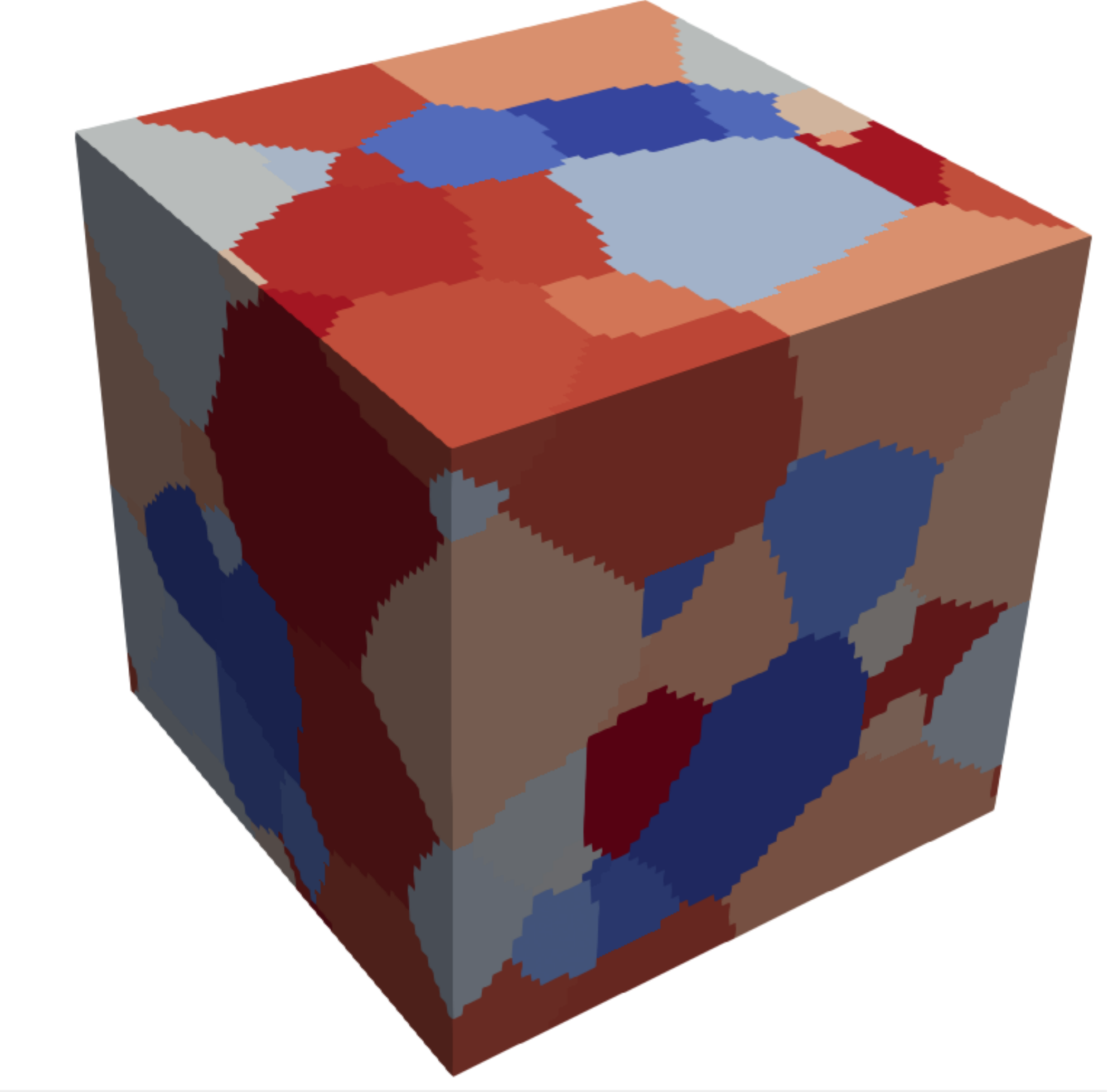}
\caption{\textcolor{black}{Al RVE 2.}}
\end{subfigure}
\begin{subfigure}[b]{0.175\textwidth}
\centering
\includegraphics[width=\textwidth]{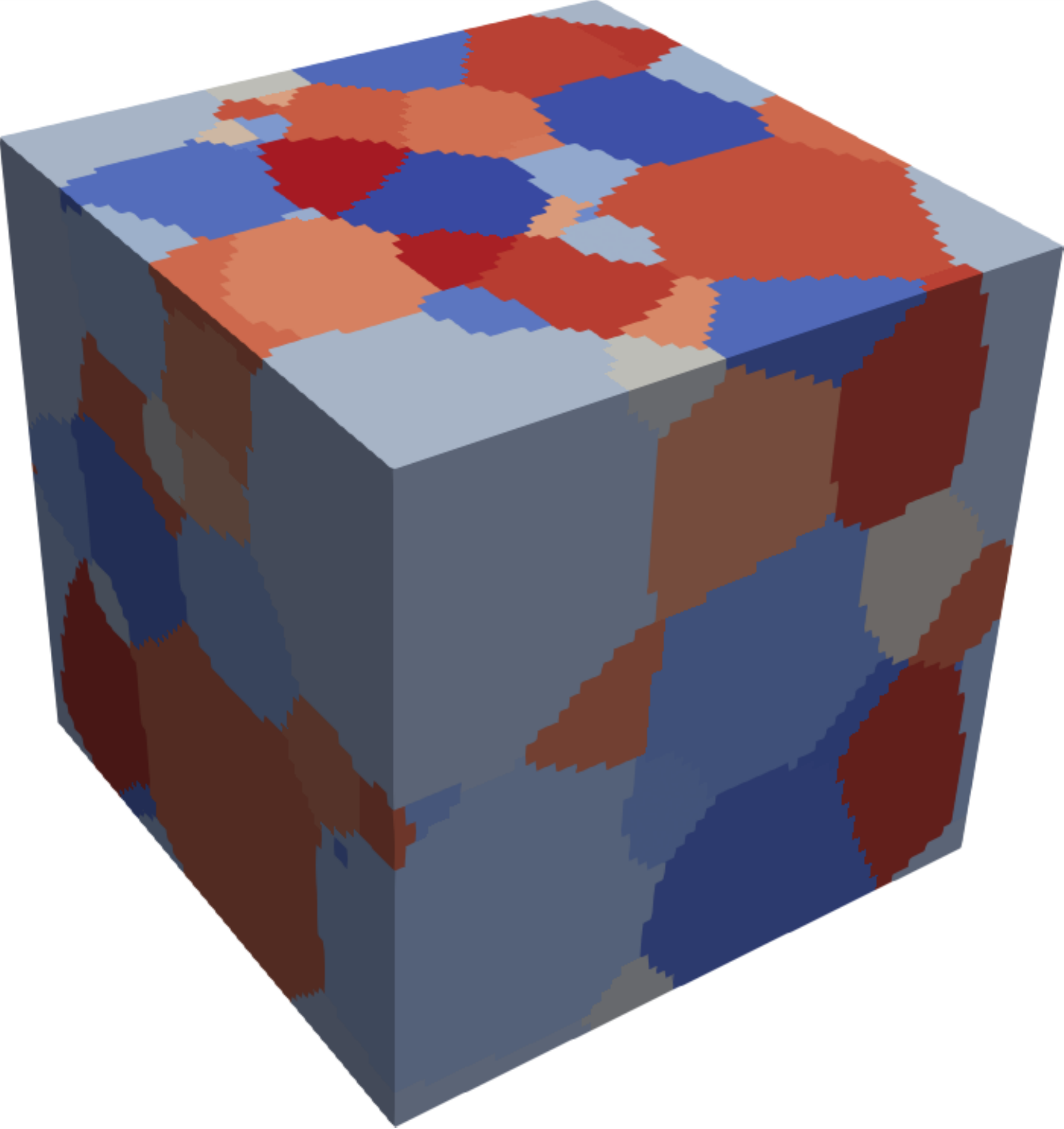}
\caption{\textcolor{black}{Al RVE 3.}}
\end{subfigure}
\begin{subfigure}[b]{0.175\textwidth}
\centering
\includegraphics[width=\textwidth]{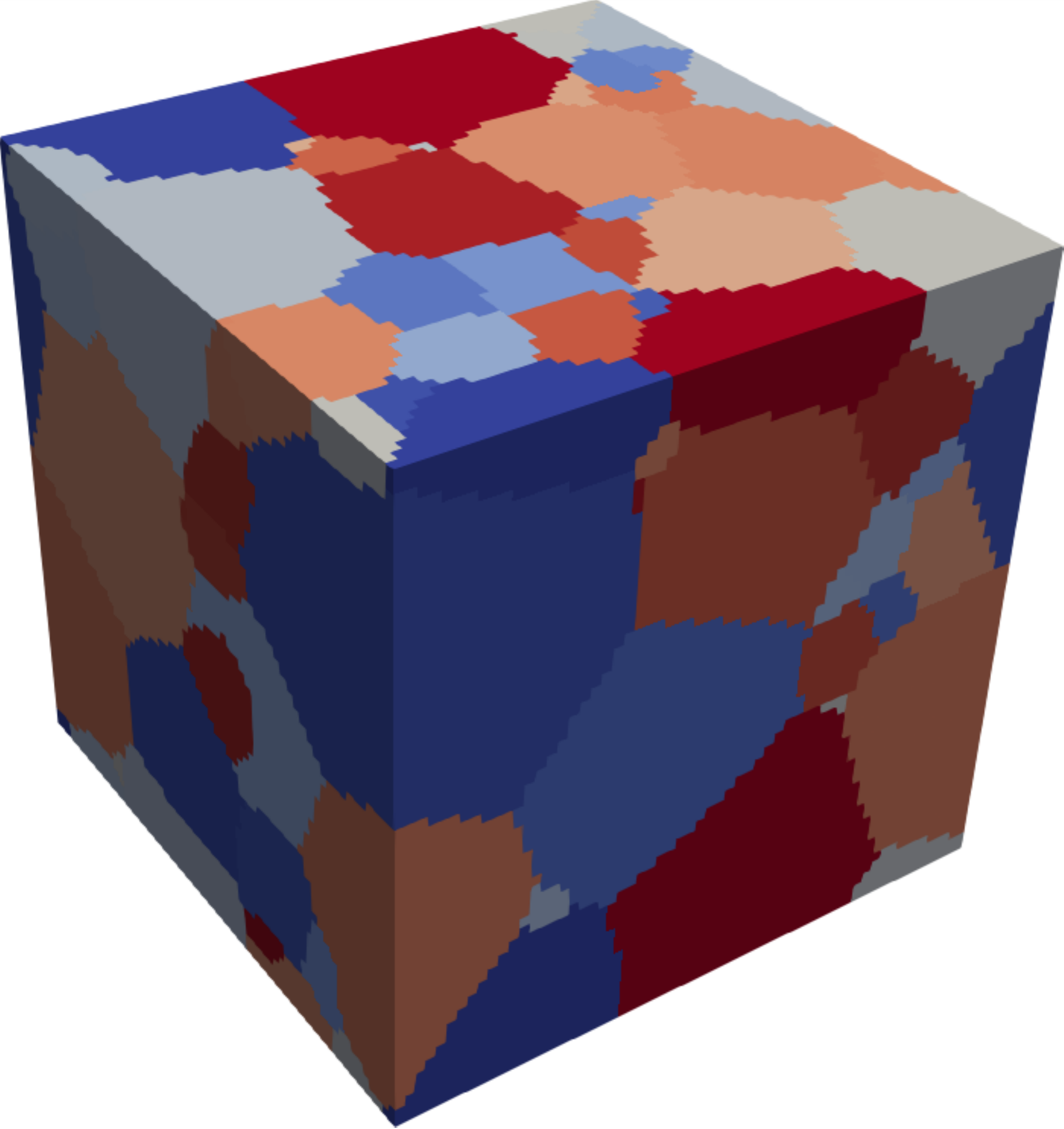}
\caption{\textcolor{black}{Al RVE 4.}}
\end{subfigure}
\begin{subfigure}[b]{0.175\textwidth}
\centering
\includegraphics[width=\textwidth]{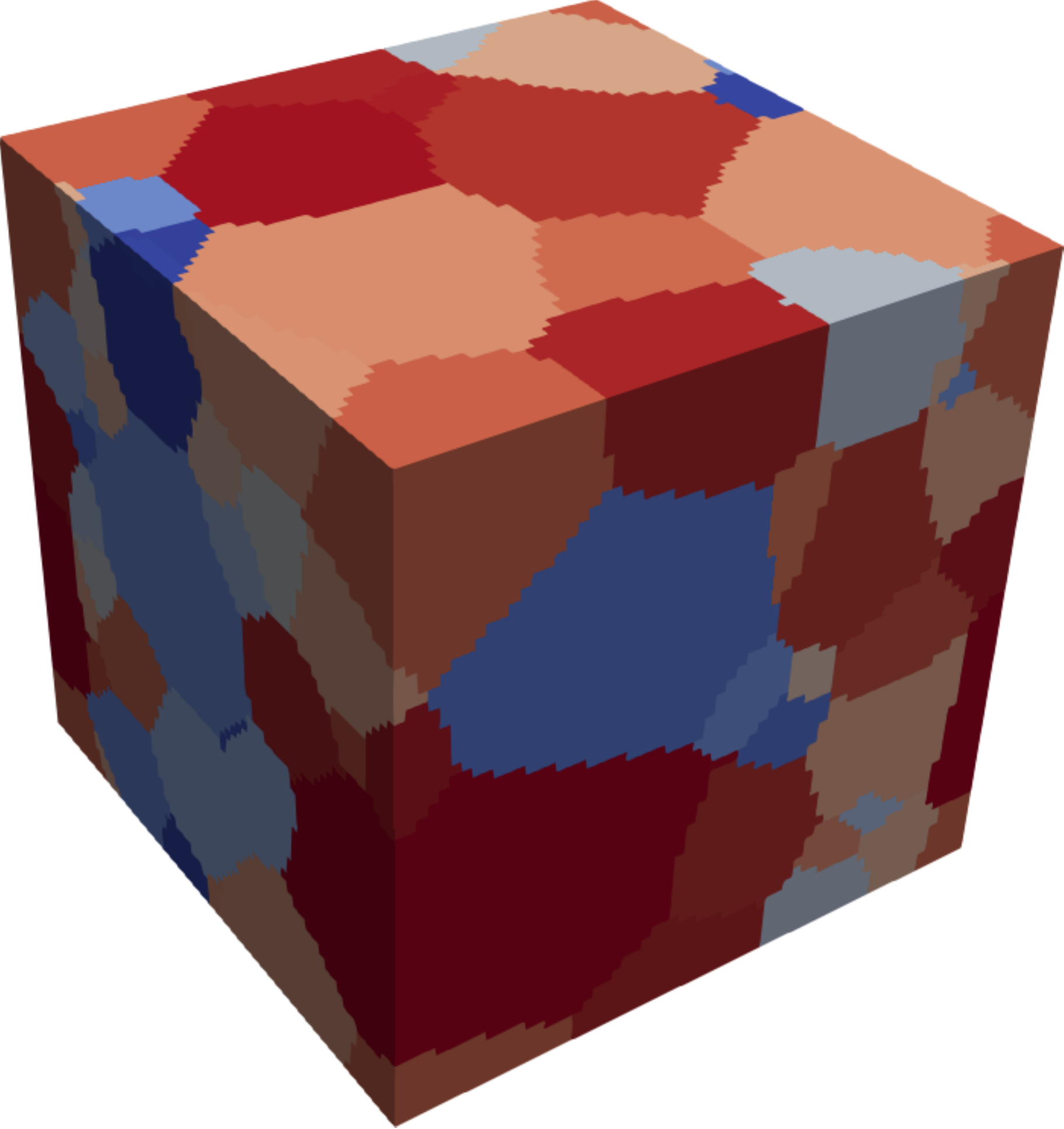}
\caption{\textcolor{black}{Al RVE 5.}}
\end{subfigure}
\caption{\textcolor{black}{An illustrative microstructure ensemble of five Aluminum RVEs.}}
\label{fig:5AlEnsemble}
\end{figure}

The grain size is described by a log-normal distribution as in \cref{eq:GrainLogNormalDistribution} with $\mu_D = 4$ and $\sigma_D = 1.2$.
The crystallographic texture for $\alpha$-Ti is shown in \cref{fig:cropped_fccTexture-45-35-65-Masked-eps-converted-to.pdf}, with the Euler angles of $(\phi_1, \theta, \phi_2) = (45,35,65)$.
Microstructure RVEs of 320$\mu$m$^3$ are considered at multiple mesh resolutions: $64^3, 32^3, 20^3, 16^3, 8^3$.
Uniaxial loading condition is applied \textcolor{black}{with $\dot{F}_{11} = 10^{-3}$s$^{-1}$.}
\textcolor{black}{Figure~\ref{fig:5AlEnsemble} presents an illustrative microstructure ensemble consisting of five Aluminum microstructure RVEs, with the aforementioned grain size and crystallographic texture. In this case study, the quantity of interest is the effective yield stress, calculated by offsetting the effective strain at 0.2\%.}



\subsection{Application of adaptive MIMC for Aluminum}
\label{sec:mimc_results}

\begin{figure}[!htbp]

\colorlet{active}{WildStrawberry}
\colorlet{maximum}{NavyBlue}

\setlength{\figurewidth}{3.5cm}
\setlength{\figureheight}{3.5cm}
\begin{tikzpicture}[%
        trim axis group left
    ]
    \begin{groupplot}[%
            group style={group size={6 by 1}, 
            horizontal sep=5pt},
            width=\figurewidth,
            height=\figureheight,
        ]
        \nextgroupplot[%
            xmin={-0.1},
            ymin={-0.1},
            xticklabel={{}},
            xmajorticks={false},
            yticklabel={{}},
            ymajorticks={false},
            axis line style={{ultra thin, draw opacity=0}},
            axis equal,
            xmax={4.1},
            ymax={4.1},
            width=6cm
        ]
        \drawsquare{0}{0}{white!90!black}
        \drawsquare{0}{1.5}{active}
        \drawsquare{0}{3}{maximum}
        \node[anchor=west] at (axis cs:0.5,0.5) {\quad\scriptsize\strut old index set};
        \node[anchor=west] at (axis cs:0.5,2) {\quad\scriptsize\strut active index set};
        \node[anchor=west] at (axis cs:0.5,3.5) {\quad\scriptsize\strut maximum profit};
        \nextgroupplot[%
            xmin={-0.1},
            ymin={-0.1},
            xticklabel={{}},
            xmajorticks={false},
            yticklabel={{}},
            ymajorticks={false},
            axis line style={{ultra thin, draw opacity=0}},
            axis equal,
            xmax={4.1},
            ymax={4.1},
            xlabel={\scriptsize $L = 0$}
        ]
        \drawsquare{0}{0}{active}
        \nextgroupplot[%
            xmin={-0.1},
            ymin={-0.1},
            xticklabel={{}},
            xmajorticks={false},
            yticklabel={{}},
            ymajorticks={false},
            axis line style={{ultra thin, draw opacity=0}},
            axis equal,
            xmax={4.1},
            ymax={4.1},
            xlabel={\scriptsize $L = 1$}
        ]
        \drawsquare{1}{0}{active}
        \drawsquare{0}{1}{active}
        \drawsquare{0}{0}{maximum}
        \nextgroupplot[%
            xmin={-0.1},
            ymin={-0.1},
            xticklabel={{}},
            xmajorticks={false},
            yticklabel={{}},
            ymajorticks={false},
            axis line style={{ultra thin, draw opacity=0}},
            axis equal,
            xmax={4.1},
            ymax={4.1},
            xlabel={\scriptsize $L = 2$}
        ]
        \drawsquare{0}{0}{white!90!black}
        \drawsquare{1}{0}{active}
        \drawsquare{0}{1}{maximum}
        \nextgroupplot[%
            xmin={-0.1},
            ymin={-0.1},
            xticklabel={{}},
            xmajorticks={false},
            yticklabel={{}},
            ymajorticks={false},
            axis line style={{ultra thin, draw opacity=0}},
            axis equal,
            xmax={4.1},
            ymax={4.1},
            xlabel={\scriptsize $L = 3$}
        ]
        \drawsquare{0}{0}{white!90!black}
        \drawsquare{0}{1}{white!90!black}
        \drawsquare{2}{0}{active}
        \drawsquare{1}{1}{active}
        \drawsquare{1}{0}{maximum}
        \nextgroupplot[%
            xmin={-0.1},
            ymin={-0.1},
            xticklabel={{}},
            xmajorticks={false},
            yticklabel={{}},
            ymajorticks={false},
            axis line style={{ultra thin, draw opacity=0}},
            axis equal,
            xmax={4.1},
            ymax={4.1},
            xlabel={\scriptsize $L = 4$}
        ]
        \drawsquare{0}{0}{white!90!black}
        \drawsquare{1}{0}{white!90!black}
        \drawsquare{0}{1}{white!90!black}
        \drawsquare{1}{1}{active}
        \drawsquare{3}{0}{active}
        \drawsquare{2}{0}{maximum}
    \end{groupplot}
\end{tikzpicture}
    \caption{Shape of the index set constructed during different iterations of the adaptive algorithm. The lower left index corresponds to index $(0, 0)$, i.e., the coarsest mesh size and the low-fidelity phenomenological model. Indices on the horizontal axes indicate a smaller mesh size, while indices in the vertical direction indicate a change in the constitutive model (from phenomenological to density-based). Notice how the high-fidelity model (i.e., using the largest grid and the dislocation-based constitutive model) is never activated.}
    \label{fig:mimc_adaptivity}
\end{figure}
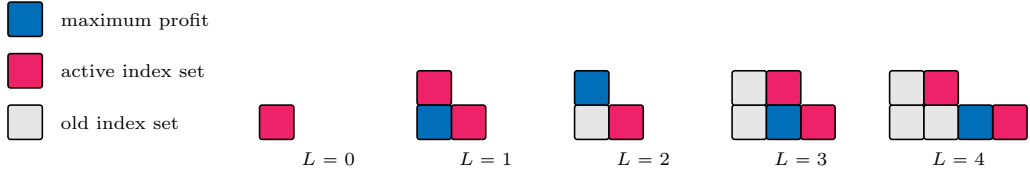

We extended the one-dimensional hierarchy of low-fidelity models based on a varying mesh size by including another constitutive model, based on phenomenological plasticity. Thus, we add another dimension for refinement or coarsening that can be exploited with the MIMC method outlined in \cref{sec:mimc}. 
We run the adaptive MIMC algorithm for the same sequence of decreasing tolerances $\varepsilon^{(k)}$, $k=1, 2, \ldots, 10$, each time using the greedy adaptive algorithm to construct the set of indices to include. In \cref{fig:mimc_adaptivity}, we show the sequence of low-fidelity hierarchies constructed by the adaptive algorithm for the target tolerance $\varepsilon^{\text{target}} = 5$. Notice how the high-fidelity model, i.e., the model using the largest grid and the dislocation-based constitutive model, corresponding to level $\ell = 4$ in the MLMC experiment from \cref{sec:mlmc_results}, is never activated. The final set of indices constructed by the adaptive algorithm contains 6 models with different levels of fidelity.

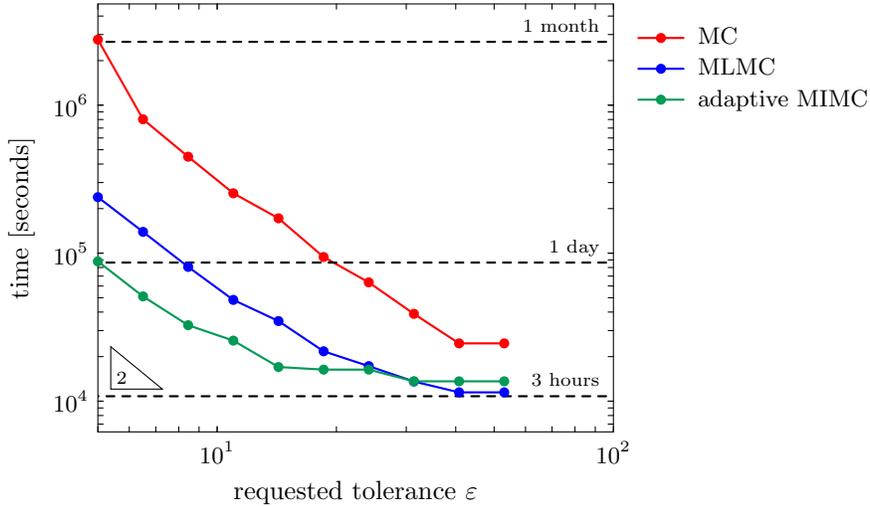
\begin{figure}[!htbp]
    \centering
    \newcommand{\LogLogSlopeTriangle}[5]
{
    \pgfplotsextra
    {
        \pgfkeysgetvalue{/pgfplots/xmin}{\xmin}
        \pgfkeysgetvalue{/pgfplots/xmax}{\xmax}
        \pgfkeysgetvalue{/pgfplots/ymin}{\ymin}
        \pgfkeysgetvalue{/pgfplots/ymax}{\ymax}

        \pgfmathsetmacro{\xArel}{#1}
        \pgfmathsetmacro{\yArel}{#3}
        \pgfmathsetmacro{\xBrel}{#1-#2}
        \pgfmathsetmacro{\yBrel}{\yArel}
        \pgfmathsetmacro{\xCrel}{\xBrel}

        \pgfmathsetmacro{\lnxB}{\xmin*(1-(#1-#2))+\xmax*(#1-#2)}
        \pgfmathsetmacro{\lnxA}{\xmin*(1-#1)+\xmax*#1}
        \pgfmathsetmacro{\lnyA}{\ymin*(1-#3)+\ymax*#3}
        \pgfmathsetmacro{\lnyC}{\lnyA+1.1*#4*(\lnxA-\lnxB)}
        \pgfmathsetmacro{\yCrel}{\lnyC-\ymin)/(\ymax-\ymin)}
        
        \coordinate (A) at (rel axis cs:\xArel,\yArel);
        \coordinate (B) at (rel axis cs:\xBrel,\yBrel);
        \coordinate (C) at (rel axis cs:\xCrel,\yCrel);

        \draw[#5]   (A)--node[pos=0.9,yshift=1ex,xshift=0.5ex] {\scriptsize #4}
                    (B)--
                    (C)-- 
                    cycle;
    }
}

\begin{tikzpicture}
\begin{axis}[ticklabel style={{font=\small}}, major tick length={2pt}, minor tick length={2pt}, every tick/.style={{black, line cap=round}}, axis on top, legend style={{draw=none, font=\small, at={(1.03,0.97)}, anchor=north west, fill=none, legend cell align=left, /tikz/every odd column/.append style={column sep=3pt}}}, xmode={log}, ymode={log}, xlabel={requested tolerance $\varepsilon$}, ylabel={time [seconds]}, xmin=5, xmax=100]
    \addplot[default line, default markers, color={red}]
        table[row sep={\\}]
        {
            \\
            53.022496865000015  24576  \\
            40.78653605000001  24576  \\
            31.374258500000003  38912  \\
            24.134045000000004  63488  \\
            18.564650000000004  94208  \\
            14.280500000000004  172032  \\
            10.985000000000003  253952  \\
            8.450000000000001  448512  \\
            6.5  802816  \\
            5.0  2772992  \\
        }
        ;
    \addlegendentry{MC}
    \addplot[default line, default markers, color={blue}]
        table[row sep={\\}]
        {
            \\
            53.022496865000015  11469.531924794  \\
            40.78653605000001  11469.5529716  \\
            31.374258500000003  13582.21158187  \\
            24.134045000000004  17251.33244051  \\
            18.564650000000004  21732.34314792  \\
            14.280500000000004  34787.371547038  \\
            10.985000000000003  48325.524031527  \\
            8.450000000000001  80784.768314001  \\
            6.5  139334.035372535  \\
            5.0  239055.66527285898  \\
        }
        ;
    \addlegendentry{MLMC}
    \addplot[default line, default markers, color={ForestGreen}]
        table[x expr=500*\thisrowno{0}, row sep={\\}]
        {
            \\
            0.10604499373000004  13621.558665658  \\
            0.08157307210000003  13621.587615821  \\
            0.06274851700000002  13621.611827409999  \\
            0.04826809000000001  16340.256666803998  \\
            0.037129300000000004  16340.292646097998  \\
            0.028561000000000007  17017.313029325996  \\
            0.021970000000000007  25661.297862397994  \\
            0.016900000000000002  32601.750294422993  \\
            0.013000000000000001  51040.162352579995  \\
            0.01  87970.792336891  \\
        }
        ;
    \addlegendentry{adaptive MIMC}
    \LogLogSlopeTriangle{0.125}{0.1}{0.1}{2}{};
    \addplot[default dashed line]
    table[row sep={\\}]
    {
        1 86400 \\
        100 86400\\
    };
    \node[anchor=south east, inner sep=0pt] at (axis cs:92.5, 86400) {\scriptsize\strut 1 day};
    \addplot[default dashed line]
    table[row sep={\\}]
    {
        1 2.6784e6 \\
        100 2.6784e6\\
    };
    \node[anchor=south east, inner sep=0pt] at (axis cs:92.5, 2.6784e6) {\scriptsize\strut 1 month};
    \addplot[default dashed line]
    table[row sep={\\}]
    {
        1 10800 \\
        100 10800\\
    };
    \node[anchor=south east, inner sep=0pt] at (axis cs:92.5, 10800) {\scriptsize\strut 3 hours};
\end{axis}
\end{tikzpicture}
    \caption{Cost of the AMIMC and MLMC methods compared to the (estimated) complexity of the single-level MC method, expressed in terms of the total simulation time in seconds, as a function of the tolerance $\varepsilon$ on the RMSE. For the target RMSE tolerance of $\varepsilon = 5$, the MLMC method is approximately $11.6$ times faster than the MC method, and the adaptive MIMC method is $2.7$ times faster than the MLMC method, resulting in a final speedup of adaptive MIMC over MC of $31.5\times$.}
    \label{fig:mlmc_cost_comparison}
\end{figure}

To investigate the performance of the adaptive MIMC algorithm for our CPFEM application, we plot the cost of the adaptively constructed MIMC estimator, expressed in wall clock time (seconds), as a function of the imposed tolerance $\varepsilon$ in \cref{fig:mlmc_cost_comparison}. Notice how the adaptive MIMC method achieves a requested tolerance $\varepsilon$ in less time, when $\varepsilon$ is small enough. For the two largest tolerances $\varepsilon^{(1)}$ and $\varepsilon^{(2)}$ considered in this experiment, the adaptive MIMC algorithm takes slightly longer compared to the MLMC method, however, it is still much faster than the predicted cost of the corresponding MC simulation for these tolerances. For the target tolerance $\varepsilon^{\text{target}} \textcolor{black}{=5}$, the adaptive MIMC simulation is approximately 2.7 times faster than the corresponding MLMC simulation. This results in an overall speedup of more than $30\times$ compared to the standard, single-fidelity MC simulation. Notice that this gain in computational effort is mainly observed in the prefactor, i.e., the cost-complexity rate of the adaptive MIMC method is still $\calO(\varepsilon^{-2})$. This is in agreement with the results reported in \cite{robbe2016dimension}.

\begin{figure}[!htbp]
\centering
\includegraphics[width=\textwidth]{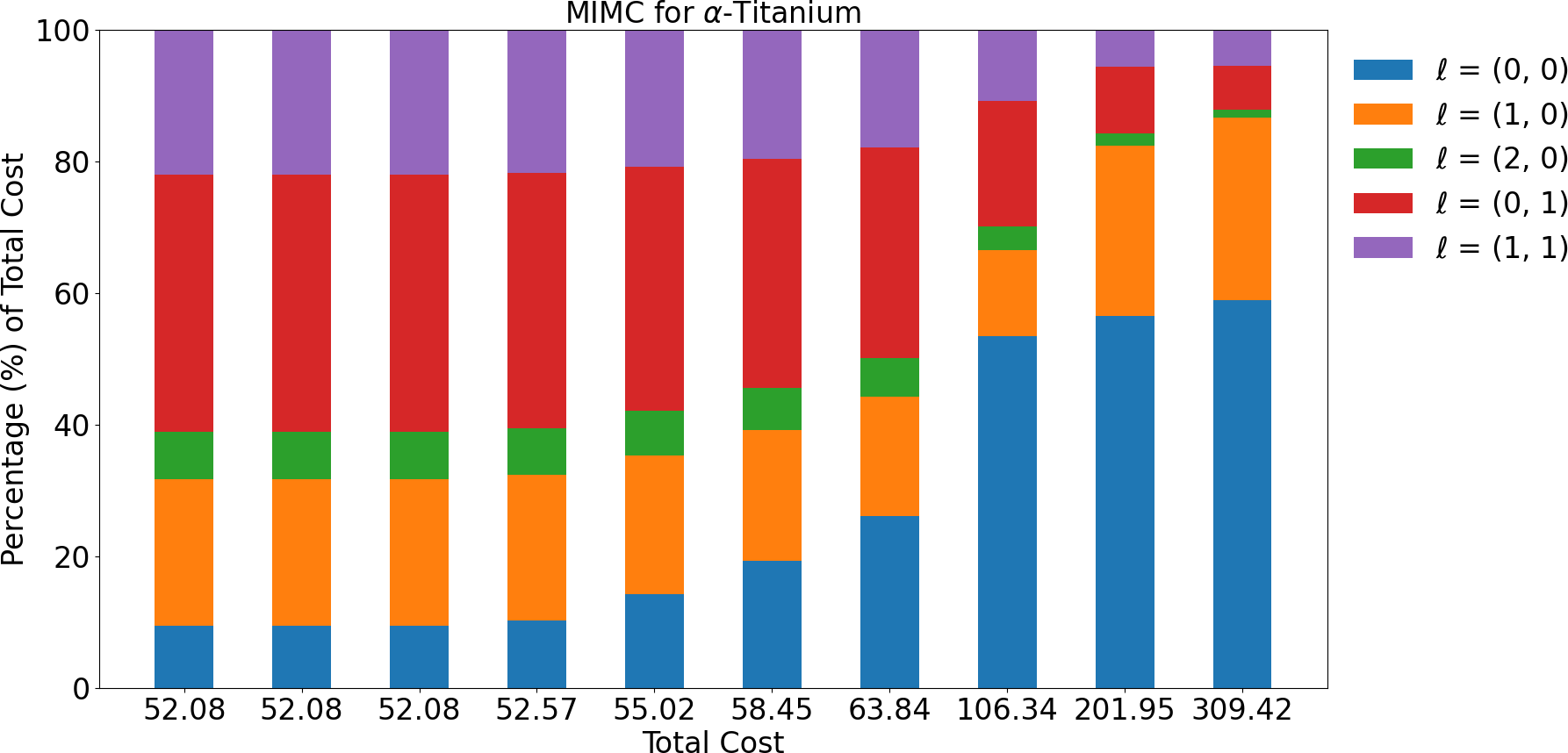}
\caption{\textcolor{black}{The distribution of the total cost across different indices. The first index $\ell_1$ of the fidelity indices corresponds to the discretization of the mesh: $8^3$ ($\ell_1=0$), $16^3$ ($\ell_1=1$), $20^3$ ($\ell_1=2$), $32^3$ ($\ell_1=3$), $64^3$ ($\ell_1=4$). The second index $\ell_2$ of the fidelity indices corresponds to the constitutive model: phenomenological ($\ell_2=0$), dislocation-density-based $(\ell_2=1)$.}}
\label{fig:cropped_mimcCostDecomp}
\end{figure}

\textcolor{black}{Figure~\ref{fig:cropped_mimcCostDecomp} show the distribution of the total cost across different indices, as a function of the tolerance $\varepsilon$ on the RMSE of the MIMC estimator. As the tolerance $\varepsilon$ decreases, the cost of the MIMC estimator increases, a larger fraction of the cost is spent on the coarser indices, and only a minor fraction of the cost is spent on the finer indices. The total cost here is measured in computational time (in seconds) spent on these indices.}

\section{Discussion}
\label{sec:Discussion}



\textcolor{black}{
Microstructures, often represented as pixelized images or voxelized volumes, are high-dimensional and by nature, intrinsically noisy. It is the intrinsic randomness of microstructure, which is the aleatory uncertainty, and the high-dimensional representation that make a practical difference between UQ in process-structure and UQ in structure-property relationship. In the process-structure relationship, the uncertainty is associated with the (high-dimensional) outputs, which UQ literature offers many tools to efficiently solve UQ forward and inverse problems. In the structure-property relationship, the uncertainty is associated with the (high-dimensional) inputs, which essentially requires a sampling approach (such as Monte Carlo estimation for an ensemble of microstructures). 
Arguably, there are more mathematical tools to solve a UQ problem with random outputs than with random inputs. 
Long story short, conceptually, the UQ problems on process-structure and structure-property relationships are of the same mathematical nature; practically, they are not, because microstructures are high-dimensional and intrinsically random. Thus, the structure-property relationship is more computationally complicated than the process-structure relationship. 
}

Because structure-property relationship is more prone to uncertainty compared to process-structure relationship, it is often desirable to impose a UQ framework to quantify both aleatory and epistemic uncertainties.
In the structure-property relationship, the aleatory uncertainty can be understood as the one induced by microstructure, whereas the epistemic uncertainty can be attributed to parametric and model-form error in general.
The proposed framework in this paper is solely dedicated for quantifying aleatory uncertainty problem in the structure-property relationship.

For certain applications where aleatory uncertainty is substantial, such as additive manufacturing~\cite{boyce2017extreme,roach2018born} or small-scale components, the proposed framework can be deployed to further accelerate the UQ process.
Jared et al.~\cite{jared2017additive} pointed out that the aleatory uncertainty associated with homogenized materials properties could be attributed to the fact that additive material properties ``can experience significant local variations, whether controlled or stochastic, based on changes in part geometry and process inputs''.
The framework proposed in this work could be used to further accelerate the qualification of additive manufactured parts, resulting in improvements of materials properties.

Fatigue applications share the same argument with additive manufacturing, where statistical effects play an even more important role when it comes to fatigue life.
The statistical notion of fatigue life has been pointed repeatedly in the literature~\cite{schijve2005statistical,pineau2016failure}, where Gumbel~\cite{gumbel1958statistics,gumbel1963parameters}, Weibull~\cite{schijve1993normal,sakin2008statistical}, and (log-)normal distributions~\cite{schijve1993normal} are often used to model the extreme value statistics.
Utilizing CPFEM to investigate microstructure-sensitiv\textcolor{black}{ity} fatigue has been studied in the last decade or so and is still very much an active field of research, notably by David McDowell and collaborators~\cite{przybyla2010microstructure,przybyla2011simulated,przybyla2012microstructure,stopka2020effects,gu2020prediction,muth2021analysis1,muth2021analysis2,stopka2022simulated}.

We should also \textcolor{black}{point} out that UQ plays a critical role in the material design process by enhancing the reliability of materials design.
The Materials Genome Initiative (MGI) was established with the ultimate goal of significantly accelerating materials design process by modernizing its approaches.
Prior to the MGI, materials were mainly designed based on experiments and theoretical analysis, which typically takes 20-50 years to develop.
To significantly reduce resource-intensive procedure~\cite{arroyave2019systems}, the MGI adopts ICME development with computational materials models and simulations, recently further leveraged by machine learning, to design materials as solving an inverse problem in the process-structure-property relationship.
Due to significant variability attributed to microstructure, it is often desirable to deploy a robust design framework that accounts for uncertainty during the process.
MLMC and MIMC stand out as a significant mathematical UQ tool that completes the UQ task conveniently.
UQ and optimization go hand-in-hand: by considering materials design as an optimization under uncertainty problem, one can robustly design materials by limiting its microstructure-sensitive behaviors.

In this work, we only consider multiple mesh resolutions and constitutive models as an example for MLMC and MIMC, respectively; however, many other parameters could also be utilized.
Notable examples include time-step and orders of numerical integrator in the underlying numerical integrator (for \texttt{DAMASK}, the underlying numerical solver used in this study is \texttt{PETSc}), multiple mesh resolutions, multiple constitutive models, element type in FEM formulation (Feather et al.~\cite{feather2021numerical}) in terms of $hp$-formulation for FEM as described by Blondeel et al.~\cite{blondeel2019h}.
A discussion of $hp$-refinement in \texttt{DAMASK} can also be found in Shanthraj et al.~\cite{shanthraj2019spectral}.

\textcolor{black}{In this paper, as in the vanilla MLMC and MIMC, we restrict the number of quantities of interest to one. While certainly these MLMC and MIMC algorithms can be extended to multiple quantities of interests to capture stress as a function of strain, in the scope of this paper, we solely focus on demonstrating the efficiency of these MLMC and MIMC algorithms over the classical MC algorithm that is still being used in the crystal plasticity finite element literature, while leaving further potential applications for future works.}

Besides MLMC and MIMC, many other methods are also available in the literature; examples include, but are not limited to, multi-fidelity Monte Carlo (MFMC)~\cite{peherstorfer2016optimal,qian2018multifidelity,peherstorfer2019multifidelity}, approximate control variate generalization of MFMC~\cite{gorodetsky2020generalized}, multigrid (quasi-) Monte Carlo~\cite{robbe2019recycling,robbe2021enhanced}, multi-index stochastic collocation~\cite{haji2016multi2,haji2016multi3,jakeman2020adaptive}.
It should be noted that machine learning predictions can play a role of low-fidelity with low computational cost and relatively high error, as demonstrated in one of our previous studies~\cite{tran2020multi}.
It is worth mentioning that sometimes it is difficult to assign the fidelity of different constitutive models, compared to the mesh resolution. However, in the case of phenomenological versus dislocation-density-based models, generally speaking, dislocation-density-models are more accurate in predicting \textcolor{black}{homogenized} behaviors.



\section{Conclusion}
\label{sec:Conclusion}

In this work, we proposed a generic MLMC and MIMC for quantifying uncertainty in structure-property relationship through a multi-fidelity framework.
The proposed approach is based on applied mathematical work of MLMC and MIMC, which views the microstructure RVE as a stochastic sample, where multiple fidelity of mesh resolutions, constitutive models, and numerical solvers are applied on the microstructure RVE to map from the materials microstructure space to the materials property space.
Our approach is demonstrated with two case studies.
In the first case study, we demonstrated the efficiency of MLMC, where multiple mesh resolutions are considered.
In the second case study, we demonstrated the efficiency of MIMC, where multiple mesh resolutions and multiple constitutive models are considered simultaneously.
\textcolor{black}{In both case, the effective yield stress is the quantity of interest. }
Compared to the classical MC method, which utilizes the microstructure ensemble approach, in the first case study, MLMC offers a 12$\times$ speedup factor; in the second case study, MIMC offers a 2.7$\times$ speedup over MLMC, whereas MLMC offers a 11.6$\times$ speedup compared to MC.
We conclude that the multi-fidelity UQ methodology proposed in this paper offers a significant reduction in computational cost for quantifying uncertainty associated with microstructure variations in the context of CPFEM. 

\section*{Acknowledgment}

The views expressed in the article do not necessarily represent the views of the U.S. Department of Energy or the United States Government. Sandia National Laboratories is a multimission laboratory managed and operated by National Technology and Engineering Solutions of Sandia, LLC., a wholly owned subsidiary of Honeywell International, Inc., for the U.S. Department of Energy's National Nuclear Security Administration under contract DE-NA-0003525.

\bibliographystyle{elsarticle-num}
\bibliography{lib}

\end{document}